\documentclass[prx,aps,notitlepage,nofootinbib,superscriptaddress,twocolumn,longbibliography,10pt]{revtex4-2}
\usepackage[normalem]{ulem}
\usepackage{svg}
\usepackage{amsmath}
\usepackage{amsthm, amssymb}
\usepackage[colorlinks=true,citecolor=blue,linkcolor=blue]{hyperref}
\usepackage{graphicx}
\usepackage{dcolumn}
\usepackage{bm}
\usepackage{color}
\usepackage{tabularx}
\usepackage{comment}
\usepackage{mathtools}
\usepackage{tikz-cd}
\usepackage{adjustbox}
\usepackage{braket}
\usepackage{dsfont}
\usepackage{subcaption}
\usepackage{ragged2e}
\usepackage[linesnumbered,ruled,vlined]{algorithm2e}
\SetKwComment{Comment}{/* }{ */}
\SetKwInput{Input}{Input}
\SetKwInput{Return}{Return}
\SetKwFor{For}{for}{do}{endfor}
\DontPrintSemicolon
\usepackage{tikz}
\usetikzlibrary{quantikz2}

\newcommand{\Tr}{\text{Tr}}
\newcommand{\sket}[1]{\ensuremath{|#1\rangle\!\rangle}}
\newcommand{\sbra}[1]{\ensuremath{\langle\!\langle #1|}}
\begin{document}

\title{Quantum resource localizability transitions in deep thermalization}

\author{Xiaozhou Feng${}^\S$}
\email{xzfengphys@gmail.com}
\affiliation{Department of Physics, The University of Texas at Austin, Austin, Texas 78712, USA}

\author{Chang Liu${}^\S$}
\affiliation{Department of Physics, National University of Singapore, Singapore 117551}

\author{Zihan Cheng}
\affiliation{Department of Physics, National University of Singapore, Singapore 117551}

\author{Wen Wei Ho}
\altaffiliation{\href{mailto:wenweiho@nus.edu.sg}{wenweiho@nus.edu.sg}}
\affiliation{Department of Physics, National University of Singapore, Singapore 117551}
\affiliation{Centre for Quantum Technologies, National University of Singapore,   Singapore 117543}

\author{Matteo Ippoliti}
\email{ippoliti@utexas.edu}
\affiliation{Department of Physics, The University of Texas at Austin, Austin, Texas 78712, USA}

\date{\today}
\begin{abstract}
We investigate the effect of quantum resource constraints on deep thermalization, the emergence of universal local wavefunction distributions from partial measurements of a quantum many-body state. Since quantum resources, such as non-stabilizerness (magic), coherence, asymmetry, imaginarity, and non-Gaussianity, are essential for  quantum information processing, constraints on their global abundance can fundamentally reshape these emergent wavefunction distributions. To address this question, we develop a unified framework for deep thermalization within general quantum resource theories (QRTs).
Our central result is that QRTs fall into two distinct classes: 
``smoothly localizable'' (SL) QRTs, where the resource content of the local post-measurement states changes continuously with the global resource density, set by the initial state and measurement basis, yielding a family of continuously tunable 
wavefunction distributions in deep thermalization; 
and ``threshold localizable'' (TL) QRTs, where the local resource content jumps discontinuously from minimal to near-maximal past a critical global resource threshold, producing a sharp phase transition between a resourceless, ``deep-ergodicity breaking'' distribution and a resourceful, maximally random one. 
We trace the SL-TL dichotomy to an information-theoretic mechanism we call block sharpening: by viewing each QRT as a form of coherence between  blocks in Hilbert space, we show that the local resource content depends on the measurement's ability  to collapse an initial superposition into a single, resourceless block. 
Our theory is analytically tractable and quantitatively predicts the phase boundaries across all studied QRTs, which we validate via extensive numerical simulations. Finally, we highlight two consequences of our framework: the discovery of a novel magic transition in zero-rate quantum error-correcting codes---previously believed to occur only at finite rates---and new implications for quantum resource certification protocols utilizing post-measurement state ensembles.
\end{abstract}

\maketitle

\def\thefootnote{$\S$}\footnotetext{These authors contributed equally to this work}\def\thefootnote{\arabic{footnote}}

\section{Introduction}
\label{sec:intro}

Experimental advances in programmable quantum systems allow for increasingly precise control of quantum dynamics and the ability to read out the global state of a system in fine-grained detail. Such capabilities have enabled exciting explorations of quantum many-body physics in new settings, featuring the controlled interplay of intrinsic scrambling dynamics and extrinsic dissipation or measurement by an outside environment or agent~\cite{bluvstein2021controlling,google2023measurement,iqbal2024nonabeliantopo,ibm2025nishimori,bloch2026swssb}. 

Deep thermalization is a novel physical phenomenon that has been recently uncovered from this interplay~\cite{choi2023preparing,cotler2023emergent,ho2022exact,claeys2022emergent,ippoliti2023dynamical,ippoliti2022solvable,chan2024projected,mark2024maximum, 2025arXiv250621061Y}. 
Here, one studies the behavior of a local subsystem of an isolated quantum many-body system {\it conditioned} on classical knowledge of the complementary region (commonly referred to as the environment or bath), obtained via measurements. 
In generic late-time quantum many-body dynamics, the collection of the different local post-measurement states---called the projected ensemble (PE)---was found to approach universal, maximally-random distributions over the Hilbert space, such as the Haar and Scrooge ensembles. Deep thermalization represents a finer-grained probe of quantum equilibration: it concerns fluctuations of a local subsystem {\it conditional on the bath}, in contrast to regular thermalization where the bath is ignored (i.e., averaged over)~\cite{srednicki1994chaos,rigol2008thermalization,nandkishore2015many,kaufman2016quantum,d2016quantum,abanin2019colloquium}.

\begin{figure*}
    \centering
    \includegraphics[width=0.95\linewidth]{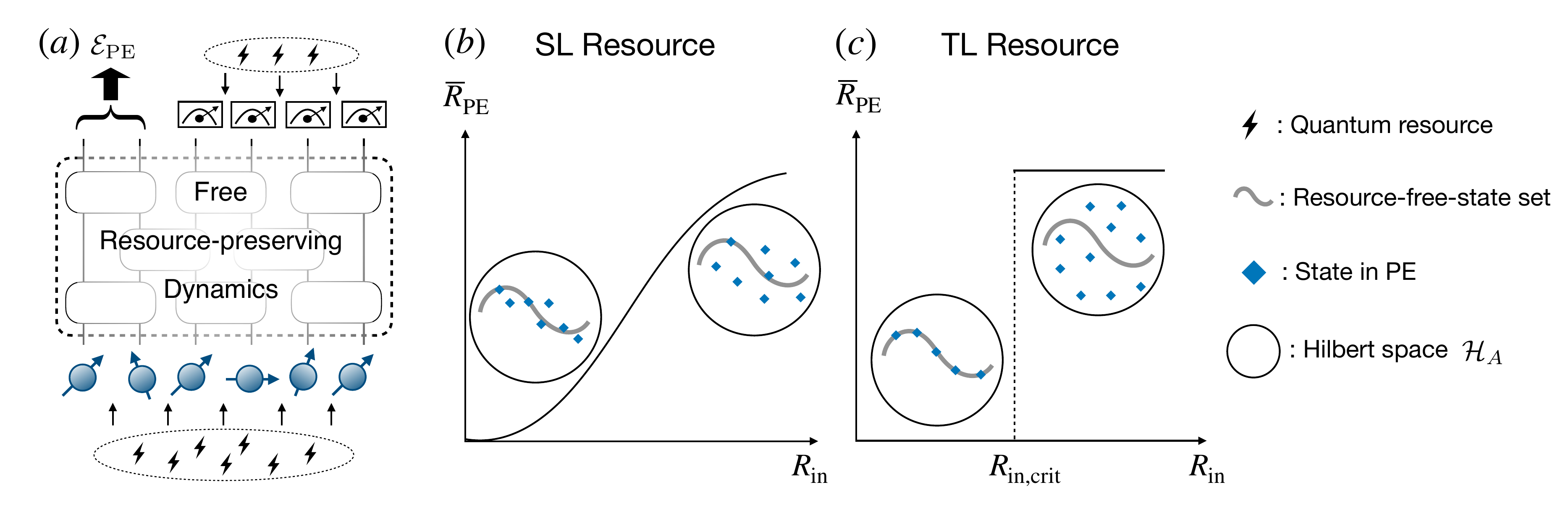}
\caption{\justifying 
(a) Setting of our work. We study deep thermalization through the lens of quantum resource theories (QRTs).
We consider systems where a quantum resource (depicted by lightning symbols) is injected through the initial state and the final measurements, and is scrambled by free (i.e., resource-preserving) unitary dynamics. 
Our work identifies two fundamentally distinct classes of QRTs: (b) ``smoothly-localizable'' (SL) QRTs, where the resource content $\overline{R}_\text{PE}$ of the PE  varies smoothly with the globally injected resource ${R}_\text{in}$,
and (c) ``threshold-localizable'' (TL) QRTs, where the resource content of the PE $\overline{R}_\text{PE}$ undergoes a sharp transition from being zero to being near maximal, past a certain  critical global resource threshold ${R}_\text{in,crit}$. 
The degree of resource in the PE $\overline{R}_\text{PE}$ is a measure of the ergodicity of its distribution over the Hilbert space $\mathcal{H}_A$ (depicted by clustering of diamonds within circles). 
}
\label{Fig:1}
\end{figure*}

Our understanding of deep thermalization is based on maximum entropy principles rooted in quantum information theory~\cite{mark2024maximum, liu2024deep, chang2025deep}. 
These principles state that generic complex quantum dynamics tend to {\it optimally} hide classical information about the bath into local quantum states, representing a strong manifestation of ergodicity in nature. 
Maximum-entropy principles correctly predict the precise form of the PE across a wide range of physical scenarios, from systems with conserved quantities like energy or charge~\cite{mark2024maximum, chang2025deep, mok2026nature},
to systems of non-interacting bosons or fermions~\cite{bhore2023deep,lucas2023generalized,liu2024deep,bejan2025matchgate}.
In all these cases, the PE takes on the maximally entropic form compatible with the relevant constraints. 

Despite their apparent success across diverse settings, maximum-entropy principles can fail.
Ref.~\cite{liu2025coherence} showed that quantum dynamics constrained to preserve global \textit{coherence} (the ``amount'' of superposition over computational basis states) can exhibit a sharp \textit{deep ergodicity breaking transition}: upon tuning the density of coherence in the system, the PE changes sharply from a minimally-entropic form, made of random computational basis states, to the maximally-entropic Haar ensemble. 
 
This discovery represents a genuine exception to the maximum entropy principles of deep thermalization: unlike previously studied constraints, which simply alter the form of the maximum-entropy distributions, 
the constraint on coherence 
can produce a fundamentally distinct limiting PE, which evades the paradigm altogether. This raises the central question behind this work: for which classes of quantum dynamics
do the maximum-entropy principles of deep thermalization hold, and conversely, when do they break down? Equivalently, what physical mechanisms can obstruct deep thermalization, and how can we understand the ensuing deep-ergodicity-breaking transitions? These are fundamental questions about how physical restrictions on dynamics give rise to distinct universality classes in deep thermalization, akin to the study of how symmetries and conservation laws shape phases of matter in many-body physics.

In this work, we address these questions by adopting a quantum information lens to view general constraints in  dynamics as the imposition of limitations on their quantum information processing power, which are hence captured by the framework of {\it quantum resource theories} (QRTs)~\cite{chitambar2019quantum}.
These encompass both ``conventional'' constraints arising in many-body physics, such as global symmetries and conservation laws or the absence of interactions in fermionic and bosonic systems (Gaussianity)~\cite{PhysRevA.65.032325,Lloyd_Quantum_1999,Bartlett_Efficient_2002,3yx4-1j27}, as well as restrictions on the ability to generate information-theoretic features like coherence~\cite{baumgratz2014quantifying,PhysRevLett.116.120404, RevModPhys.89.041003,Kelly_Coherence_2023} and non-stabilizerness (also known as magic) ~\cite{bravyi2005universal,Leone_quantumchaos_2021,Liu_Manybody_2022,tirrito2024anticoncentration,dowling2025magic,Bravyi_Improved_2016,Howard_application_2017,Liu_Manybody_2022,Leone_Stabilizer_2022,turkeshi2025magic}.
We develop a unifying theoretical framework for deep thermalization under dynamical constraints described by a large family of QRTs, and show that the nature of the PE is fundamentally altered by the type of QRT and the amount of quantum resources present: the emergent randomness in the PE depends crucially on the ability of measurements to transfer the initial resource---which is preserved by the dynamics, although scrambled in space---from the global system into a local subsystem, a notion called ``resource localization'' in quantum information theory~\cite{verstraete2004entanglement,PhysRevA.71.042306,HAMMA2021127264}; see Fig.~\ref{Fig:1}(a).

We map the question of resource localizability to a learning or inference problem that we term ``block sharpening''. 
Associating each QRT to a form of coherence between subspaces (``blocks'') of free states, an initial resourceful state is characterized by superposition between distinct blocks. Measurements on the bath may then collapse the superposition down to a single, resourceless block, or may fail to uniquely pin down a block, leaving behind a resourceful PE state. The outcome depends on the structure of the QRT's free-state blocks and on the learning power of measurements. 
This picture generalizes the idea of charge sharpening studied in measurement-induced phase transitions (MIPTs)~\cite{PhysRevX.12.041002,Li_QuantumZeno_2018,Barratt_Transitions_2022,Agrawal_Observing_2024,Feng_chargespin_2025,Choi_Quantum_2020,Gullans_Dynamical_2020,Fisher_random_2022,Potter_Entanglement_2022}, which likewise corresponds to a learning or inference problem. 

Our mapping to block sharpening reveals that QRTs fall into two distinct classes, which we call ``smoothly localizable'' (SL) and ``threshold-localizable'' (TL), see Figs.~\ref{Fig:1}(b) and (c). In the former, which includes resources like asymmetry, imaginarity, and non-Gaussianity, 
the block sharpening problem is generically in a ``fuzzy'' phase where the inference fails, leading to a family of resourceful wavefunction distributions that change smoothly with the density of resource in the system. 
Conversely, in the TL class, which includes resources like coherence and magic, the inference problem undergoes a sharp transition from being under- to over-determined at a critical resource threshold.
This results in a deep thermalization transition between a resourceless, minimally-random distribution to a resourceful, maximally-random distribution, i.e., a resource-localizability transition. 
This greatly generalizes the phenomenology of deep thermalization transitions, going beyond the coherence-induced transition in random permutation dynamics~\cite{liu2025coherence}, to general resource-induced deep thermalization transitions realized in wider classes of constrained quantum dynamics.

Our results substantially enrich the landscape of deep thermalization and deep ergodicity breaking: they reveal novel continuous families of ergodic wavefunction distributions (for SL QRTs) and novel exceptions to the maximum-entropy principle induced by resource constraints (for TL QRTs). 
Furthermore, they conceptually unify diverse physical constraints under the common information-theoretic mechanism of block sharpening, revealing unexpected links between seemingly unrelated phenomena. 
As an important example, our theoretical framework sheds new light on the nature of recently studied magic phase transitions in encoding-decoding dynamics~\cite{Niroula_PT_2024,sierant2026theorymagicphasetransitions}: we quantitatively explain prior observations and predict a novel magic transition in zero-rate quantum error-correcting codes, previously believed to occur only at finite rates. 

Beyond its fundamental interest, our work  also carries practical implications for quantum information science, notably in quantum state certification~\cite{Huang_2025_certifying}. Recent work~\cite{du2025certifying} has proposed an efficient method to certify the presence of a quantum resource in a global system by probing the amount of resource contained within the projected ensemble of a local subsystem. The key to the success of the protocol is precisely resource localization.
However, as our results show, such a property is only guaranteed for SL resources. For TL resources, the localizability transition predicted by our theory obstructs certification below the threshold, signaling a fundamental limitation of the protocol (and more generally, of any protocol using post-measurement states as a proxy to detect properties of the global state~\cite{varikuti2025resources}).

The remainder of the paper is structured as follows. 
In Sec.~\ref{sec:setup} we review the phenomenon of deep thermalization, lay out the essential formalism for QRTs studied in this work, and introduce the relevant models of dynamics with quantum resource constraints.
Sec.~\ref{sec:generaltheory} presents our general theory of resource localizability based on ``block sharpening'' and the ensuing classification of TL and SL resources. In Sec.~\ref{sec:transitions} and \ref{sec:crossover}, we investigate in detail concrete examples of TL and SL resources respectively. We separately investigate the QRTs of non-stabilizerness and non-Gaussianity, which are most naturally discussed in operator space rather than state space, in Sec.~\ref{sec:heisenberg}. 
Lastly, in Sec.~\ref{sec:discuss} we conclude by summarizing our results, discussing their implications, and highlighting open questions for future research.

\section{Overview of deep thermalization, 
quantum resource theories, 
and models}
\label{sec:setup}

We first briefly review the phenomenon of deep thermalization, the emergence of universal wavefunction distributions in the projected ensemble (PE) at late times in dynamics~\cite{cotler2023emergent,choi2023preparing,ho2022exact,claeys2022emergent,ippoliti2023dynamical,ippoliti2022solvable,mark2024maximum, liu2024deep, chang2025deep, mok2026nature,bhore2023deep,lucas2023generalized,bejan2025matchgate,chakraborty2025fast,zhang2025holographic,yu2025mixed,sherry2025mixed, dcdc-pxn3}.
Knowledgeable readers may skip to Sec.~\ref{sec:review_resources}, where we introduce the new ingredient considered in this work: the imposition of physical constraints on the dynamics and their interpretation in the language of quantum resource theories (QRTs).
There, we also provide a high-level overview of the structure of several example QRTs and explain how they can affect the resulting universality in deep thermalization through the notion of resource localizability.
Finally, in Sec.~\ref{sec:review_models} we introduce the concrete models of dynamics studied in this work. 

\subsection{Background: Deep thermalization \label{sec:review_pe}} 

Consider a quantum many-body state $\ket{\Psi}$ on a bipartite system $AB$ of qubits (or spins). We will typically be interested in the situation where $A$ is a small  local subsystem of fixed size, while the complementary subsystem $B$ is taken to be large, and thus serves as a ``bath'' allowing $A$ to equilibrate. Suppose that one measures $B$ projectively in a basis $\{\ket{\Phi_\nu}_B\}_\nu$, and obtains the measurement outcome $\nu$. For a local product basis, this is a string of binary local outcomes which denotes a particular classical configuration of the bath in the direction of the measurement basis. Then, the state on $A$ will be collapsed to the pure state  $\ket{\psi(\nu)}_A = (\mathbb{I}_A \otimes \bra{\Phi_\nu}_B)\ket{\Psi}/\sqrt{p(\nu)}$, which occurs with Born probability $p_\nu = \|(\mathbb{I}_A \otimes \bra{\Phi_\nu}_B)\ket{\Psi}\|^2$. 
The collection of all possible post-measurement states $\ket{\psi(\nu)}_A$ (also called conditional or projected states) with their respective probabilities defines an ensemble of quantum states
\begin{equation}
    \mathcal{E} = \left\{ (p_\nu, \ket{\psi(\nu)}_A) \right\}_\nu,
\end{equation}
which is called the {\it projected ensemble} (PE)~\cite{choi2023preparing,cotler2023emergent}.

The PE defines a probability distribution on the Hilbert space of subsystem $A$. It physically captures (quantum) information of the local subsystem $A$ {\it correlated} with (classical) information of the bath $B$. This information includes, in particular, the average state on $A$: this is obtained by simply ignoring the classical information of $B$, i.e., taking the first moment of the distribution
\begin{align}
    \rho^{(1)}_\mathcal{E} 
    & := \mathbb{E}_{\psi \sim \mathcal{E}} \left[ \ket{\psi}\bra{\psi} \right]
    = \sum_\nu p_\nu \ket{\psi(\nu)}_A\bra{\psi(\nu)}_A \nonumber \\
    & = {\rm Tr}_B\left[ \ket{\Psi}\bra{\Psi}\right] 
    \equiv \rho_A,
\end{align}
which we see is nothing but the familiar reduced density matrix $\rho_A$ on $A$. 
Conversely, fluctuations on $A$ conditioned on the state of $B$ are captured by higher statistical moments of the distribution, 
\begin{align}
\rho^{(k)}_\mathcal{E} 
    & := \mathbb{E}_{\psi \sim \mathcal{E}} \left[ \ket{\psi}\bra{\psi}^{\otimes k}\right] \nonumber \\
    & = \sum_\nu p_\nu (\ket{\psi(\nu)}_A\bra{\psi(\nu)}_A)^{\otimes k},
\end{align}
for integer $k \geq 2$. 
This conditional information present in the PE goes strictly beyond the information contained in the average state $\rho_A$. 

Deep thermalization is the emergence of universality in the PE for many-body states at late times in dynamics:
\begin{align}
|\Psi_t \rangle = U_t |\Psi_0\rangle,
\label{eq:initial_state}
\end{align}
where $U_t$ is the unitary time evolution operator arising from e.g. local Hamiltonian or unitary circuit dynamics for time $t$~\cite{choi2023preparing,cotler2023emergent,ho2022exact,ippoliti2023dynamical,ippoliti2022solvable}.
Concretely, it was found that under generic complex quantum many-body dynamics, the PE tends to limiting forms (in the thermodynamic limit and at late-time limit, taken in that order) which are 
(i) independent of precise microscopic details, such as the exact nature of the initial state, and 
(ii) characterized by the property of being ``maximally random'', i.e., maximizing a notion of entropy suitable for quantum state ensembles, see Refs.~\cite{mark2024maximum,liu2024deep, chang2025deep}. 
Prominent examples of these maximum-entropy ensembles include the (unitarily-invariant) Haar ensemble\footnote{Here $\mathcal{E}_\text{Haar} = \{ d\psi, |\psi\rangle_A\}$ with $d\psi$ the Haar or Fubini-Study measure on the Hilbert space $\mathcal{H}_A$.} $\mathcal{E}_\text{Haar}$
for dynamics $U$ without explicit conservation laws or at infinite temperature,
as well as the so-called ``generalized Scrooge ensemble'' (GSE) $\mathcal{E}_\text{GSE}$ in the presence of conservation laws such as energy or charge, see Refs.~\cite{mark2024maximum,chang2025deep,mcginley2025scrooge, mok2026nature} for details.  The Scrooge ensemble $\mathcal{E}_\text{Scrooge}$ (derived first in quantum information theory~\cite{jozsa1994lower} and then in quantum statistical mechanics under the name of ``Gaussian-adjusted projected (GAP)'' ensemble \cite{Goldstein2006,Goldstein_Universal_2016}) can be understood as a continuous deformation of the Haar ensemble to account for the presence of different values of conserved quantities\footnote{Specifically, the Scrooge ensemble is the maximally information-stingy unraveling of a density matrix $\rho$ into pure states, defined as $\mathcal{E}_\text{Scrooge}[\rho] = \{D_A \langle \psi| \rho |\psi \rangle d\psi,  \frac{ \rho^{1/2} | \psi \rangle }{ \langle \psi| \rho|\psi\rangle^{1/2}} \} $, with $D_A$ the Hilbert space dimension. This corresponds to a distortion of the Haar measure $d\psi$ and a deformation of the sampled wavefunction $\ket{\psi}$, both of which effectively ``bias'' the Haar measure toward directions in Hilbert space where $\rho$ has larger eigenvalues. This could mean, e.g., a bias toward low-energy states if $\rho$ is a thermal density matrix.}.
It has the special property of having the least accessible information among all ensembles with the same mean~\cite{jozsa1994lower}.
The GSE is a statistical mixture of Scrooge ensembles that account for different levels of knowledge of the conserved quantity obtained by different measurement outcomes on the bath.

Finally, we remark on the relationship and differences between conventional thermalization and deep thermalization. 
The former refers to the equilibration of local observable expectation values, described by the reduced density matrix $\rho_A$. This is understood to equilibrate to a universal form (the Gibbs state) at late times, which is dictated by a maximum entropy principle and controlled by few parameters (e.g., temperature, chemical potential) set by the values of conserved quantities in the system~\cite{PhysRev.106.620, srednicki1994chaos,rigol2008thermalization,nandkishore2015many,kaufman2016quantum,d2016quantum,abanin2019colloquium}. 
Since the reduced density matrix $\rho_A$ is the first moment of the PE, $\rho_{\mathcal E}^{(1)}$, deep thermalization poses a strictly more demanding condition---equilibration of the whole wavefunction ensemble, not just of its first moment. This requires a generalization of the maximum entropy principle and yields a more refined notion of equilibration in many-body systems, sensitive not just to local features but also to global patterns of conditional information in the many-body state.

\subsection{Quantum resources and resource localizability \label{sec:review_resources}}

The central question we ask in this work is: what are the different universality classes that can arise in deep thermalization, due to different classes of dynamics? 
As reviewed above, physical constraints from conserved quantities like energy and charge 
lead to a continuous family of limiting wavefunction distributions for the PE---the generalized Scrooge ensemble (GSE). Still, this family is smoothly connected to the Haar ensemble, and hence falls within the same maximum entropy paradigm. However, recent work~\cite{liu2025coherence} has in contrast found that a qualitatively different outcome is possible from a different kind of constraint: dynamics $U_t$ restricted to preserve quantum coherence (the ``amount'' of superposition over computational basis states) leads to sharp transitions in the PE between a maximum-entropy ensemble and a minimum-entropy one, representing a genuine breakdown of deep thermalization.

In this work, we address this question by describing a wide variety of physical constraints in the unified language of {\it quantum resource theories} (QRTs)---an established framework in quantum information theory that formalizes operational restrictions on quantum dynamics~\cite{chitambar2019quantum}. Indeed, quantum coherence~\cite{baumgratz2014quantifying,PhysRevLett.116.120404, RevModPhys.89.041003,Kelly_Coherence_2023} is a necessary resource for quantum computing, while conservation of energy and charge can be cast as constraints on the resources of athermality~\cite{brandao_2013_qrttherm,gour_2015_qrtthermodynamics} and asymmetry~\cite{marvian_extending_2014} respectively. Other well-known examples of QRTs include non-stabilizerness (also known as magic), imaginarity, and non-Gaussianity. We note that the landscape of QRTs is vast and we do not address all possible QRTs in this work. Instead, we will focus on a large subclass of QRTs that have mathematically simple structure, yet are highly relevant to many-body physics: 
these are resources that can be cast as {\it subspace coherence}\footnote{
Some prominent examples of QRTs, like quantum entanglement, lie outside this category. 
Our theory still applies in some of these cases, specifically for QRTs that are ``refinements'' of a suitable subspace coherence QRT: i.e., resources that require subspace coherence as a necessary but insufficient ingredient. Notable examples of this kind are non-stabilizerness and non-Gaussianity, discussed in Sec.~\ref{sec:heisenberg}.  
}~\cite{aberg2006quantifyingsuperposition,mani_2024_subspacecoherence}, for a certain subspace decomposition of the state or operator Hilbert space. 

Let us now describe the general structure of subspace coherence QRTs, 
and explain how they may affect the phenomenon of deep thermalization. A subspace coherence resource is defined by a particular block decomposition of the  Hilbert space $\mathcal{H}$ (taken to be either the vector space of pure states or the vector space of operators) as
\begin{equation}
    \mathcal{H} = \bigoplus_i \mathcal{B}_i,
    \label{eq:block_decomposition}
\end{equation}
such that elements within each block $\mathcal{B}_i$ are regarded as ``free''.
The resource content of a Hilbert space element is then quantified by the amount of superposition between distinct blocks, which is a generalized form of coherence. 
This can be quantified by a function denoted generally by $R$ called a resource monotone; the precise expression for $R$ depends on the resource in question, and we will explicitly state when we consider particular resources in  Sec.~\ref{sec:transitions}, \ref{sec:crossover}, and \ref{sec:heisenberg}. A QRT is further defined by a set of ``free operations'', which are quantum operations acting on the Hilbert space $\mathcal{H}$ that never increase the resource content $R$ of the system. 
Non-free operations are deemed resourceful. 
We will also assume that the subspace coherence QRT in question can be defined consistently for any system size $N$, i.e., that a decomposition like Eq.~\eqref{eq:block_decomposition} exists for each system size.
This in particular ensures that the Hilbert space $\mathcal{H}_A$ of the local subsystem hosting the PE also admits a valid instance of the QRT, so that we can meaningfully define the resource both in the global many-body state and in the PE states. 
Compatibility of the subspace decompositions for different system sizes is discussed in more detail in Appendix~\ref{app:blocks}. 

By definition, a dynamical constraint represented by a QRT means that the unitary dynamics $U_t$ in Eq.~\eqref{eq:initial_state} (whose deep thermalization we aim to study) is free with respect to the given QRT. 
Free unitaries form a subgroup $\mathcal{G}_R$ of the full unitary group on the many-body Hilbert space, which in terms of the block decomposition in Eq.~\eqref{eq:block_decomposition} is generated by 
(i) {\it internal} unitary transformations of each block $\mathcal{B}_i$ and
(ii) {\it permutations} between blocks of the same dimension. These operations cannot create (nor destroy\footnote{In general, free operations are defined only by their property of being resource {\it non-increasing}, but since we restrict to free operations which are unitary and hence reversible, they necessarily do not change the resource monotone $R$.}) a superposition between subspaces, which limits their computational power. 
Since they cannot create resource, the only resource ever present in the system is due to either the initial state $|\Psi_0\rangle$, or the choice of measurement basis $\{ |\Phi_\nu\rangle_B\}_\nu$. Both can be thought of as ``injecting'' resource into the global system, which is then scrambled by the free dynamics $U_t$. 

With this framework, the central question posed in this work then becomes the following quantitative one:
how does the average resource content 
\begin{equation}
\overline{R}
=\mathbb{E}_{\ket{\psi}\sim\mathcal{E}}
\!\left[R(\ket{\psi})\right]
\label{eqn:local_resource}
\end{equation}
of the PE $\mathcal{E}$ on the local subsystem $A$ depend on the amount of resource $R_\text{in} = R(|\Psi\rangle) + R(|\Phi_\nu\rangle_B)$ injected in the global system $AB$? 
Fig.~\ref{Fig:1}(a) illustrates our set-up. 
As an important limiting case, it is clear that if both the initial state and measurement basis are free (i.e., resourceless), such that $R_\text{in} = 0$, then the resource content of the PE is necessarily vanishing too: $\overline{R} = 0$.
This implies that the PE's distribution cannot possibly cover the entire Hilbert space $\mathcal{H}_A$ of $A$: the states must be confined with the blocks of the relevant subspace-coherence QRT on $A$. The nontrivial question is thus about the evolution of the PE's resource $\overline{R}$ as the injected global resource $R_\text{in}$ increases: whether it evolves smoothly or exhibits a threshold behavior (i.e., a discontinuous jump). 
In other words, this is a question about {\it resource localizability}---a concept from quantum information theory describing the ability of measurements on a multi-partite state to concentrate resource into an unmeasured subsystem~\cite{verstraete2004entanglement,PhysRevA.71.042306,HAMMA2021127264}. This concept maps naturally to deep thermalization, where the  resource present in the global system may or may not become localized in the PE. Our theory, presented in Sec.~\ref{sec:generaltheory}, quantitatively predicts that
there are QRTs for which $\overline{R}$ evolves smoothly [``smoothly-localizable'' (SL) resources], and QRTs for which $\overline{R}$ exhibits threshold behavior [``threshold-localizable'' (TL)], see Figs.~\ref{Fig:1}(b) and (c). 

Before proceeding to our specific models of dynamics and theory, it is helpful to present a few illustrative examples of subspace coherence resources that we will study in this work, in order to make concrete the aforementioned abstract discussions on QRTs. 
\begin{itemize}
    \item {\bf Coherence}: the most natural instance of a subspace coherence QRT is that of coherence itself~\cite{baumgratz2014quantifying}. 
    Here, the Hilbert space  $\mathcal{H}$ of $N$-qubit pure states breaks down into one-dimensional blocks, each spanned by a single computational basis state: $\mathcal{B}_{\mathbf z} = \text{Span}(\ket{\mathbf z})$, for $\mathbf z\in \{0,1\}^N$, such that 
    free states are those that are block diagonal in the computational basis. 
    The resource of coherence for a generic state then consists of superpositions between different blocks, i.e., between different computational basis states, and can be quantified by the {\it relative entropy of coherence}: $R_C(\rho) := S(\rho_D)-S(\rho)$, where $S$ is the von Neumann entropy and $\rho_D = \sum_{\mathbf z}|\mathbf z\rangle \langle \mathbf z|\rho|\mathbf z\rangle \langle \mathbf z|$ is the completely-dephased state in the computational basis. 
    Free unitaries  act as scramblers within each block (i.e., as 1$\times$1 unitary matrices or $U(1)$ phase factors) or by permuting any of the blocks, all of which have the same dimension. 
    Thus the subgroup of free unitaries $\mathcal{G}_R$ in this case is the product of permutation unitaries and diagonal unitaries\footnote{We note that Ref.~\cite{liu2025coherence} considered the more restricted setting of random permutation dynamics, which does not include the diagonal phase unitaries.}~\cite{Aldana_2011, PhysRevB.102.224311, PRXQuantum.2.010329,Gopalakrishnan_2018}.
    
    \item{\bf Asymmetry}: In many-body settings, it is common to consider dynamics that obeys a symmetry associated with a group of unitaries $G$, where the time-evolution operator $U_t$ obeys $[U_t,V] = 0$ for all $V\in G$~\cite{bartlett2007reference,gour2009measuring,ares2023entanglement,summer2026mpembaresource}.
    These symmetry constraints can also be framed in the language of QRTs: while the symmetry prevents $U_t$ from achieving universal quantum computation by itself, any asymmetry present in the system may enhance its computational power. Hence, such a QRT is called the resource theory of {\it asymmetry}.
    Here we describe the simplest example of a unitary $\mathbb{Z}_2$ symmetry acting on an $N$-qubit system with generator $ \mathcal{P}=\bigotimes_{i=1}^N \sigma^z_i$, measuring the total parity of computational basis states. 
    As $\mathcal{P}^2 = \mathbb{I}$, this defines a subspace coherence QRT whose  Hilbert space decomposition, Eq.~\eqref{eq:block_decomposition}, consists of two blocks $\mathcal{B}_\pm$: the $\pm 1$ eigenspaces of the symmetry $\mathcal{P}$, i.e., the even and odd parity sectors, of equal dimension $2^{N-1}$. 
    The free states are stochastic mixtures of fixed-parity states, 
    while coherence between the two parity sectors constitutes the resource.
    The group $\mathcal{G}_R$ of free unitaries is generated by $\mathbb{Z}_2$-symmetric dynamics, acting internally in the two blocks, and a parity flip, e.g. $\sigma^x_1$, which swaps the two blocks. 
    
    \item {\bf Non-stabilizerness or magic}: 
    Non-stabilizerness, or magic, is an important QRT in quantum information theory and increasingly relevant to quantum many-body physics as well~\cite{bravyi2005universal,Bravyi_Improved_2016,Howard_application_2017,Liu_Manybody_2022,Leone_Stabilizer_2022}. It characterizes the limitations on quantum information processing posed by the Gottesman-Knill theorem on efficient classical simulability~\cite{PhysRevA.70.052328}. 
    Here, free states are so-called stabilizer states 
    and free operations are given by the Clifford group---the discrete subgroup of unitaries whose adjoint action normalizes the Pauli group.
    Achieving universal quantum computation requires resourceful operations outside this group, known as ``magic gates''. 
    While nonstabilizerness is not a subspace-coherence QRT, we will find that a block decomposition of the $N$-qubit operator Hilbert space in terms of blocks $\mathcal{B}_P = \text{Span}(P)$ for each of the $4^N$ Pauli strings $P$ (tensor products of Pauli matrices on each qubit) provides a useful perspective into magic. 
    In particular, we can define a QRT of ``Pauli coherence'' for which  superposition between the Pauli blocks $\mathcal{B}_P$ is a necessary but {\it insufficient} condition for magic. The free unitaries are again Clifford unitaries, which act as permutations between different blocks and as signs within the blocks. 
    This auxiliary QRT of ``Pauli coherence''  fits within our subspace coherence QRT framework and usefully informs the localizability of magic in deep thermalization, as we will discuss in Sec.~\ref{sec:magic}. 
\end{itemize}

\subsection{Models \label{sec:review_models}}

The concrete models of dynamics studied in this work are as follows. Unless stated otherwise, we consider an $N$-qubit system bipartitioned into subsystems $A$ and $B$  with $N_A$ and $N_B$ qubits respectively. We take $A$ to be fixed in size, and investigate the limiting form of the PE in the thermodynamic limit $N_B \to \infty$.
We also fix a subspace coherence QRT to study, which has $\mathcal{G}_R$ as its group of free unitaries. 

We assume the system is initialized in a state $\ket{\Psi_0}$ that may be resourceful: $R(\ket{\Psi_0}) \geq  0$, where $R$ is the resource monotone associated with the QRT in question. 
Then, we evolve the system under unitary dynamics $U_t$ which is free, i.e., resource-preserving, see Fig.~\ref{Fig:1}(a). 
To simplify our analysis, we will study the case of {\it typical} free dynamics associated with the QRT: we take $U_t$ to be a resource-free unitary $U$ drawn uniformly randomly from  $\mathcal{G}_R$ according to its  Haar measure\footnote{We assume the Haar measure for $\mathcal{G}_R$ exists. This is guaranteed if the group $\mathcal{G}_R$ is compact, regardless of whether it is discrete or continuous.}. 
This is a standard random matrix theory simplification in quantum many-body dynamics where we replace late-time dynamics generated by a particular Hamiltonian or unitary circuit with a random matrix with analogous desired properties.  
Note since $U$ is free, it satisfies $R(U\ket{\Psi_0}) =  R(\ket{\Psi_0})$.

To generate the projected states $|\psi(\nu)\rangle_A$ on $A$, subsystem $B$ is measured in a local product basis $\{\ket{\Phi_\nu}_B\}$, 
which may itself carry resource. 
We will consider two models of local bases for the input states and measurements (used also in \cite{liu2025coherence}), which offer different advantages in our theory in analyzing the resource content of the PE: 
\begin{itemize}
    \item {\bf Mixed-basis model}. Some number $\alpha_0 N$ of the $N$ qubits are prepared in a maximally resourceful local basis; the remaining $(1-\alpha_0)N$ qubits are prepared in a resource-free local basis. 
    Similarly, some number $\alpha_m N$ of the $N_B$ qubits on $B$ are measured in a maximally resourceful basis; the remaining $(1-\alpha_m)N-N_A$ qubits on $B$ are measured in a resource-free local basis.  
    As an example, for the QRT of coherence, a mixed-basis product state is $\ket{+}^{\otimes \alpha_0 N} \otimes \ket{0}^{\otimes (1-\alpha_0)N}$. The parameters $\alpha_0 \in [0,1]$ and $\alpha_m \in [0,1]$ control the density of resource injected by the input state and measurement basis, respectively. An advantage of the mixed basis is its analytical tractability due to the flat distribution of wavefunction amplitudes in the computational basis.

    \item {\bf Tilted-basis model}. Each qubit is prepared (respectively, measured) in the same local product basis, which is ``tilted'' away from the resource-free basis by an angle $\theta_0$ (respectively, $\theta_m$). As an example, for the QRT of coherence, a tilted-basis initial product state is $(\cos(\theta_0/2)\ket{0} + \sin(\theta_0/2)\ket{1})^{\otimes N}$, while a tilted-basis measurement on a given qubit corresponds to measuring along the directions $\{\cos(\theta_m/2)|0\rangle+\sin(\theta_m/2)|1\rangle, \sin(\theta_m/2)|0\rangle-\cos(\theta_m/2)|1\rangle \}$. The angles $\theta_0, \theta_m$ control the density of resource injected by the input state and measurement basis, respectively, with $\theta_0=0,\theta_m=0$ being the free limits. While usually less analytically tractable than the mixed basis, the tilted basis is advantageous in finite-sized numerical simulations due to the greater freedom in tuning the resource content via the continuous angles $\theta_0, \theta_m$. 
\end{itemize}
Ultimately, these two choices (or any other choice of local basis) are expected to give rise to the same universal phase structure, controlled by the density of resource, regardless of microscopic details; see Appendix~\ref{app:basis} for an explanation of how to map between the mixed- and tilted-basis models. 

Lastly, we note for subspace coherence QRTs defined on the operator Hilbert space, analogous models of dynamics (on operator space) including the mixed- and tilted-basis initial state and measurements can be considered. The key is the Choi-Jamio{\l}kowski isomorphism allowing us to view operator dynamics as state dynamics but in a doubled Hilbert space~\cite{choi1975completely,jamiolkowski1972linear}. This is discussed more in  Sec.~\ref{sec:heisenberg} where we analyze deep thermalization within the QRTs of magic and non-Gaussianity.

\section{Theory of resource localization as block sharpening}
\label{sec:generaltheory}

In this section, we present our theory of resource localization in deep thermalization for subspace-coherence QRTs.
We first develop the theory focusing on the  microscopic mechanism behind resource localization, which yields quantitative results; then later provide a high-level, information-theoretic interpretation of the results in Sec.~\ref{sec:block_sharpening}. 

Our aim is to estimate the magnitude of the coefficient of an {\it unnormalized} projected state 
\begin{align}
\langle \mathbf z_A|\tilde{\psi}(\nu)\rangle_A := (\langle \mathbf z_A|\otimes \langle \Phi_\nu|_B) U|\Psi_0\rangle
\label{eqn:matrix_element0}
\end{align}
in the computational basis $|\mathbf z_A\rangle$ of $A$. 
From this quantity, we will then reason about the nature of the projected state, and subsequently the resource content of a typical projected state of the PE. 
Above, $|\Psi_0\rangle$ is the initial global state, $|\Phi_\nu\rangle_B$ is the state on $B$ associated with the measurement outcome $\nu$, and $U$ is a random free (resource-preserving) unitary of a given subspace coherence QRT. 
Recalling the structure of free operations\footnote{To remind the reader, the free subgroup $\mathcal{G}_R$ associated to this QRT is generated by arbitrary unitaries within each block, $\bigoplus_i U(\mathcal{B}_i)$, and transpositions between any two blocks $\mathcal{B}_i$, $\mathcal{B}_{i'}$ of the same dimension, $d_i = d_{i'}$.} discussed in Sec.~\ref{sec:review_resources}, it proves useful to decompose $U = \pi V$, with $\pi$ a block permutation and $V$ a direct sum of random unitaries within each block. Defining $|\Psi_m\rangle := \pi^\dagger (\ket{\mathbf z_A} \otimes \ket{\Phi_\nu}_B)$, we can thus rewrite the key quantity of our analysis, the overlap in Eq.~\eqref{eqn:matrix_element0}, simply as 
\begin{equation}
    \langle \mathbf z_A|\tilde{\psi}(\nu)\rangle_A = \bra{\Psi_m} V \ket{\Psi_0}. 
    \label{eqn:matrix_element}
\end{equation}

Computing Eq.~\eqref{eqn:matrix_element} for a {\it particular} subspace coherence QRT is possible, but entails a case-by-case analysis, as it may depend sensitively on the exact block partitioning of Hilbert space, choice of input state, and choice of measurement basis.
To make progress, we will assume that the QRT in question is {\it typical}, in the sense that it has a block decomposition $\{\mathcal{B}_i\}$ of the global Hilbert space $\mathcal{H}$ obtained by randomly selecting for each block $\mathcal{B}_i$, a set of $d_i$ computational basis elements $ |\mathbf z\rangle$.
For simplicity, we also focus on the mixed-basis model, though we expect our analysis applies to the tilted  basis model too---we comment on the connection in Appendix~\ref{app:basis}. 
Explicitly, we take the initial state to be
\begin{align}
|\Psi_0\rangle=|+\rangle^{\alpha_0 N} \otimes |0\rangle^{(1-\alpha_0)N},
\label{eqn:initial_state}
\end{align}
so that the $\alpha_0 N$ qubits which are prepared along the $X$ axis are resourceful, while the $(1-\alpha_0)N$ qubits which are prepared along the $Z$ axis are free, with $\alpha_0\in [0,1]$.
Similarly, the measurement basis $\{ |\Phi_\nu\rangle_B\}$ is taken to be a tensor product of $\alpha_m N$ qubits ($\alpha_m \in [0,1]$) aligned or anti-aligned with the $X$ axis, which are resourceful;
and $(1 - \alpha_m)N - N_A$ qubits aligned or anti-aligned with the $Z$ axis, which are resourceless. 
For example, if the measurement outcome is 
\begin{equation}
\nu = (\underbrace{+-++\cdots}_{\alpha_m N} \underbrace{010110\cdots}_{(1-\alpha_m)N - N_A}),
\end{equation}
then 
\begin{align}
|\Phi_\nu\rangle_B = |\underbrace{+-++\cdots}_{\alpha_m N} \underbrace{010110\cdots}_{(1-\alpha_m)N - N_A}\rangle_B.
\label{eqn:measurement_basis}
\end{align}

\begin{figure*}
    \centering
    \includegraphics[width=0.99\textwidth]{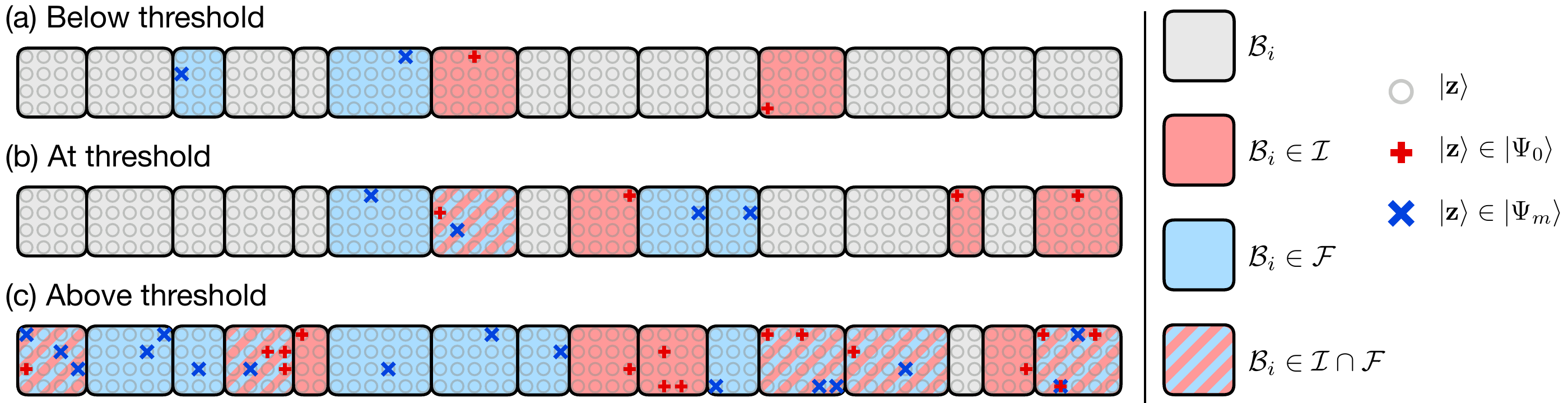}
    \caption{\justifying 
    Schematic illustration of the block sharpening mechanism. 
    The Hilbert space $\mathcal{H}$ is depicted as a collection of bitstrings (circles) arranged into blocks $\mathcal{B}_i$ defined by a QRT [Eq.~\eqref{eq:block_decomposition}] (black contours). The ket $\ket{\Psi_0}$ and bra $\bra{\Psi_m}$ in Eq.~\eqref{eqn:matrix_element} contain a set of bitstrings (`$+$' and `$\times$' symbols respectively) that populate two subsets of the blocks, $\mathcal{I}$ (shaded red) and $\mathcal{F}$ (shaded blue). 
    The free dynamics fully scrambles each bitstring within its block. 
    The overlap [Eq.~\eqref{eqn:matrix_element}] receives contributions only from blocks that belong to both $\mathcal{I}$ and $\mathcal{F}$ (hatched).
    (a) Resourceless phase. 
    Below the resource threshold, the subsets $\mathcal{I}$ and $\mathcal{F}$ are so sparse that they typically do not intersect.
    (b) Critical point. At the resource threshold, the ``birthday paradox'' kicks in, and the subset intersection $\mathcal{I}\cap\mathcal{F}$ becomes typically finite even though each subset is sparse.
    (c) Resourceful phase. Above the resource threshold, the subset intersection $\mathcal{I}\cap\mathcal{F}$ proliferates, giving many random contributions to the amplitude in Eq.~\eqref{eqn:matrix_element} and thus random-looking PE states. This behavior persists as the blocks become more densely populated.
    }
    \label{fig:blocks}
\end{figure*}

Our strategy to compute the coefficient Eq.~\eqref{eqn:matrix_element} will be to view it  as arising from the following process: 
the ket $|\Psi_0\rangle$ provides $A_0: = 2^{\alpha_0 N}$ distinct global bit-strings $\mathbf{z}$, each of which falls into some block $\mathcal{B}_i$ of the Hilbert space defined by the QRT. These are depicted in Fig.~\ref{fig:blocks} as red `$+$' symbols falling into different boxes.
We denote the set of blocks that $|\Psi_0\rangle$ populates as $\mathcal{I} \subseteq \{\mathcal{B}_i\}$. Likewise, the bra $\langle \Psi_m|$  provides $A_m := 2^{\alpha_m N}$ global bit-strings $\mathbf{z'}$, which also populate some of the blocks, illustrated in Fig.~\ref{fig:blocks} as blue `$\times$' symbols falling into different boxes. We denote this subset of blocks as $\mathcal{F} \subseteq \{ \mathcal{B}_i\}$. The unitary $V$ then scrambles or delocalizes the support of each bit-string within its respective block (but not across blocks), shown in Fig.~\ref{fig:blocks} as a shading of the boxes $\mathcal{I}$ and $\mathcal{F}$ in their respective color\footnote{%
Note that, due to the left invariance of the Haar measure from which $V$ is drawn, we may replace $V$ in Eq.~\eqref{eqn:matrix_element} by a product $V'V$, and have $V$ and $V'$ act on the ket and bra respectively. This highlights the symmetry of the problem and justifies the shading of both $\mathcal I$ and $\mathcal F$ blocks in Fig.~\ref{fig:blocks}. 
}.

Our key observation is that a given block $\mathcal{B}_i$ contributes to the coefficient Eq.~\eqref{eqn:matrix_element} if and only if it is in {\it both} $\mathcal{I}$ and $\mathcal{F}$. This describes a (generalized) {\it birthday problem}~\cite{von1939aufteilungs}\footnote{
The classic birthday problem or ``birthday paradox'' deals with the probability of two people in a group sharing the same birthday. The underlying ``blocks'' there are the 365 days of the calendar year, with a uniform probability distribution. Our setting generalizes this to the case of a non-uniform probability distribution. 
},
or more formally,  a {\it random subset intersection} problem.  
In Sec.~\ref{sec:block_sharpening}, we also discuss how this can be framed alternatively as a {\it learning} problem, 
in the sense that information acquired by the measurements on the bath can collapse an initially resourceful superposition into a single free state. We term this concept ``block sharpening'', in analogy with the phenomenon of charge sharpening from measurement-induced phase transitions (MIPT)~\cite{PhysRevX.12.041002,Barratt_Transitions_2022,Li_QuantumZeno_2018,Agrawal_Observing_2024,Feng_chargespin_2025,Choi_Quantum_2020,Gullans_Dynamical_2020,Fisher_random_2022,Potter_Entanglement_2022} which can likewise be framed as a learning problem.

\subsection{Estimation of the amplitude as a random subset intersection problem \label{sec:amplitude}}

Having mapped the question of PE wavefunction amplitudes [Eq.~\eqref{eqn:matrix_element0}] to a random subset intersection problem, we proceed to solve this problem. 
First, the probability\footnote{The probability space here is over the choice of random resources.} that a given block $\mathcal{B}_i$ is {\it not} populated by the initial state $|\Psi_0\rangle$ can be computed as follows:
\begin{align}
\mathsf{Prob}(\mathcal{B}_i \notin \mathcal{I}) & = \prod_{j=0}^{A_0-1}\left(1-\frac{d_i}{2^N-j}\right) \nonumber \\
& \approx (1-p_i)^{A_0} \nonumber \\
& \approx \exp(-A_0 p_i).
\end{align}
Above, $p_i := d_i/2^N$ is the probability that a single bit-string lands in block $\mathcal{B}_i$, which we call the `block probability'; also,  $A_0 = 2^{\alpha_0 N}$ is the number of bitstrings present in $\ket{\Psi_0}$. 
(The first expression arises from asking that each of the $A_0$ distinct bit-strings  must fall outside the block; the second line is an approximation valid when $\alpha_0 < 1$). 
Then, the probability that block $\mathcal{B}_i$ {\it is} populated  by $|\Psi_0\rangle$ is
\begin{align}
    \mathsf{Prob}(\mathcal{B}_i \in \mathcal{I}) & \approx 1 - \exp(-A_0 p_i) \nonumber  \\
    & \approx \begin{cases}
      A_0 p_i  & \text{if } A_0p_i\ll 1
    \\ 1 & \text{otherwise}.
    \end{cases}
    \label{eq:prob_populated}
\end{align}
In other words, if $d_i \ll 2^{(1-\alpha_0)N}$, which can always be met by a suitably small choice of $\alpha_0$ so long as $d_i = O(2^{cN})$ for some constant $c<1$, the probability that block $\mathcal{B}_i$ is populated by $|\Psi_0\rangle$ is small.  In this regime,  with high probability each bitstring from the input state lands in a distinct block,
so that $\alpha_0$ acquires the meaning of a density of resource. By analogous reasoning, the probability that a block $\mathcal{B}_i$ is populated by the ``final'' state in Eq.~\eqref{eqn:matrix_element}, $\bra{\Psi_m}$, is
\begin{align}
    \mathsf{Prob}(\mathcal{B}_i \in \mathcal{F}) & 
    \approx \begin{cases}
      A_m p_i  & \text{if } A_m p_i\ll 1
    \\ 1 & \text{otherwise},
    \end{cases}
    \label{eq:prob_compatible}
\end{align}
and $\alpha_m$ is also interpretable as a density of resource. 

It is interesting to focus on the regime of sparse block populations for both $\ket{\Psi_0}$ and $\ket{\Psi_m}$, 
corresponding to the condition $A_0 p_i \ll 1$ and $A_m p_i \ll 1$ for all $i$, see Fig.~\ref{fig:blocks}(a).
Combining Eq.~\eqref{eq:prob_populated} and Eq.~\eqref{eq:prob_compatible} and treating the events $\mathcal{B}_i \in \mathcal I$ and $\mathcal{B}_i \in \mathcal F$ as independent, we find that the probability that the populated blocks $\mathcal{I}$ and $\mathcal{F}$ do not intersect at block $\mathcal{B}_i$  is
\begin{equation}
    \mathsf{Prob}(\mathcal{B}_i \notin \mathcal{I} \cap \mathcal{F}) \simeq 1 - A_0 A_m p_i^2. 
    \label{eq:prob_notboth}
\end{equation}
The probability that they do not intersect on {\it any} blocks is then, treating the events for different $i$ as independent,
\begin{align}
    \mathsf{Prob}(\mathcal I \cap \mathcal F = \emptyset)
    & \simeq \prod_i \mathsf{Prob}(\mathcal{B}_i \notin \mathcal{I} \cap \mathcal{F}) \nonumber \\
    & \simeq \prod_i (1-A_0 A_m p_i^2) \nonumber \\
    & \simeq \exp\left( -A_0 A_m \sum_i p_i^2 \right) \nonumber \\
    & = \exp\left(-2^{(\alpha_0  + \alpha_m) N - H_2(\{p_i\}) } \right),
    \label{eq:prob_nointersection}
\end{align}
where 
\begin{align}
H_2(\{p_i\}) := -\log_2\left(\sum_ip_i^2\right)
\label{eqn:Renyi}
\end{align}
is the second R\'enyi entropy\footnote{In mathematics, the second R\'enyi entropy is also called the collision entropy, which aptly characterizes the physics underlying Eq.~\eqref{eq:prob_nointersection}, the probability of no block intersections.} of the block probabilities $\{ p_i \}$ defined by the QRT. 

Eq.~\eqref{eq:prob_nointersection} is a crucial outcome of our theory: it identifies a resource threshold, given by
\begin{align}
 \alpha_\text{crit}:=\lim_{N \to \infty} \frac{H_2(\{p_i\})}{N},
 \label{eqn:alpha_crit}
\end{align}
determining the behavior of the PE wavefunction amplitudes, Eq.~\eqref{eqn:matrix_element}. 
When $\alpha_0+\alpha_m < \alpha_\text{crit}$,  the probability of non-intersection is $\mathsf{Prob}(\mathcal I \cap \mathcal F = \emptyset) \to 1$, meaning the block subsets $\mathcal I$ and $\mathcal F$ are so sparse in Hilbert space that they are unlikely to overlap at all, $|\mathcal I \cap \mathcal F | = 0$, see Fig.~\ref{fig:blocks}(a). Thus, for any given $\mathbf{z}_A$,  the coefficient $\langle \mathbf z_A|\tilde{\psi}(\nu)\rangle_A$ is zero with high probability. 
However, note that our discussion so far has considered a {\it forced} measurement outcome $\nu$ (or equivalently, the uniform measure over $\nu$). In reality, the Born rule takes care of avoiding null vectors in the PE, and selects rare measurement outcomes $\nu$ where the wavefunction $|\tilde{\psi}(\nu)\rangle$ is nonvanishing. This
ensures that there is at least one coefficient $\langle \mathbf z_A|\tilde{\psi}(\nu)\rangle_A$ with a non-empty subset intersection $\mathcal{I}\cap\mathcal{F}\neq \emptyset$. 
Since, as shown, subset intersections are exponentially rare and we are in the sparse limit $A_0 p_i \ll 1, A_m p_i \ll 1$, with high probability there will be a {\it single} such $\mathbf{z}_A$ where $|\mathcal I \cap \mathcal F | = 1$. 
This means that the projected state $|\psi(\nu)\rangle$ is with overwhelming probability just a single bit-string state on $A$:
\begin{align}
|\psi(\nu)\rangle \approx |\mathbf{z}_A(\nu)\rangle,
\end{align}
which is clearly a free state of the QRT (recall we are considering subspace-coherence QRTs defined in the computational basis). 
In other words, the PE is composed of an ensemble of random bit-string states, and is hence resourceless: $\overline{R} = 0$. 

Conversely, when $\alpha_0+\alpha_m > \alpha_\text{crit}$, the probability of an empty intersection in Eq.~\eqref{eq:prob_nointersection} is small (asymptotically close to 0), signaling that the block subsets $\mathcal I$ and $\mathcal F$ occupy a large fraction of Hilbert space and are extremely unlikely to avoid each other, see Fig.~\ref{fig:blocks}(c). Thus their  intersection is likely to be exponentially large, $|\mathcal I \cap \mathcal F| \sim 2^{\gamma N}$ where $\gamma = \alpha_0 + \alpha_m - \alpha_\text{crit} > 0$. 
In this regime, the magnitude of a {\it typical} coefficient Eq.~\eqref{eqn:matrix_element0} (considered either over $\mathbf{z}_A$ or $\nu$)  can be safely estimated (due to self-averaging) as
\begin{align}
|\langle \mathbf z_A|\tilde{\psi}(\nu)\rangle_A|^2  \sim \frac{1}{A_0 A_m} \sum_{i} |Z_i|^2
\label{eqn:coeff_estimate}
\end{align}
where $Z_i$ is the contribution due to bitstrings landing within block $\mathcal{B}_i$. This in turn is estimated as 
\begin{align}
|Z_i|^2 \sim \frac{1}{d_i}(A_0 p_i)(A_m p_i),
\end{align}
which has an intuitive meaning: the $1/d_i$ factor is the average squared overlap between two scrambled bitstrings in block $\mathcal{B}_i$ (since $V$ acts Haar-randomly within the block), while $A_0 p_i$ and $A_m p_i$ correspond to the expected number of bit-strings from $|\Psi_0\rangle$ and $\bra{\Psi_m}$ respectively. 
Overall, Eq.~\eqref{eqn:coeff_estimate} reduces to
\begin{align}
|\langle \mathbf z_A|\tilde{\psi}(\nu)\rangle_A|^2  & \sim \sum_i 
\frac{p_i^2}{d_i} = \frac{1}{2^N},
\label{eqn:universal_scaling1}
\end{align}
since $\sum_i p_i=1$ and $p_i/d_i = 2^{-N}$. 
Importantly, one sees that the initial state and measurement basis dependent quantities $A_0, A_m$  completely drop out in this calculation, giving rise to a universal estimate of the magnitude of the coefficient. 
Since this scaling is true for a typical bit-string $\mathbf{z}_A$, it means that the {\it normalized} projected state $|\psi(\nu)\rangle_A$ has coefficient with magnitude
\begin{align}
|\langle \mathbf z_A|{\psi}(\nu)\rangle_A| \sim \frac{1}{\sqrt{2^{N_A}}},
\label{eqn:universal_scaling2}
\end{align}
whose uniformity (anticoncentration) describes a vector delocalized in the Hilbert space $\mathcal{H}_A$. In other words, when $\alpha_0+\alpha_m > \alpha_\text{crit}$, we can expect each PE state $|{\psi}(\nu)\rangle_A$ to behave like a typical random vector over $\mathcal{H}_A$, which is a resourceful state. 

Note that in the above analysis, what we have argued for is the universality of the scaling of the coefficients in Eq.~\eqref{eqn:universal_scaling1} and Eq.~\eqref{eqn:universal_scaling2} with respect to $N$ and $N_A$ respectively; however, there could be non-universal pre-factors depending on the QRT in question, such as the precise distribution of $\{p_i\}$, which can affect the nature of the phase transition and bias the randomness in the PE away from {\it uniform} randomness as we shall see below.

It is worth noting that despite the different roles of $\alpha_0$ and $\alpha_m$ entering in the definition of the PE (the former characterizes the number of bit-strings in the initial state $|\Psi_0\rangle$ while the latter the number of bit-strings in the measurement basis $|\Phi_\nu\rangle_B$), they appear on exactly equal footing in Eq.~\eqref{eq:prob_nointersection}, consistent with prior~\cite{liu2025coherence} and upcoming~\cite{us_2026_replica} results on the transition for the QRT of coherence.

Lastly, the astute reader may ask what happens if one or both of the conditions of ``sparsity'' in the initial and measurement basis, $A_0 p_i \ll 1, A_m p_i \ll 1$ for all $i$, are violated, such that at least one block $\mathcal{B}_i$ is densely populated by either $\ket{\Psi_0}$ or $\ket{\Psi_m}$. In any of these cases, intersections of $\mathcal{I}$ and $\mathcal{F}$ still proliferate, leading to the universal scaling forms Eq.~\eqref{eqn:universal_scaling1} and Eq.~\eqref{eqn:universal_scaling2}. 
In other words, the assumption of sparsity is necessary to obtain a transition into a resourceless phase, but not for a resourceful phase.

We next apply our theory to QRTs with different block structures, and show that it leads to a classification of QRTs into different classes giving rise to different emergent behaviors of the PE. 

\subsection{Extensive resources: presence of a localizability threshold} 
\label{sec:extensive}

Let us first focus on resources that are extensive according to the R\'enyi-2 entropy, i.e., QRTs with block probabilities $p_i$ such that
\begin{equation} 
H_2(\{p_i\})= \sigma N
\end{equation} 
with $\sigma > 0$. 
Referring to Eq.~\eqref{eqn:alpha_crit}, we see that
\begin{align}
\alpha_\text{crit} = \sigma >0,
\end{align}
which leads to the prediction of a {\it finite-density} localizability threshold for QRTs with an extensive block entropy: 
if $\alpha_0 + \alpha_m < \sigma$ then the PE tends to a resourceless phase; if $\alpha_0 + \alpha_m > \sigma$ then it tends to a phase with extensive resource, $\overline{R}\sim N_A$. 
A sharp phase transition is predicted at the finite threshold
\begin{align}
\alpha_0+\alpha_m = \sigma. 
\label{eq:threshold}
\end{align} 
In other words, the injected global resource must exceed a finite threshold density before any of it can get localized in the PE.
We therefore term such QRTs ``threshold-localizable'' (TL). 

For the QRT of coherence (superposition between computational basis states), Eq.~\eqref{eq:threshold} recovers the results of Ref.~\cite{liu2025coherence}: a deep thermalization transition at $\alpha_0 + \alpha_m = 1$. 
Indeed, for coherence, all blocks are one-dimensional, $p_i = 2^{-N}$, yielding $\alpha_\text{crit}=\sigma = 1$ in accordance with our theory. 
Eq.~\eqref{eq:threshold} vastly generalizes that result by predicting a host of novel deep thermalization transitions for QRTs characterized by extensive entropy. 
In Sec.~\ref{sec:transitions} and Sec.~\ref{sec:magic} we numerically and analytically investigate some of these transitions in other extensive QRTs (like ``syndrome coherence'' and magic), finding excellent agreement with the prediction in Eq.~\eqref{eq:threshold} across diverse resources. This corroborates the universality of our theory. 

\subsection{Intensive resources: absence of a localizability threshold}

Conversely, suppose we consider QRTs for which the block entropy is intensive, i.e.,
\begin{align}
H_2(\{p_i\})= O(1)
\end{align}
in system size $N$.
This captures QRTs with for example a constant number of blocks, such as $\mathbb{Z}_2$ asymmetry. 
Then, 
\begin{align}
\alpha_\text{crit} = 0,
\end{align}
indicating that {\it any} non-zero value of $\alpha_0+\alpha_m$ renders the PE {\it always} in the resourceful phase\footnote{Note that an intensive resource, with blocks of probability $p_i = O(1)$ in $N$, violates the sparsity assumption $2^{\alpha_0 N} p_i \ll 1$ for any $\alpha_0 > 0$ (and similarly for $\alpha_m$), so the calculation leading up to Eq.~\eqref{eq:prob_nointersection} does not apply. Nevertheless, it is easy to see that the subset intersection problem is trivial in this case: {\it any} divergent number of bit-strings $A_0, A_m$ results in dense coverage of the blocks $\mathcal{B}_i$ and yields a resourceful PE. 
}. In other words, the PE does not exhibit any finite resource density threshold transition.  

On the other hand, suppose we scale $\alpha_0, \alpha_m \sim \frac{1}{N}$, so that $A_0$ and $A_m$ are independent of system size $N$. Then it is possible that the precise amount of resource localized in the PE changes smoothly.
We hence call such QRTs ``smoothly-localizable'' (SL).

\subsection{Subextensive resources: marginal localizability threshold}
\label{sec:marginal}

Lastly, we consider an interesting marginal case of resources featuring a divergent but subexponential number of blocks. 
For example, we may assume a polynomial number of blocks of comparable dimension, giving $\sum_i p_i^2 = N^{-\rho}$ or 
\begin{equation}
    H_2(\{p_i\})= \rho \log_2N
\end{equation}
for some $\rho > 0$. The entropy associated with this QRT is thus  divergent, but subextensive (entropy scalings other than logarithmic, e.g., sublinear power laws, could be addressed similarly). 

In this case, it is clear that an exponential number of input bitstrings, $A_0 = 2^{\alpha_0 N}$ with constant $\alpha_0\in (0,1]$, is enough to populate all the blocks and give near-maximal resource; the same reasoning applies to the measurement basis choice. Thus, just like the case of intensive resources, there is no finite density threshold behavior of the PE: any non-zero $\alpha_0 + \alpha_m$ results in a resourceful PE.

Nevertheless, a phase transition in the thermodynamic limit $N \to \infty$ can be realized for subextensive resources if we carefully tune the resource injected into the system.
If we let $A_0 = N^{\beta_0}$ and $A_m = N^{\beta_m}$, the same reasoning used for Eq.~\eqref{eq:prob_nointersection} yields in this case
\begin{equation}
    \mathsf{Prob}(\mathcal I \cap \mathcal F = \emptyset)
    = \exp\left[ -N^{\beta_0 + \beta_m - \rho} \right],
    \label{eq:marginal_transition}
\end{equation}
which results in a threshold determined by  
\begin{equation}
    \beta_0 + \beta_m = \rho.
    \label{eq:subextensive_threshold}
\end{equation}

Despite the superficial similarity with Eq.~\eqref{eq:threshold}, however, this result carries different physical implications. 
In this case, tuning the input resource in the relevant range requires a family of input states where the resourceful basis is chosen only on a  number of qubits that grows {\it subextensively} with system size $N$:
\begin{equation}
    \ket{\Psi} = \ket{0}^{\otimes N - \ell} \otimes \ket{+}^{\otimes \ell}, 
    \qquad 
    \ell = \beta_0 \log_2(N). 
    \label{eq:mixed_basis_subextensive}
\end{equation}
Converting the above result to a translationally invariant ``tilted basis'' initial state $(\cos(\theta/2) \ket{0} + \sin(\theta/2) \ket{1})^{\otimes N}$
using the mapping of Appendix~\ref{app:basis}, 
this corresponds to a tuning of Bloch angle $\theta$ as $\theta \sim N^{-\beta_0}$.

Therefore, the transition in the quantity $\mathsf{Prob}(\mathcal I \cap \mathcal F = \emptyset)$ 
is ``hidden'' if we were to tune local, intensive parameters (such as the tilt angle $\theta$) in a system-size-independent manner, the usual setting to observe for phase transitions in the thermodynamic limit.  
To resolve the transition one  must instead ``zoom in'' around the resourceless point with  a careful scaling of parameters in system size, e.g., $\theta\sim N^{-\beta_0}$, which vanishes with system size. 
For this reason, while subextensive resources may technically have a threshold in the sense of Eq.~\eqref{eq:subextensive_threshold}, they in practice exhibit SL behavior, just like intensive resources. 

An important example of a divergent but subextensive resource is $U(1)$ asymmetry~\cite{bartlett2007reference}, where typical blocks have probability $p_i \sim N^{-1/2}$. This scenario was considered already in Ref.~\cite{chang2025deep} in the context of deep thermalization under $U(1)$ conserving dynamics, and we revisit it numerically in Sec.~\ref{sec:asymmetry} through our lens of the QRT of $U(1)$ asymmetry, verifying the prediction of SL behavior. There we also uncover a subtle obstruction to the mechanism behind the ``hidden'' threshold of Eq.~\eqref{eq:subextensive_threshold}, related to the structure of blocks $\mathcal{B}_i$ for this QRT. Specifically, because blocks are defined by a charge (the Hamming weight) with a local density, local product bases such as Eq.~\eqref{eq:mixed_basis_subextensive} populate far fewer blocks than a na\"ive counting of bitstrings would suggest, undermining the derivation of Eq.~\eqref{eq:subextensive_threshold}.
We further show how this obstruction can be circumvented by using long-range-entangled initial states instead of local product states (App.~\ref{app:u1}).

\subsection{Interpretation as block sharpening}\label{sec:block_sharpening}

Our theory of resource localizability based on the vanishing or proliferation of intersections between random subsets 
admits a broader information-theoretic  interpretation as a learning or inference problem. This  perspective is valuable not just for providing useful intuition, but also in making a conceptual connection to recent results on measurement-induced phase transitions (MIPTs) in systems with conserved quantities, where such learning or inference problems naturally arise too. 

The key insight is to think of the resource as uncertainty, or ``fuzziness'', of the block label $i$ in the system. Measurements can reveal information that reduces this uncertainty, progressively ``sharpening'' the value of the block label, potentially down to a deterministic answer---in which case the post-measurement state is completely resourceless. We term this mechanism ``block sharpening'' as a generalization of ``charge sharpening'' in MIPTs~\cite{PhysRevX.12.041002,Li_QuantumZeno_2018,Barratt_Transitions_2022,Agrawal_Observing_2024,Feng_chargespin_2025,Choi_Quantum_2020,Gullans_Dynamical_2020,Fisher_random_2022,Potter_Entanglement_2022}, where measurements of a local charge density may collapse the system into a specific sector of a $U(1)$ (or other) global charge. 
There, the problem of inferring the value of global charge, given some initial uncertainty, abruptly transitions from under-determined to over-determined. 
The same thing happens, {\it mutatis mutandis}, in our resource localizability transitions. A crucial difference between the setups, however, is that charge sharpening transitions arise in monitored dynamics with bulk mid-circuit measurements, whereas in the framework of deep thermalization we only consider final measurements\footnote{This also explains why $U(1)$ asymmetry is SL in our framework: final measurements represent a vanishing space-time density, placing the dynamics in the ``fuzzy'' phase and precluding a transition.}.

The inference problem of block sharpening is framed precisely as follows. 
An initial state is prepared in a superposition of blocks $\mathcal{I}\subseteq \{\mathcal{B}_i\}$;
it is scrambled by free dynamics, $U = \pi V$, with $\pi$ a block permutation and $V$ a block-diagonal unitary; 
then it is measured in a mixed basis: 
first a subsystem $B_Z$ is measured in the $Z$ (computational) basis, then $B_X = B\setminus B_Z$ is measured in the $X$ basis. 
Consider an observer, Eve, who is given the following classical data: $\mathcal{I}$ (labels of the blocks populated by the input state), $\pi$ (permutation part of the dynamics), and the measurement outcome on $B_Z$; we assume that $B_X$ has not been measured yet. Can Eve guess which block $\mathcal{B}_i$ the global state is in? 
Or more precisely, can she predict the outcome of a global measurement of the block label\footnote{The block label is a (global) observable. It is the outcome of a POVM with effects $\{ \Pi_{\mathcal{B}_i}\}$, with $\Pi_{\mathcal{B}_i}$ the projector on block $\mathcal{B}_i$.}?

To answer this question,
Eve takes the set of global bitstrings that are compatible with the known measurement outcome on $B_Z$ (i.e., all possible bitstrings obtained by completing the measurement outcome on $B_Z$ to an $N$-bit string with arbitrary assignments of bits to $A$ and $B_X$) and permutes them with $\pi^\dagger$. She thus obtains a set of blocks\footnote{
Note that the subset $\mathcal{F}$ here is slightly different from the $\mathcal{F}$ used in Sec.~\ref{sec:amplitude}, as it contains {\it all} possible bitstring completions on $A$ rather than a single one; but since $A$ is finite, the thermodynamic-limit behavior does not change.
} $\mathcal{F}\subseteq \{\mathcal{B}_i\}$ in which these bit-strings land. 
These are all the blocks that the system {\it could} have occupied at the beginning of the dynamics, based on her current knowledge of the system, coming from initial knowledge of the permutation part of the dynamics $\pi$ and updated knowledge from the measurement outcome. 
At the same time, Eve also has knowledge of the set $\mathcal I$ of initially populated blocks. 
The question is then whether these two pieces of information suffice for her to uniquely pin down a block label: does the intersection $\mathcal{I}\cap\mathcal{F}$ reduce to a single block $\mathcal{B}_i$? 
If so, then the inference succeeds with certainty---Eve can take this solution for the initial block $\mathcal{B}_i$ and evolve it forward in time using $\pi$, thus predicting with high probability the outcome of a final block measurement. 
If instead the subset intersection contains a large number of blocks, then Eve can only do as well as randomly guessing  a block $\mathcal{B}_i$ from this large set, and the inference fails with high probability. 
By setting up the problem this way,  we see that the analysis proceeds with the same formalism discussed in Sec.~\ref{sec:amplitude} for the PE wavefunction amplitudes and illustrated in Fig.~\ref{fig:blocks}.

The physical consequence for deep thermalization is as follows. 
If Eve's inference fails, it means that the block label remains ``fuzzy'', and the global state resourceful, even after the $Z$ measurements. Since $X$ measurements do not reveal information about the blocks, one expects the local projected states of the PE to inherit this ``fuzziness'' in their respective, local QRT, i.e., to the resource being successfully localized. This corresponds to a resourceful phase of the PE. Conversely, if Eve's inference succeeds with high probability, it means that the resource of the input state has been entirely destroyed by the measurement on $B_Z$, paving the way for the emergence of a resourceless phase in the PE.
In particular, if the condition of sparse block populations $A_mp_i \ll 1$ is met, then when Eve uniquely identifies the block $\mathcal{B}_i$, she also uniquely identifies a bitstring. This implies that the PE is made of bitstring states and thus resourceless.
Otherwise, the PE's resource content may in general depend on other features of the QRT\footnote{
One needs to account for the effect of resourceful measurements on $B_X$, which may introduce resource in $A$ even when $AB_X$ is resourceless. This depends on whether distinct combinations of local blocks on $A$ and $B_X$ can fuse into the same block on the composite system $AB_X$, a QRT feature that we call ``non-injective block fusion'' (see Appendix~\ref{app:blocks}). 
None of the TL QRTs studied in this work exhibit this feature, so that in practice the random subset intersection mechanism correctly identifies the existence of a resourceless PE phase. 
}.

\section{Threshold-localizable resources}
\label{sec:transitions}

To corroborate our theory of resource localization in Sec.~\ref{sec:generaltheory}, we will next study  in detail deep thermalization within several concrete QRTs, like syndrome coherence, purity, imaginarity, asymmetry, magic, and non-Gaussianity to verify their TL or SL behaviors. We begin with extensive resources for which it is expected from our theory that they have threshold-localizable behavior.

\subsection{Syndrome coherence}
\label{sec:rand_stab}

The simplest example of an extensive resource is coherence itself (the presence of superposition between the computational basis states $|\mathbf{z}\rangle$), whose QRT structure was detailed in Sec.~\ref{sec:review_resources}. Being extensive, it is predicted to be threshold localizable. Indeed, as summarized in Sec.~\ref{sec:extensive}, this was the setting in which the first instance of a resource-induced deep thermalization transition was found~\cite{liu2025coherence} and a major motivation of this work. 
Concretely, Ref.~\cite{liu2025coherence} (i)  studied deep thermalization under {\it random permutation dynamics} (RPD)~\cite{Aldana_2011, bertini2025permutation, szasz2025entanglement,bertini2025permutation_chaotic}, which in our framework can be understood as a subset of free unitary operations for coherence; (ii) considered initial states and measurement basis   of both the mixed and tilted-basis types, which injected the requisite global coherence\footnote{Importantly, they also provided complex phases, another resource called imaginarity (see Sec.~\ref{sec:imaginarity}) that RPD cannot create.}; 
and (iii) found a phase transition between the minimally ergodic classical bit-string ensemble and maximally ergodic complex Haar random ensemble.

As a new application of our general theory (Sec.~\ref{sec:generaltheory}), we consider here the QRT of a generalized form of coherence we call ``syndrome coherence'', defined by a block decomposition of the Hilbert space $\mathcal{H} = \bigoplus_{\mathbf s} \mathcal{B}_{\mathbf s}$, where each block $\mathcal{B}_{\mathbf s}$ is a {\it syndrome subspace} of a classical parity check code~\cite{Nielsen_Chuang_2010}. 
Like coherence, this resource is also extensive and hence exhibits a TL behavior, but whose threshold condition is variable and depends on the size of the codespace in question. 

Specifically, we consider a classical code defined by a random parity-check matrix $H \in \mathbb{F}_2^{\sigma N\times N}$. Each row of $H$ specifies a parity check, i.e., a $Z$-type stabilizer generator, $g_i := \prod_{j=1}^{N} Z^{H_{ij}}_j$. Assuming $\sigma N$ independent parity checks, the code has $(1-\sigma)N$ logical bits.  
It has $2^{\sigma N}$ distinct syndromes $\mathbf s$, each labeling a subspace $\mathcal{B}_{\mathbf s}$ containing $2^{(1-\sigma)N}$ computational-basis states. 
There are thus $2^{\sigma N}$ blocks, each of dimension $d = 2^{(1-\sigma)N}$, so that the block probability distribution is uniform: $p_{\mathbf s} = 2^{-\sigma N}$ for all $\mathbf s$. 
We thus expect a threshold behavior in deep thermalization at $\alpha_0 + \alpha_m = \sigma$. 
We note that this family of resources reduces to the previously studied case of ordinary coherence if we take $\sigma = 1$ and $H = I$ (identity matrix on $\mathbb{F}_2^{N\times N}$); this can be viewed as a trivial code with no logical bits, where every bitstring is its own syndrome subspace. 
It is also interesting to observe that the QRT of syndrome coherence described here can equivalently be cast as the QRT of $\mathbb{Z}_2^{\sigma N}$ asymmetry, since the (extensively many) stabilizers $\{g_i\}_{i=1}^{\sigma N}$ are mutually commuting and square to the identity. 

As described in Sec.~\ref{sec:review_resources}, the block structure described above defines a resource monotone~\cite{aberg2006quantifyingsuperposition}
\begin{align}
R_S(\rho):= S\left(\sum_\textbf{s} \Pi_{\mathcal{B}_\textbf{s} }\rho \Pi_{\mathcal{B}_\textbf{s} }\right) - S(\rho),
\label{eqn:Rsyndrome}
\end{align}
where $\Pi_{\mathcal{B}_\textbf{s}}$ is the projector onto the syndrome subspace $\mathcal{B}_\textbf{s}$ and $S$ is the von Neumann entropy. It captures the amount of superposition between different syndrome subspaces in a given many-body state (hence the name ``syndrome coherence''). 
Qualitatively, the structure of the QRT is as follows:  states which are block diagonal in $\mathcal{B}_s$ are free (this can be easily seen from Eq.~\eqref{eqn:Rsyndrome}). Logical operations are free (since they preserve the block diagonal structure). Errors (or dynamics) that simply ``flip'' syndromes are also free; while errors (or dynamics) that create {\it coherent} superposition across different syndrome subspaces are resourceful. 
The subgroup $\mathcal{G}_R$ of free unitaries for this resource is given by unitaries of the form
\begin{equation} 
U = U_\pi \left(\bigoplus_{\mathbf s} V_{\mathbf s} \right), \label{eq:freeU_stabilizer}
\end{equation}
where $U_\pi$ denotes a permutation of syndrome blocks, and $V_{\mathbf s}$ denote generic unitaries acting inside each block $\mathcal{B}_{\mathbf s}$.

Given any globally defined parity check matrix $H$, it is also possible to define the resource on a subsystem: we simply restrict the parity check matrix $H$ to the subsystem, retaining only the relevant column indices ($H|_A \in \mathbb{F}_2^{\sigma N \times N_A}$), then define the syndrome blocks and the resource on $A$ accordingly.
Note that, since $H$ is maximum-rank ($\sigma N$) with high probability, on any subsystem $A$ smaller than $\sigma N$ the reduced resource is simply standard coherence, i.e., $H|_A$ is equivalent (under row reduction) to the identity matrix. 

\begin{figure}
    \centering
    \includegraphics[width=1.0\linewidth]{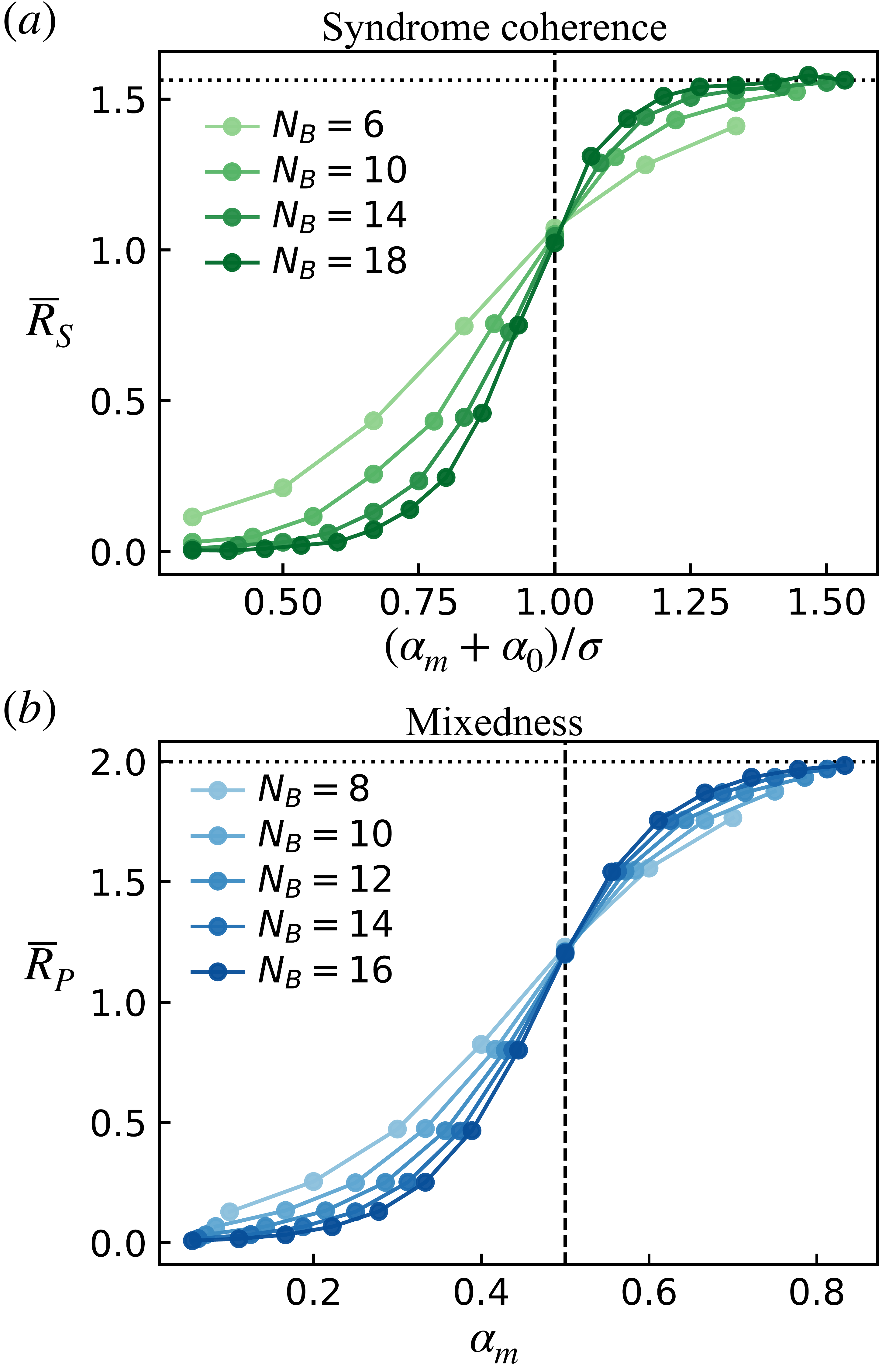}
    \caption{\justifying 
    Localizability transitions in the QRTs of syndrome coherence and mixedness / classical randomness. 
    (a) Syndrome coherence. Average resource of PE states $\overline{R}_{s}$ as a function of parameters $\alpha_{0,m}$, within one instance of a random parity-check code of rate $1-\sigma$. We choose $N_A=2$, $\alpha_0=1/4$ and $\sigma=3/4$, while $\alpha_m$ is varied; every data point is averaged over $1000$ realizations; each realization uses a random independently sampled parity check code. The vertical dashed line is the predicted critical point $\alpha_m+\alpha_0=\sigma$, the horizontal dotted line is the value for the Haar ensemble. (b) Mixedness/classical randomness. 
     Average PE resource $\overline{R}_{P}=-\mathbb{E}_{\mathcal{E}_{\mathrm{PE}}}\log_2\left[\sum_{\mathbf z} p_{\mathbf z}^2 \right]$ as a function of $\alpha_m$, for fixed $N_A=2$ and $\alpha_0=0.5$. Each data point is obtained by averaging over $500$ realizations. The vertical dashed line indicates the predicted critical point $\alpha_m+\alpha_0=1$ ($\alpha_m=0.5$), while the horizontal dotted line is the maximum possible value (2 bits).
     }
\label{fig:threshold_resource}
\end{figure}

We numerically investigate deep thermalization within a typical instance of a syndrome coherence QRT by randomly drawing $H$, then
 simulating random free dynamics of mixed-basis states (specified by parameter $\alpha_0$) before  performing  mixed-basis measurements (specified by parameter $\alpha_m$). In both the initial state and final state, the local resourceless state is a qubit prepared in the $Z$ direction, while the local resourceful state is a qubit prepared in the $X$ direction. In Fig.~\ref{fig:threshold_resource}(a), we plot the ensemble-averaged local syndrome coherence $\overline{R}_S$ within the PE as a function of the globally injected syndrome coherence density, captured by the combination of parameters $\alpha_0 + \alpha_m$.
We see excellent agreement with our theoretical prediction that there exists a threshold behavior at the predicted value $\alpha_0 + \alpha_m = \sigma$: 
the plots of $\overline{R}_S$ versus $(\alpha_0+\alpha_m)/\sigma$ for different system sizes $N_B$ (with $N_A$ fixed) get progressively sharper and step-like, eventually tending to a step-function where below the threshold $\overline{R}_S = 0$, while above the threshold it saturates to the syndrome coherence of a Haar random state (dotted horizontal line in Fig.~\ref{fig:threshold_resource}(a)).
These signal that the PE tends to the classical bit-string and Haar ensemble respectively, in the thermodynamic limit.

\subsection{Mixedness and classical randomness} \label{sec:mixedness}

Next we consider the localizability of {\it mixedness}, i.e., classical uncertainty in a quantum state. 
Viewing mixedness as a resource means that {\it classical randomness}, or noise, is likewise resourceful (this can be useful, for example, in cryptographic applications). As a consequence, the set of free states (pure states) is not closed under convex combinations, making this QRT non-convex~\cite{chitambar2019quantum}. 
A more common approach in the literature is to view {\it purity} as the resource, which gives rise to a convex QRT~\cite{Streltsov_Maximal_2018,Horodecki_Reversible_2003}.
However, our block sharpening framework (Sec.~\ref{sec:generaltheory}) leads naturally to mixedness as the resource. 

Mixedness of a state $\rho$ can be quantified by 
\begin{align}
    R_P(\rho) := -\log_2\left(\Tr(\rho^2)\right). 
\end{align}
Since mixedness is invariant under general unitary dynamics, the free subgroup $\mathcal{G}_R$ for this QRT is the entire unitary group $U(2^N)$ itself. 
On the other hand it can be reduced by measurements and increased by noisy channels. It is natural to ask whether mixedness can exhibit a localizability transition in our setting. Concretely, one may consider an input state with tunable mixedness (for instance a local product state comprising some pure qubits and some maximally mixed qubits), evolve it under a random unitary, and finally apply measurements of tunable strength to $B$.
Can this setting lead to a mixedness transition in the PE? 
We find that, generically, the answer is negative: in Appendix~\ref{app:purity} we show that if the scrambling dynamics forms a unitary $2$-design, then a robust resourceless (i.e., pure) phase of the PE is impossible.
This means that quantum dynamics generically localizes classical randomness in the PE.

It is natural, then, to ask the same question of {\it classical} dynamics: does classical dynamics similarly localize randomness in the PE? 
To address this question, we restrict to classical states, i.e., diagonal density matrices 
\begin{equation}
    \rho = \sum_{\mathbf z} p_{\mathbf z} |\mathbf z\rangle \langle \mathbf z| , \label{eq:classical_state}
\end{equation}
with $p_{\mathbf z} \geq 0$ a probability distribution.
Its mixedness can be measured by $R_P=-\log_2\left(\sum_{\mathbf{z}}p_{\mathbf{z}}^2\right)$. 
Free states are pure, i.e., deterministic distributions $p_{\mathbf z} = \delta_{\mathbf z, \mathbf z_0}$; free operations are those that preserve the set of pure states, i.e., permutations of the bitstrings. 
We thus take random permutation dynamics (RPD) as the relevant class of free dynamics. Note that despite the word `random', {\it each instance} of RPD is deterministic, mapping each input $\mathbf z$ to a unique output $\pi(\mathbf z)$ for some permutation $\pi \in S_{2^N}$. RPD is thus randomness-preserving in this sense. 

The question of classical randomness in this setup maps very directly to the question of coherence addressed previously. 
Indeed, through the vectorization mapping $|\mathbf z\rangle\langle \mathbf z| \mapsto \sket{\mathbf z}$ we may formally reinterpret Eq.~\eqref{eq:classical_state} as a quantum state, in such a way that classical randomness maps to quantum coherence. 
A maximally mixed one-bit state, $p_0=p_1=1/2$, is formally analogous to a resourceful $X$-basis state in the case of coherence: $|0\rangle\langle 0| + |1\rangle\langle 1| \mapsto |0\rangle\!\rangle + |1\rangle\!\rangle := |+\rangle\!\rangle$; tracing out a qubit at the end corresponds to projecting on $\langle\!\langle +|$, analogous to an $X$-basis measurement. 
Through these identifications, the block sharpening picture leading to the coherence transition readily applies. 

Specifically, we consider the initial state
\begin{equation} 
\rho_0 = \left( \frac{\mathbb{I}_2}{2} \right)^{\otimes \alpha_0N} \otimes (|0\rangle\langle0|)^{\otimes(1-\alpha_0)N},
\end{equation}
corresponding to probability distribution $p_{\mathbf z} = 2^{-\alpha_0N}$ if $z_j=0$ for all $j>\alpha_0N$, $p_{\mathbf z} = 0$ otherwise.
Under the vectorization mapping this corresponds to $|+\rangle\!\rangle^{\otimes \alpha_0 N} \otimes |0\rangle\!\rangle^{\otimes (1-\alpha_0) N}$.
We evolve the system under RPD, then measure $(1-\alpha_m)N$ qubits in the $Z$ basis. The subsystem $A$ is chosen from the unmeasured qubits and hosts the PE. Qubits that are not in $A$ and are not measured are traced out, corresponding to projection onto $\langle\!\langle +|$. Our theory for coherence predicts a classical randomness transition in this setting determined by $\alpha_0 + \alpha_m = 1$, with a (resourceful) maximally mixed classical state above threshold and a (resourceless) pure state ensemble below threshold---the ``classical bitstring'' ensemble already encountered in the case of coherence~\cite{liu2025coherence}.

We numerically verify this classical randomness transition by computing the ensemble-averaged mixedness  $\overline{R}_{P}=-\mathbb{E}_{\mathcal{E}_{\mathrm{PE}}}\log_2\left[\sum_{\mathbf z} p_{\mathbf z}^2 \right]$ of subsystem $A$ while fixing the fraction of randomized input bits $\alpha_0=0.5$ and varying the fraction of unmeasured bits $\alpha_m$. As shown in Fig.~\ref{fig:threshold_resource}(b), the averaged randomness of classical PE states $\overline{R}_p$ undergoes a sharp transition near $\alpha_m=0.5$, consistent with the analytical prediction.

Additionally, this result readily generalizes to more restricted types of classical dynamics, where not all bitstring permutations are allowed. For instance, permutations that act trivially inside various {\it blocks} of bitstrings (but map bit-strings in one block to bit-strings of another block) can be seen to also yield a randomness transition at $\alpha_0 + \alpha_m = \alpha_{\rm crit}$, with $\alpha_{\rm crit} \leq 1$ determined by the block dimensions as in Sec.~\ref{sec:generaltheory}.
The classical randomness transition described in this section has focused on the special case $\alpha_{\rm crit} = 1$. 

To summarize our results in this section, we have tested our block sharpening theory for different QRTs with extensive entropies, predicted to be in the TL class: the ``syndrome coherence'' family of QRTs (which includes standard coherence as studied in Ref.~\cite{liu2025coherence}), and the QRT of mixedness or classical randomness.
All cases confirm the existence of a finite localizability threshold given by its predicted value, $\alpha_0 + \alpha_m = \sigma$, with $\sigma$ the second R\'enyi entropy density of the block probability distribution $p_i$. In all cases, the classical bitstring ensemble emerges below threshold, while a resourceful phase emerges above threshold (Haar-random for syndrome coherence, maximally-mixed state for mixedness).

\section{Smoothly-localizable resources} \label{sec:crossover}

We now move to a study of smoothly-localizable resources. 
Here, we analyze two prominent examples. The first is the intensive resource of imaginarity, discussed in Sec.~\ref{sec:imaginarity}. The second is the subextensive resource of $U(1)$ charge asymmetry, discussed in Sec.~\ref{sec:asymmetry}. In both cases, we will find the resource content of the PE varies smoothly as the resource injected through the initial state and measurement basis is tuned, in agreement with the prediction of our  theory in Sec.~\ref{sec:generaltheory}.

\subsection{Imaginarity} \label{sec:imaginarity}

We begin with  the resource associated with imaginarity. 
Quantum mechanics fundamentally operates within a complex Hilbert space, a mathematical feature that is key to the power of quantum information processing. What happens, then, if we restrict the theory to a purely real vector space, and treat the injection of imaginary elements as a physical asset? This question is rigorously formalized through the QRT of imaginarity~\cite{wu2021resource}. 
Under this framework, defined relative to a fixed choice of a reference basis, the free elements are strictly real-valued states and operations. Any deviation into the complex plane is treated as resourceful. In our context of deep thermalization under constraints on imaginarity, we consider the following situation: we prepare an initial state of $N$ qubits with some amount of imaginarity; 
then we apply a free scrambling unitary---a real orthogonal matrix, which does not introduce more imaginarity;
finally, we construct the PE by measuring $B$ in bases with tunable imaginarity, and we ask about the imaginarity content $\overline{R}_I$ of the PE on the remaining local subsystem $A$.

Concretely, let us fix the reference basis as the computational basis. This basis is taken to be real; it is the basis on which the operation of complex conjugation is defined.
For any pure state $\ket{\psi}$, by applying a suitable overall phase (a gauge choice), we can impose the condition $\langle \psi| \psi^\ast\rangle \in \mathbb{R}$. 
It is then easy to see that, in this gauge, there is a unique decomposition into real and imaginary components as 
\begin{align}
|\psi\rangle = a_R \ket{\psi_R} + i a_I \ket{\psi_I}, \label{eq:canonical_decomposition}
\end{align}
with real coefficients $a_R\geq a_I\geq0$ and real normalized vectors $\ket{\psi_R},\ \ket{\psi_I}$ obeying $\langle \psi_R|\psi_I\rangle = 0$. Eq.~\eqref{eq:canonical_decomposition} will henceforth be known as the ``canonical decomposition''. 
Note that $0 \leq a_I \leq 1/\sqrt{2}$. 
Free pure states are those whose canonical decompositions are purely real (i.e., $a_I = 0$), while resourceful states are complex-valued ones (i.e., $a_I>0$; there exists no global phase rotation rendering the state purely real). A measure of imaginarity valid for pure states\footnote{
More generally, a free mixed state is any convex combination of free pure states, and the resource monotone $R_I$ described above can be extended to mixed states via the so-called ``convex-roof construction''~\cite{uhlmann_2010_convexroof,chitambar2019quantum}. We mention this only for completeness; in this work we only consider pure states.
}
is given by $R_{I}(\ket{\psi}) = 1 - |\langle \psi |\psi^\ast \rangle|^2$~\cite{Haug2025pseudorandom} which takes values in the interval $[0,1]$. 
Using the canonical decomposition Eq.~\eqref{eq:canonical_decomposition}, this can also be expressed as $R_{I}(\ket{\psi}) = 1 - |a_R^2 - a_I^2|^2 = 4a_I^2(1-a_I^2)$, where we also used the normalization condition $a_R^2 + a_I^2 = 1$. One sees that this vanishes precisely if $a_I = 0$, while it attains its maximum value if $a_I = 1/\sqrt{2}$. 

Imaginarity admits a natural interpretation as a subspace coherence QRT as described in Sec.~\ref{sec:review_resources}, though formulated in a quantum mechanical system defined over a {\it real} Hilbert-space. It therefore fits within the scope of our theory in Sec.~\ref{sec:generaltheory}. Indeed, by introducing an ancillary         ``rebit''~\cite{caves_rebit_2001} (i.e., real-valued qubit), an $N$-qubit complex state $a_R \ket{\psi_R} + i a_I \ket{\psi_I}$ can be mapped to a $(N+1)$-qubit real state $a_R \ket{\psi_R}\ket{0} + a_I \ket{\psi_I}\ket{1}$. In this representation, the Hilbert-space decomposition in Eq.~\eqref{eq:block_decomposition} is realized with two $2^N$-dimensional blocks, $\mathcal{B}_0$ and $\mathcal{B}_1$, corresponding to the ancilla rebit being in the states $\ket{0}$ and $\ket{1}$, respectively. 
The imaginarity content of a state is then directly associated with superpositions between these two blocks, and can therefore be viewed as coherence of the ancilla degree of freedom. 
The group of free unitary operations $\mathcal{G}_R$ corresponds to the orthogonal group, which acts as internal transformations within each block. We do not consider the transposition of the two blocks, as this corresponds merely to multiplication by $i$ which is physically trivial. 

To characterize the localizability of imaginarity from the global system to the PE, we first consider the minimal setting in which the resource is injected solely through the initial state, while the measurement basis is chosen to be resource-free. This setting is sufficient for diagnosing whether even a small amount of initial imaginarity can become detectable in the PE. Once such detectability is established, allowing for general resourceful measurements would only provide an additional source of imaginarity, and cannot produce threshold behavior. 

We consider an $N$-qubit input state (sticking to the language of the original complex Hilbert space),
\begin{equation} 
\ket{\Psi_0} =   \cos(\alpha)\ket{0}^{\otimes N} + i \sin(\alpha)\ket{1}^{\otimes N} , 
\end{equation}
where the parameter $\alpha$ continuously tunes the imaginarity: $R_{I}(\ket{\Psi_0}) = \sin^2(2\alpha) =: R_{I,\text{in}}(\alpha)$. After scrambling by a Haar-random real orthogonal matrix $U_O$, subsystem $B$ is measured in the computational ($z$) basis, which is real and therefore resource-free. We now use the rebit representation introduced above as a diagnostic tool to analyze the imaginarity of the PE. Under this embedding, the real and imaginary parts of the initial state occupy two ancillary-rebit sectors, $\mathcal{B}_0$ and $\mathcal{B}_1$. Since $U_O$ acts only on the physical $N$-qubit system, it acts identically within the two sectors and does not mix them. Therefore, for a measurement outcome $\nu$, the $\mathcal{B}_0$ and $\mathcal{B}_1$ components of the projected state $|\psi(\nu)\rangle_A$  are given by
\begin{align}
    &\frac{\cos(\alpha)}{\sqrt{p(\nu)}}(\mathbb{I}_A \otimes\langle \nu|)U_O|0\rangle^{\otimes N}\otimes|0\rangle,\\
    &\frac{\sin(\alpha)}{\sqrt{p(\nu)}}(\mathbb{I}_A \otimes\langle \nu|)U_O|1\rangle^{\otimes N}\otimes |1\rangle,
\end{align}
with Born probability $p(\nu) = \left|(\mathbb{I}_A \otimes \bra{\nu})U_O\ket{\Psi_0}\right|^2$. 

Within each block $\mathcal{B}_{0,1}$, the projected component takes the form $(\mathbb{I}_A \otimes \langle \nu|)U_O\ket{\psi}$, where $\ket{\psi}$ is a real state. These components are proportional to real Haar-random states, since the orthogonal unitary $U_O$ fully scrambles them in the real Hilbert space. Consequently, the projected state $|\psi(\nu)\rangle_A$  can be written as $a_0'\ket{\psi_{r_0}}\ket{0} +  a_1'\ket{\psi_{r_1}}\ket{1}$, where $\ket{\psi_{r_0}}$ and $\ket{\psi_{r_1}}$ are independent real Haar-random states, not necessarily orthogonal, while $a_0', a_1'$ are real coefficients
\begin{align}
a'_0 & = \cos(\alpha)\frac{\sqrt{\bra{0}^{\otimes N} U_O^T (\mathbb{I}_A \otimes |\nu\rangle\langle\nu|) U_O \ket{0}^{\otimes N}}}{\sqrt{p(\nu)}} \nonumber \\
a'_1 & = \sin(\alpha)\frac{\sqrt{\bra{1}^{\otimes N} U_O^T (\mathbb{I}_A \otimes |\nu\rangle\langle\nu|) U_O \ket{1}^{\otimes N}}}{\sqrt{p(\nu)}}
\end{align}

The imaginarity of the projected state can thus be expressed as
\begin{align}
    R_I(|\psi(\nu)\rangle_A) & = 1 - |\langle \psi(\nu)|\psi(\nu)^*\rangle_A|^2\nonumber \\
    & = 
    4(a_0'a_1')^2 [1-\langle \psi_{r_0}|\psi_{r_1}\rangle^2].
\end{align}
It is easy to see that there should be no threshold for imaginarity in the PE even for small $\alpha$.
Indeed, the typical overlap squared between real Haar-random states on $A$ is $\approx 2^{-N_A}$, so generically the term $1-\langle \psi_{r_0}|\psi_{r_1}\rangle^2$ is positive and not vanishing. 
Moreover, focusing on $\alpha \ll 1$, one finds 
\begin{equation}
    (a_1')^2 \simeq \alpha^2 \frac{\bra{1}^{\otimes N} U_O^T (\mathbb{I}_A \otimes |\nu\rangle\langle\nu|) U_O \ket{1}^{\otimes N}}{p(\nu)} 
\end{equation}
where the numerator in the fraction is simply the probability $q(\nu)$ of measuring $\nu$ on the real Haar-random state $U_O|1\rangle^{\otimes N}$. 
On average over $\nu \sim p(\nu)$, this gives 
\begin{equation}
    \mathbb{E}_{\nu \sim p} (a_1')^2 
    \simeq \alpha^2 \sum_\nu p(\nu) \frac{q(\nu)}{p(\nu)} 
    = \alpha^2. 
\end{equation}
Similar analysis yields $\mathbb{E}_{\nu \sim p} (a_0')^2 \sim 1$ for small $\alpha$. 
Since $R_{I,\text{in}} \sim 4\alpha^2$ in this regime, we see that the imaginarity content of the projected state increases smoothly from zero with the injected global imaginarity from the initial state $\ket{\Psi_0}$, 
with the relation
\begin{align}
\overline{R}_I \approx (1-2^{-N_A}) R_{I,\text{in}}
\label{eqn:prediction_imaginarity}
\end{align}
for small $R_{I,\text{in}}$. 

\begin{figure}
    \centering
    \includegraphics[width=1.0\linewidth]{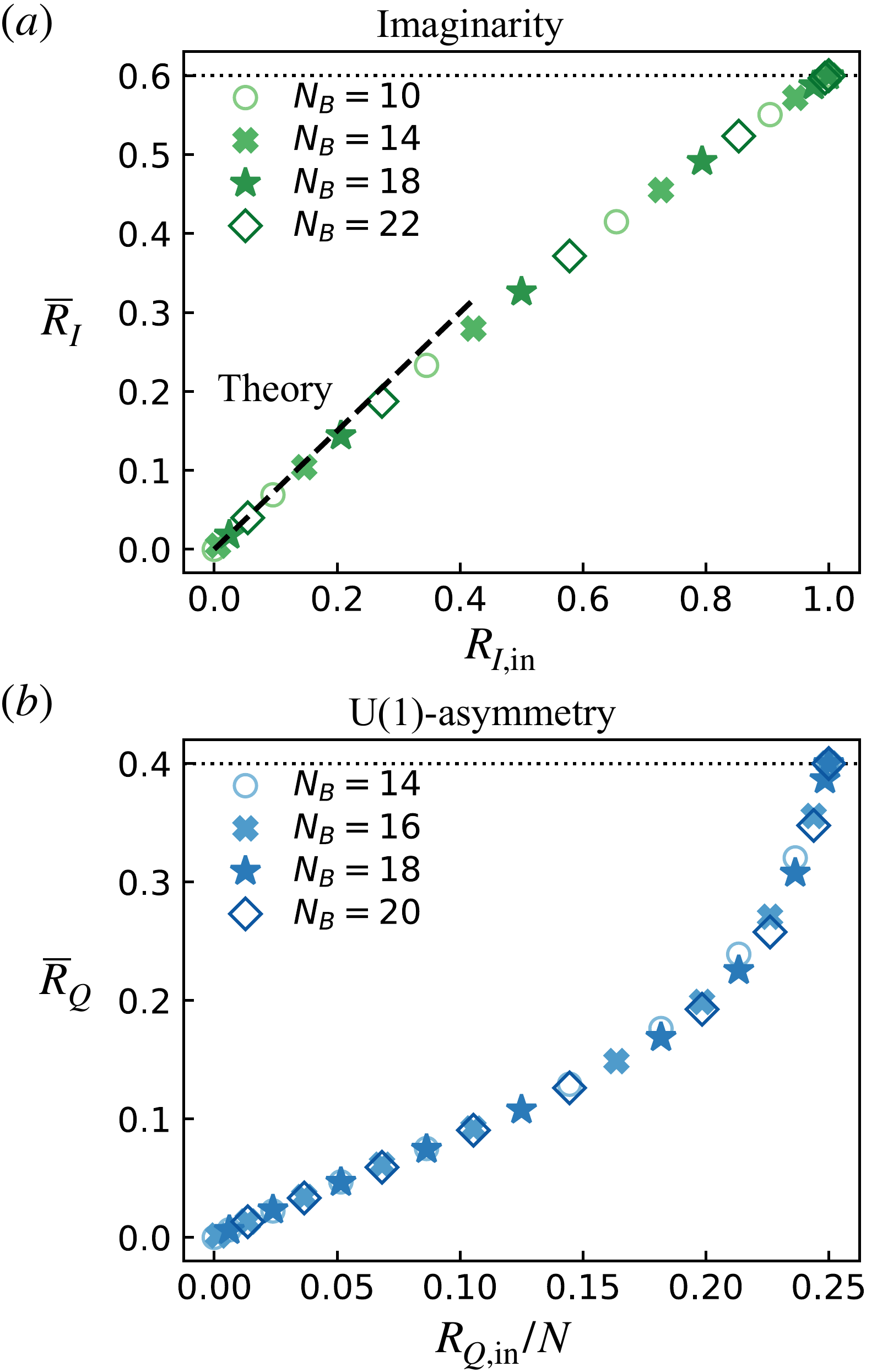}
    \caption{\justifying 
    Smooth localizability of imaginarity and $U(1)$-asymmetry. All results are for $N_A=2$. 
    (a) Average imaginarity of the PE, $\overline{R}_I$, as a function of the imaginarity $R_{I,\text{in}}$ of the initial state under Haar-random real orthogonal dynamics (see Sec.~\ref{sec:imaginarity}). The theoretical prediction Eq.~\eqref{eqn:prediction_imaginarity} at small $\alpha$ is depicted by the dashed line (here the slope is $3/4$), showing good agreement. The average imaginarity of a Haar random ensemble is indicated by the horizontal dashed line.
    (b) Averaged charge variance $\overline{R}_Q$ of the PE versus charge-variance density $R_{Q,\text{in}}/N$ of the input state under random asymmetry-free dynamics (see Sec.~\ref{sec:asymmetry}). The horizontal dotted line is the corresponding Haar-random value. }
    \label{fig:crossover_resource}
\end{figure}

To verify this prediction, we numerically evaluate the ensemble-averaged imaginarity in the above setup.
As shown in Fig.~\ref{fig:crossover_resource}(a), $\overline{R}_{I}$ increases smoothly from zero with a slope predicted by Eq.~\eqref{eqn:prediction_imaginarity}, upon the  injection of global imaginarity $R_{I,\text{in}}$. As the  curves of various system sizes $N_B$ lie on top of each other, this confirms the presence of a crossover rather than a sharp transition even in the thermodynamic limit.

This result is consistent with the prediction of our general theory in Sec.~\ref{sec:generaltheory} that intensive resources should exhibit a crossover, rather than a sharp transition, in the PE resource. Since imaginarity corresponds to the coherence of a single rebit, any initial state carrying an intensive amount of resource significantly populates both of the blocks; likewise, any measurement that is not a real basis measurement is compatible with both blocks (i.e., it is always {\it a priori} possible for unmeasured qubits in $A$ to have imaginarity); since both blocks are populated by the input state and compatible with the measurement outcome, the subset intersection problem is always trivial, ruling out a sharp transition. 
A complementary analytical approach based on Weingarten calculus confirms the same result~\cite{us_2026_replica}. 

\subsection{Asymmetry \label{sec:asymmetry}}

Another well-studied quantum resource is asymmetry, which quantifies coherence between symmetry sectors. The case of $\mathbb{Z}_2$ symmetry (parity), briefly illustrated in Sec.~\ref{sec:setup}, is closely related to imaginarity. It is an intensive resource defined by a decomposition of the Hilbert space into two blocks of dimension $2^{N-1}$ each (even- and odd-parity bitstrings, respectively); the only difference with respect to imaginarity is that the Hilbert space is complex, which does not qualitatively change the outcome. 
$\mathbb{Z}_2$-asymmetry is also an instance of ``syndrome coherence'', Sec.~\ref{sec:rand_stab}, for a code with a single parity check $Z_1 Z_2 \cdots Z_N$ (the symmetry itself). The absence of a finite threshold then also follows from our general prediction for this family of resources, which is the threshold condition $\alpha_0 + \alpha_m = \sigma$ with $\sigma \to 0$. 

In the following we focus on the case of $U(1)$ symmetry~\cite{bartlett2007reference}, which exhibits distinct physics. 
In the QRT of $U(1)$ asymmetry, the Hilbert space decomposes into charge sectors
\begin{equation}
\mathcal{B}_Q = \mathrm{Span}\left\{ \ket{\mathbf z} : \sum_{i=1}^N z_i = Q \right\}, \quad Q=0,\dots,N,
\end{equation}
with dimensions $d_Q = \dim(\mathcal{B}_Q)=\binom{N}{Q}$. Resourceful states are those that exhibit coherence between different charge sectors. 
The group of free unitaries consists of arbitrary unitaries within each sector (i.e., $U(1)$-symmetric unitaries), together with allowed permutations between sectors of equal dimension. Because the sector dimensions differ, the only nontrivial permutations are exchanges $Q \leftrightarrow N-Q$, corresponding physically to a particle-hole transformation. Combining all these together, the corresponding resource-free operations take the form 
\begin{equation} 
U = U_\pi   \bigoplus_{Q=0}^N V_Q , 
\label{eq:freeU_charge}
\end{equation}
where $V_Q$ is a generic unitary acting within charge sector $\mathcal{B}_Q$ and $U_\pi$ is an allowed permutation between charge sectors, i.e., for each $Q$ it may independently exchange $Q\leftrightarrow N-Q$. 
This operation does not conserve charge but does not generate asymmetry.

The block probability, $p_Q = d_Q/2^N = 2^{-N} \binom{N}{Q}$, scales as $\sim N^{-1/2}$ for the largest blocks (those near charge neutrality: $Q = N/2 \pm O({N}^{1/2})$) and is exponentially small otherwise. 
The entropy of resource Eq.~\eqref{eqn:Renyi} scales logarithmically, $\sim \frac{1}{2} \log(N)$, making this resource subextensively divergent. 
Based on this scaling, our theory (Sec.~\ref{sec:marginal}) predicts SL behavior for deep thermalization under a $U(1)$ asymmetry QRT, with the possibility of a ``hidden'' threshold.

Deep thermalization under $U(1)$-conserving dynamics---which are a subset of free dynamics in the QRT of $U(1)$ symmetry---had been investigated in Ref.~\cite{chang2025deep}. There they found that the PE is generally described by the so-called {\it generalized Scrooge ensemble} (GSE), which encompasses as two extreme limits, the fully Haar-random ensemble, and the resourceless ``diagonal ensemble'' composed of random states of definite charge. Here we extend the analysis to those of {\it general} free dynamics with respect to the QRT of $U(1)$ asymmetry (thus going beyond just charge-conserving dynamics). We verify that these two limiting forms of the PE seen in Ref.~\cite{chang2025deep} are obtained as endpoints of a smooth crossover in the amount of global $U(1)$ asymmetry resource injected, without a sharp transition.

We first consider the tilted-basis initial state 
\begin{equation}
\ket{\Psi_0(\theta)} =\left(\cos(\theta/2)\ket{0} +\sin(\theta/2)\ket{1}\right)^{\otimes N} 
\end{equation}
and evolve it under a random free unitary $U$ as in Eq.~\eqref{eq:freeU_charge} (with each $V_Q$ chosen Haar-randomly and each $Q\leftrightarrow N-Q$ transposition applied with $1/2$ probability). 
Finally we measure $B$ in the (resourceless) $Z$ basis to form the PE. A valid resource monotone of a given state $\rho$ is captured by its 
so-called ``skew information''~\cite{Girolami_Observalbe_2014}:
\begin{align}
R_Q(\rho) & := -\frac{1}{2}\text{Tr}\left( [\sqrt{\rho},Q]^2 \right) \nonumber \\
& = \text{Tr}(\rho Q^2) - \text{Tr}(\sqrt{\rho}Q\sqrt{\rho}Q).
\end{align}
In the case of a pure state, this reduces to its variance of charge
\begin{equation}
   R_Q(\ket{\psi}) = \langle \psi|Q^2|\psi\rangle - \langle \psi|Q|\psi\rangle^2. 
\end{equation}
As shown in Fig.~\ref{fig:crossover_resource}(b), the local ensemble-averaged variance of charge 
$\overline{R}_{Q}$ within the PE changes smoothly with the  global charge variance density $R_{Q,\mathrm{in}}/N=0.25\sin^2(\theta)$ set by the initial state $|\Psi_0(\theta)\rangle$, confirming the absence of a sharp transition. 

This outcome is consistent with the known phenomenology of {\it charge sharpening} in $U(1)$-symmetric monitored dynamics~\cite{PhysRevX.12.041002,Barratt_Transitions_2022}. 
``Charge sharpening'' refers to the loss of superposition between charge sectors over the course of quantum dynamics featuring measurements of the local charge density (e.g., $Z_i$ in our case) at a finite rate. It is, in the present language, a dynamical transition in $U(1)$ asymmetry. 
Charge sharpening is known to occur when measurements are applied everywhere in space-time at a rate that exceeds some finite critical value. Below the critical rate, one has a ``fuzzy'' phase that preserves asymmetry for extensive time; above threshold, a ``sharp'' phase, where asymmetry is lost in constant or logarithmic time. 
From the standpoint of charge sharpening, the absence of a transition in the PE is unsurprising: the scrambling dynamics in our case is purely unitary, and measurements appear only at the final time---a boundary of the spacetime, corresponding to a vanishing bulk density. Thus the PE is always in the ``fuzzy'' (i.e., resourceful) phase, with a nonuniversal amount of resource. 

This analysis confirms the expected SL behavior of $U(1)$ asymmetry.
However, in Sec.~\ref{sec:generaltheory} we also argued that subextensive resources should have a ``hidden'' threshold, visible only at vanishing density of injected resource. 
Looking for this threshold in the case of $U(1)$ reveals a surprise. The prescription in Sec.~\ref{sec:generaltheory}, to use a local product basis with a vanishing tilt angle ($\theta \sim N^{-\beta_0}$) or a mixed basis with resource localized in a logarithmically sized subsystem, does not work in this case. 
The reason is that the (relatively) small number of bitstrings produced by these input states {\it are not} spread sufficiently widely across blocks, such that the mechanism of random subset intersection we identified in our theory in Sec.~\ref{sec:generaltheory} is not applicable. Indeed, the bit-strings cluster into a very small number of blocks. As an example, take a mixed-basis input state with the resourceful $\ket{+}$ state on $\beta_0 \log_2 N$ qubits and resourceless $|0\rangle$ state on the remaining qubits; this yields $N^{\beta_0}$ bitstrings, but only populates $1 + \beta_0 \log_2(N)$ blocks! This is due to the massive degeneracy among Hamming weights of bitstrings that differ only on a small subregion. It is a manifestation of the local structure of the Hilbert space blocks, which do not resemble the typical random blocks considered in Sec.~\ref{sec:generaltheory}. 
Thus, starting from a local product basis, $U(1)$ asymmetry does {\it not} show a threshold, not even at vanishing density. 

The mechanism of this obstruction also points to a natural solution to achieve a transition: to inject $U(1)$ asymmetry not in a local product basis, but instead in a long-range-entangled basis. As we discuss in Appendix~\ref{app:u1}, by instantiating a ``cat-like state'' with $\sim N^{\beta_0}$ bitstrings  ($\beta_0 < 1/2$) each in a distinct charge sector, one lifts the locality obstruction and recovers the predicted hidden threshold, verifying the final piece of our general theory.

\section{Heisenberg picture resources}
\label{sec:heisenberg}

Thus far, we have focused on QRTs based on a block decomposition of the {\it state} Hilbert space. Here we consider QRTs where the relevant block decomposition arises instead in the {\it operator} Hilbert space. 
This setting allows us to extend our analysis of resource localizability to the important QRTs of non-stabilizerness (magic)~\cite{bravyi2005universal,Bravyi_Improved_2016,Howard_application_2017,Liu_Manybody_2022,Leone_Stabilizer_2022} and non-Gaussianity~\cite{Hebenstreit_AllPure_2019,3yx4-1j27,PhysRevA.97.062337,PhysRevA.98.052350,PhysRevA.97.062337,PhysRevA.98.052350}. 
While strictly speaking these are not subspace-coherence QRTs themselves, their relevant free operations (Clifford unitaries and fermionic Gaussian unitaries, respectively) do give rise to invariant blocks in operator space. These invariant blocks will inform our understanding of resource localizability of such QRTs, in line with our general theory of block sharpening (Sec.~\ref{sec:generaltheory}), but now in operator space. 

The key point is that the framework of Sec.~\ref{sec:generaltheory}, developed in state-space, can be readily adapted to the  operator-space setting via the Choi--Jamio{\l}kowski isomorphism~\cite{jamiolkowski1972linear,choi1975completely}. 
An operator $O$ can be mapped to a Choi state $|O\rangle\!\rangle$ in a doubled Hilbert space,
while its Heisenberg-picture evolution $O \mapsto U O U^\dagger$, with $U$ a resource-free unitary, is mapped to the state evolution $|O\rangle\!\rangle \mapsto (U \otimes U^*) |O\rangle\!\rangle $. Measurements can likewise be performed by projection on a basis of operators for subsystem $B$, ${}_B\sbra{\Phi_\nu}$.

\begin{figure*}
    \centering
    \includegraphics[width=1.0\linewidth]{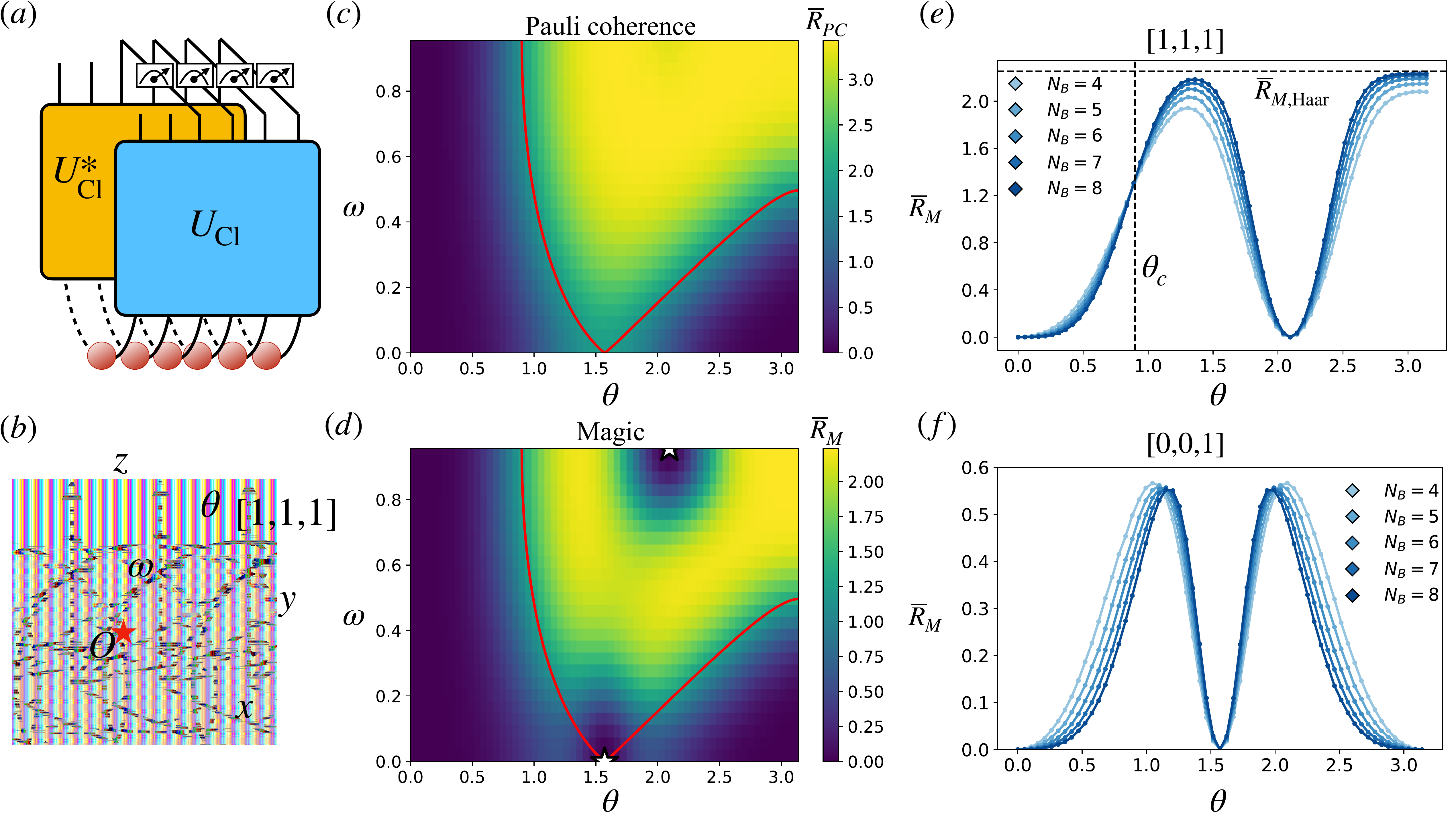}
    \caption{\justifying (a) Physical realization of our operator evolution setup in 1D as qubits arranged in a ladder geometry, depicted for the case of magic. An initial operator $\sket{u(\theta,\mathbf{n})}^{\otimes N}$ (each $u(\theta,\mathbf{n})$ corresponds to a red circle)
    is evolved under a random Clifford unitary channel $U_{\rm Cl} \otimes U^{\ast}_{\rm Cl}$, then the bath $B$ is measured in the ``phase basis'', Eq.~\eqref{eq:phase_basis}.
    (b) Illustration of the angles $\theta$ and $\omega$ used to parametrize $u(\theta,\mathbf n) = e^{-i\frac{\theta}{2} \mathbf{n} \cdot \boldsymbol{\sigma} }$. The unit vector $\mathbf{n}$ moves along the great circle (red) from direction $[0,0,1]$ to direction $[1,1,1]/\sqrt{3}$, forming an angle $\omega$ with $[0,0,1]$. 
    (c) Average Pauli coherence $\overline{R}_{PC}$ of the PE for $N_A=2$ and $N_B=8$. The critical boundary predicted by Eq.~\eqref{eq:magic_crit} is shown in red. 
    (d) Average magic of the PE $\overline{R}_M$ for $N_A=2$ and $N_B=8$. The critical boundary (red curve) is the same as in (c). The two Clifford points $(\theta,\omega)=(\pi/2,0)$ and $(2\pi/3,\arccos(1/\sqrt{3}))$ are marked by white stars. The data in (c) and (d) are obtained by averaging over $10240$ realizations. 
    (e) Average PE magic $\overline{R}_{M}$ versus $\theta$ for $\mathbf{n} = [1,1,1]/\sqrt{3}$ (i.e., $\omega = \arccos(1/\sqrt{3})$), $N_A=2$, and variable $N_B$. The low magic regime near the Clifford point $\theta=2\pi/3$ shrinks as $N_B$ increases and converges to the Haar random value (horizontal dashed line). The critical point $\theta_c$ (marked by the vertical line) is obtained by solving Eq.~\eqref{eq:magic_crit} with $\omega = \arccos(1/\sqrt{3})$. 
    (f) Average PE magic $\overline{R}_{M}$ versus $\theta$ for $\mathbf{n} = [0,0,1]$ (i.e., $\omega = 0$), $N_A=2$, and variable $N_B$.  The magic features near the Clifford point $\theta = \pi/2$ shrink as $N_B$ increases, indicating the absence of a stable magic phase anywhere along this slice of the phase diagram.}
    \label{fig:bell_phase_diagram}
\end{figure*}

Importantly, viewing Heisenberg-picture operator evolution as Schr\"odinger-picture  state evolution in a doubled Hilbert space is not merely a mathematical trick; it can be realized as a concrete physical setup. Consider for simplicity a one-dimensional many-qubit system, in which there are {\it two} columns of qubits placed side by side, denoted ``left'' and ``right'' respectively. 
This describes a ladder geometry. 
Then, each column (i.e., each leg of the ladder) represents one of the two Hilbert space copies.
Given an operator $O$, 
its Choi state $\sket{O}$ is physically realized in this system by preparing the Bell state $\ket{00} + \ket{11}$ on each rung of the ladder (corresponding to the vectorization of the identity operator, $\sket{\mathbb{I}_2}$, up to normalization), and then acting with $O$ (assumed unitary throughout this section) on the left qubit of the rung: $(O\otimes \mathbb{I}_2) \sket{\mathbb{I}_2} = \sket{O}$. Likewise, measurements in the doubled Hilbert space are implemented as measurements in a local two-qubit basis on each rung. See Fig.~\ref{fig:bell_phase_diagram} for an illustration.

\subsection{Non-stabilizerness (magic) \label{sec:magic}}

The QRT of non-stabilizerness (or magic) is an especially interesting application of our general framework. Several recent works have investigated, theoretically and experimentally, the possibility of magic phase transitions~\cite{Niroula_PT_2024,sierant2026theorymagicphasetransitions,Cheng_Emergent_2025} especially in the setting of quantum error correcting codes. This setting carries important implications for applications in quantum information science: a magic phase transition presents the possibility of controlling and preparing useful resource states~\cite{Niroula_PT_2024,sierant2026theorymagicphasetransitions} or even computationally universal unitary gates~\cite{Cheng_Emergent_2025}, which are crucial to fault-tolerant quantum computation.  Below, we study deep thermalization within constraints set by the QRT of magic.

\subsubsection{Threshold prediction from Pauli coherence}

As reviewed in Sec.~\ref{sec:review_resources}, the QRT of magic has free unitary operations composed of the Clifford group, whose adjoint action permutes Pauli operators. 
Thus, let us build a subspace coherence QRT in operator space based on a decomposition of operator space into Pauli blocks $\bigoplus_P \mathcal{B}_P$, with each block $\mathcal{B}_P$ the one-dimensional span of a Pauli $P$; there are $4^N$ such blocks. Now, a free element consists of a single Pauli, while a sum of Paulis is resourceful. In the doubled Hilbert space language, this means a state like $|P\rangle\!\rangle$ is free while a  state like $|O\rangle\!\rangle=\sum_P c_P |P\rangle\!\rangle$ with arbitrary coefficients $c_P$ is generically not. 
The resource is coherence between Pauli operators,
\begin{align}
R_{PC}(|O\rangle\!\rangle) := -\sum_P |c_P|^2\log_2 |c_P|^2
\label{eq:pc_monotone_def}
\end{align}
that we will refer to as ``Pauli coherence''.

As mentioned, free unitary scrambling dynamics of the  Pauli coherence QRT consists of random Clifford unitaries $U_\text{Cl}$ acting by conjugation on the input operator $O$ (and thus acting as $U_\text{Cl}\otimes U_\text{Cl}^\ast$ on its Choi state: $U_\text{Cl}\otimes U_\text{Cl}^\ast |P\rangle\!\rangle = |P'\rangle\!\rangle$).  
These map each Pauli to another Pauli: $P' = U_\text{Cl}PU_\text{Cl}^\dagger$, describing permutations of the Pauli blocks.
Unlike RPD for bitstrings though, such permutations cannot be completely general---for example, they must preserve the algebra of Paulis: $(PQ)' = P' Q'$. 
Nevertheless, it is reasonable to conjecture that these permutations carry enough randomness to justify our random subset intersection picture discussed in Sec.~\ref{sec:generaltheory}. Since Pauli coherence is an extensive resource with exponentially many blocks, our theory predicts TL behavior, i.e., there can be a Pauli coherence phase transition in deep thermalization. 

What is the relation of the localization of Pauli coherence to the localization of magic in deep thermalization? 
Strictly speaking, Pauli coherence and magic are not identical: coherence between Pauli blocks in operator space is {\it necessary} but {\it insufficient} for magic. 
For example, the absence of Pauli coherence implies a single Pauli operator and thus the absence of magic; however, Clifford operators such as $H = (X+Z)/\sqrt{2}$ have Pauli coherence but do not harbor magic.
Nevertheless, the concept of Pauli coherence is useful because its threshold behavior implies the existence of a robust {\it non-magic} phase, which suffices to establish that magic is also a TL resource. 
As we shall see in our numerical investigations below, the threshold we derive for Pauli coherence in fact matches the threshold of magic generically, except at isolated points in parameter space where dynamics is Clifford.

Let us now describe in detail our model to study the localizability of magic in deep thermalization. 
As described at the beginning of this section, we work in the doubled Hilbert space picture of operator evolution, physically realized as two columns of qubits arranged in a ladder geometry; see Fig.~\ref{fig:bell_phase_diagram}(a). We prepare qubits in a uniform product state $|\Psi_0 \rangle \! \rangle = |O\rangle \!\rangle^{\otimes N}$ of local Bell states $|O\rangle\!\rangle$ defined on rungs, with the choice of operator $O$ allowing us to tune the amount of Pauli coherence; thus this corresponds to the tilted-basis model, see Sec.~\ref{sec:review_models}. It is convenient to take $O$ to be a unitary and define the orthonormal basis $(\sigma^{\alpha}\otimes\mathbb{I}_2) \sket{O}$. 
Decomposing $O = \sum_i c_\alpha\sigma_\alpha$, each element of this basis has Pauli coherence
\begin{equation}
    R_{PC}[O] = -\sum_{\alpha = 0}^3 |c_\alpha|^2 \log_2 |c_\alpha|^2,
    \label{eqn:PC_op}
\end{equation}
which can take on any value in $[0,2]$. For example, $O = \mathbb{I}_2$ yields the Pauli basis $\{ |\mathbb{I}_2\rangle\!\rangle, |X\rangle\!\rangle, |Y\rangle\!\rangle, |Z\rangle\!\rangle \}$, corresponding to the two-qubit Bell states
$\frac{1}{\sqrt{2}}\{ |00\rangle+|11\rangle, |01\rangle+|10\rangle, -i|01\rangle+i|10\rangle,|00\rangle-|11\rangle \}$. This basis is free, i.e., has zero Pauli coherence. 
Taking $O$ to be the phase gate $S := \frac{1}{\sqrt 2} (\mathbb{I}_2 + i Z)$ instead yields the ``phase basis''
\begin{equation}
    \{ \sket{S}, \sket{SX}, \sket{SY},\sket{SZ}\},
    \label{eq:phase_basis}
\end{equation}
with exactly one bit of Pauli coherence per rung.
Finally, an example of a basis of rung states with the maximum Pauli coherence of two bits per rung is obtained by setting $O$ to $C := \frac{1}{2}(\mathbb{I}_2 + iX+iY+iZ)$, a Clifford gate with equal amplitude on all four Paulis. 
Note that while all three examples presented above are Clifford, a generic choice of $O$ will result in a basis with magic. 

Because $|\Psi_0\rangle\!\rangle$ is a uniform product state,
we can  define its {\it density} of Pauli coherence $\alpha_0 \in [0,2]$ via $\alpha_0 = R_{PC}[O]$.  
Similarly, a measurement basis defined by  a tensor product of local Bell states of the form $(\sigma^{\alpha}\otimes\mathbb{I}_2) \sket{O'}$, specified by $O'$, yields a Pauli coherence density  $\alpha_m = R_{PC}[O']\in [0,2]$. 
Since there are $4^N = 2^{2N}$ blocks in operator Hilbert space,
the threshold condition Eq.~\eqref{eq:threshold} for deep thermalization within the QRT of Pauli coherence is  predicted to be 
\begin{equation}
    \alpha_0 + \alpha_m = 2. 
    \label{eq:pc_predicted_threshold}
\end{equation}
Namely, if $\alpha_0 + \alpha_m > 2$, we predict a resourceful, highly Pauli-coherent phase; 
if $\alpha_0 + \alpha_m < 2$, we predict a resourceless, Pauli-incoherent deep-ergodicity-breaking phase. The latter is a PE composed of projected states $|P\rangle\!\rangle$ obtained from random Pauli operators $P$, analogous to the classical bitstring ensemble for the case of the QRT of coherence. 

However, whether the proliferation of Pauli coherence within the resourceful phase coincides with the proliferation of magic is a nontrivial question. To investigate this, we turn to numerical simulations.

\subsubsection{Numerics}

We let the operator $O$ in the initial state be 
\begin{align}
u(\theta,\mathbf n) = e^{-i\frac{\theta}{2} \mathbf{n} \cdot \boldsymbol{\sigma} }
\label{eq:initstate_magic},
\end{align}
whose action is a rotation by $\theta$ around the axis $\mathbf{n}$, so that 
\begin{equation}
\sket{\Psi_0} = \sket{u(\theta,\mathbf n)}^{\otimes N}. 
\end{equation}
Above, $\theta \in[0,\pi]$ and $\mathbf n$ is a unit vector. We will tune the latter  continuously from  $[0,0,1]$ to  $[1,1,1]/\sqrt{3}$, moving along the smaller arc of the great circle connecting the two points, which can be conveniently captured by a single parameter $\omega \in [0,\arccos(1/\sqrt{3})]$ denoting the angle from the north pole of the Bloch sphere to $\mathbf{n}$. 
Therefore, the parameter space we consider is the two-dimensional $\theta$-$\omega$ plane. 
By moving in this plane, we can continuously tune the Pauli coherence density of the initial state $\alpha_0$ within the entire range $[0,2]$. 

For measurements on the $B$ subsystem, we choose to measure each rung in the ``phase basis'', Eq.~\eqref{eq:phase_basis}, which has zero magic but exactly 1 bit of Pauli coherence per rung; see  Fig.~\ref{fig:bell_phase_diagram}(a). Hence, $\alpha_m = 1$. 
This generates projected states on the remaining $N_A$ rungs (which has $2N_A$ qubits).
Our theoretical prediction for the threshold, Eq.~\eqref{eq:pc_predicted_threshold}, therefore reduces to
\begin{equation}
    R_{PC}[u(\theta,\omega)] = 1.   
    \label{eq:magic_crit}
\end{equation}
Eq.~\eqref{eq:magic_crit} is a key result of our analysis: it describes the critical curve  separating the Pauli coherent phase and Pauli incoherent phase in the $\theta$-$\omega$ plane.

We numerically confirm this result. We compute  the ensemble-averaged Pauli coherence $\overline{R}_{PC}$ of the PE over the parameter space, where $R_{PC}$ is given by Eq.~\eqref{eq:pc_monotone_def} (in that expression, $|O\rangle\!\rangle$ is taken to be a projected state $|\psi_A\rangle \!\rangle$  and $P$ is taken a Pauli operator $P_A$ acting on $N_A$ qubits; there are $4^{N_A}$ such operators).
As shown in Fig.~\ref{fig:bell_phase_diagram}(c), the phase diagram of Pauli coherence features extended resource-free and resourceful regions. Moreover,  the observed phase boundaries are   captured excellently by the expression Eq.~\eqref{eq:magic_crit} (red curves in the figure), validating our theory of threshold behavior for Pauli coherence.

To compare to magic, we further compute the ensemble-averaged second stabilizer Rényi entropy (SSRE) $\overline{R}_M$~\cite{Leone_Stabilizer_2022,Haug2023stabilizerentropies,Leone_Stabilizer_2024}, where 
\begin{equation}
R_M(\sket{\psi_A})
:= -\log_2\!\left[
\sum_{P_A,Q_A}
\frac{ |\sbra{\psi_A} (P_A \otimes Q_A)  \sket{\psi_A} |^4}{4^{N_A}}  
\right],
\end{equation}
with $P_A$ and $Q_A$ each ranging over the $4^{N_A}$ Paulis which act on $N_A$ qubits.
The SSRE is a commonly used magic monotone valid for pure states. 

Intriguingly, the phase diagram of magic, shown in Fig.~\ref{fig:bell_phase_diagram}(d), appears identical to that of Pauli coherence, with the exception of small regions around special points $(\theta = \pi/2$, $\mathbf n=[0,0,1])$ and $(\theta = 2\pi/3$, $\mathbf n = [1,1,1] / \sqrt{3})$ in which Pauli coherence is high but magic is small (white stars in the figure).
The origin of the lack of magic at such  points can easily be  understood: at these special points, the unitaries $u(\theta,\omega)$ are  Clifford gates. Indeed, in the former, $u(\theta,\omega$) is the phase gate $S = \frac{1}{\sqrt 2}(\mathbb{I}_2+iZ)$, and has $R_{PC}(S) = 1$ (thus the system sits exactly at the threshold of localization), while in the latter, it is the Clifford gate $C = \frac{1}{2}(\mathbb{I}_2 + iX+iY+iZ)$, with $R_{PC}(C) = 2$ (which places the system deep in the Pauli coherent phase).
At these points, no magic is injected by either the (stabilizer) input state, the (stabilizer) measurement basis, or the (Clifford) unitary dynamics, all of which are free; thus the resulting PE cannot have magic, even if it is  Pauli coherent.

An immediate question is whether  these special Clifford points are surrounded by robust low-magic regions in the thermodynamic limit (in the sense that there is a finite ball in parameter space around them which retains low or no magic).
We find this is not the case.
A finite-size scaling of $\overline{R}_{M}(\theta)$ around the Clifford point $(\theta= 2\pi/3,\mathbf{n} = [1,1,1]/\sqrt{3})$, plotted as a function of $\theta$,  is shown in Fig.~\ref{fig:bell_phase_diagram}(e). 
We see that the low-magic region around $\theta = 2\pi/3$ shrinks with increasing $N_B$ as $\Delta\theta\sim N_B^{-1/2.4}$, 
such that the magic of the PE for any point near but not at $\theta = 2\pi/3$ increases, consistent with an approach to the Haar random value $\overline{R}_{M,\mathrm{Haar}}$ in the thermodynamic limit. 
Similar proliferation of magic in the thermodynamic limit is seen when we move away from the special point in the $\omega$ direction (results not shown). 
In Fig.~\ref{fig:bell_phase_diagram}(f), we perform a finite-size scaling analysis of the other special Clifford point at $(\theta=\pi/2,\mathbf n= [0,0,1])$, plotting $\overline{R}_M$ also along the $\theta$ direction.
We see for finite system sizes $N_B$, magic vanishes at $\theta=\pi/2$ (the Clifford point) but exhibits peaks on either side. 
With increasing system size $N_B$, we see the peaks approach the $\theta = \pi/2$ Clifford point and their widths shrink toward zero. Together, these suggest that $\overline{R}_M \to 0$ as $N_B \to \infty$ for {\it any} point $\theta$ along the $\omega = 0$ slice of parameter space. Indeed, this is predicted by our theory:
because the no-magic point $\theta = \pi/2$ is marginal for Pauli coherence, sitting exactly at the threshold of localizability, any $\theta$ away from $\pi/2$ drives the system into the Pauli-incoherent phase. Since Pauli coherence is a necessary condition for magic,  the PE cannot host magic, in line with the above numerical observations. 
On the other hand, when we move  away from the special point $(\theta=\pi/2,\mathbf n= [0,0,1])$ in the positive $\omega$ direction, we find that magic immediately proliferates  (results not shown). 
All together, our results strongly suggest that the no-magic (Clifford) points are {\it fine-tuned}, and that barring such exceptions, the proliferation of Pauli coherence immediately leads to the proliferation of magic. 

Lastly, we probe the nature of the magic critical point $\theta_c$ encountered along the $\omega = \arccos(1/\sqrt{3})$ slice of the phase diagram (i.e., where $\mathbf{n} = [1,1,1]/\sqrt{3}$), obtained by solving $\theta_c$ in $R_{PC}[u(\theta_c,\arccos(1/\sqrt{3}))] = 1$, see Fig.~\ref{fig:bell_phase_diagram}(e). 
Details of the finite-size scaling analysis are presented in Appendix~\ref{appendix:data-collapse}, from which we extract an estimate of the critical exponent $\nu = 2.3^{+0.6}_{-0.2}$.
This value is close to both the theoretical prediction $\nu=2$ in Ref.~\cite{us_2026_replica} (upcoming work) and the numerical observations in Ref.~\cite{sierant2026theorymagicphasetransitions} studying magic phase transitions in a related model.

\subsubsection{Relationship to encoding-decoding magic transitions}

Here, we comment on an interesting and important relationship between our result on magic localization transitions in deep thermalization and magic phase transitions recently observed in quantum error correction (QEC) circuits~\cite{Niroula_PT_2024,sierant2026theorymagicphasetransitions}. 
In those works, a stabilizer input state $\ket{0}^{\otimes N}$ is evolved under a random global Clifford unitary $U_\text{Cl}$, viewed as the encoder circuit for a random error correction code. Then it is subjected to ``coherent noise'' in the form of single-qubit magic rotations $u(\theta)=\exp(-i\theta Z/2)$ applied uniformly across the system. Lastly, it is evolved under the inverse of the encoding unitary  $U_\text{Cl}^\dagger$, interpreted as the decoder circuit. Subsystem $B$ is then measured in the computational basis (see Fig.~\ref{fig:normal_to_fold}
in Appendix~\ref{appendix:noisy_QEC}) and the outcome recorded; this corresponds to a syndrome measurement for the code, leaving behind a state on subsystem $A$ dependent on the measurement outcome. 

References \cite{Niroula_PT_2024,sierant2026theorymagicphasetransitions} reported that a transition in the magic of the post-measurement state occurs in the limit $N_B\to\infty$ with a {\it finite} ratio $r:=N_A/N_B$, known as the code rate. 
This means that the size of subsystem $A$ (the logical qubits) is asymptotically comparable to $B$ (the syndrome qubits). 
In the {\it zero-rate} limit instead, the encoding-decoding setup was not found to have a stable magic phase\footnote{With the exception of a study by Ref.~\cite{Cheng_Emergent_2025}, which however has other constraints on the nature of the code and the coherent errors.}---though Ref.~\cite{Niroula_PT_2024} saw tantalizing hints of an  onset of magic near a Clifford point. 
On the other hand, we have shown above that deep thermalization magic transitions between {\it robust} magic and non-magic phases do appear in subsystems $A$ which are finite in size while $B$ is scaled to infinity (i.e., precisely at ``zero-rate''). These apparently contrasting observations on the stability of magic phases in finite versus zero-rate codes call for a physical explanation. 

Interestingly, the error correction setup of Refs.~\cite{Niroula_PT_2024,sierant2026theorymagicphasetransitions}  can be understood  as a special case of our framework. The key idea is to ``fold'' the tensor network describing the encoding-decoding dynamics so that it maps to our system of qubits arranged in a ladder geometry capturing  the doubled Hilbert space of operators; 
see Appendix~\ref{appendix:noisy_QEC} for more details on the mapping. 
The coherent noise operator $u(\theta)=\exp(-i\theta Z/2)$ which is applied uniformly to every qubit maps to a product state of rung Bell pairs $|u(\theta)\rangle\!\rangle^{\otimes N}$; while the input state of the QEC circuit becomes a particular (forced) measurement on $B$. The syndrome measurements on $B$ in the original circuit remain as measurements in the mapped circuit. 
A nontrivial subtlety involved in the mapping pertains to the distribution of measurement outcomes (forced measurements versus actual Born-rule measurements on half of the qubits), but we do not expect this to change any qualitative features of the problem.

Now, since the input states and measurements in the QEC setup are both in the computational basis, in the operator-evolution setup they carry exactly 1 bit of Pauli coherence\footnote{The 4 computational basis states on 2 qubits are given by $\sket{\mathbb{I}_2} \pm \sket{Z}$ and $\sket{X} \pm i \sket{Y}$, which are equal superpositions of exactly two Paulis.}. 
This corresponds in our language to a Pauli coherence density of $\alpha_m = 1$, the same as the ``phase basis'' chosen in our previous discussion, so that the predicted magic transition threshold is again given by Eq.~\eqref{eq:magic_crit}, namely one bit of Pauli coherence per $u(\theta)$ gate. 
However, the Pauli coherence $R_{PC}[u(\theta)]$ of a $Z$ rotation $u(\theta) = e^{-i\theta Z/2}$ is the binary entropy of $\cos^2(\theta/2)$, {\it always} below the threshold value of $1$ bit except for the point $\theta = \pi / 2$ (phase gate) where it is exactly marginal, $R_{PC}[u(\pi/2)] = 1$. 
Thus,  it is {\it impossible} for the post-measurement states on $A$ to exhibit a magic phase when subsystem $A$ is finite. This is corroborated by Fig.~\ref{fig:bell_phase_diagram}(f): the slice $\omega=0$ through the  phase diagram  (which corresponds to $Z$-axis rotations) lies entirely outside the magic phase. 
For the range of system sizes $N_B$ plotted though, we do observe a bimodal spike in magic near the Clifford point $\theta = \pi/2$, in line with Ref.~\cite{Niroula_PT_2024}'s observation of an ``onset'' of magic near it. 
However, as argued before, such magic features are not stable and do not survive in the thermodynamic limit.

Our theory thus provides a quantitative explanation of why Ref.~\cite{Niroula_PT_2024} saw the absence of a robust magic phase in {\it zero-rate codes}: the choice of coherent noise being $Z$ rotations provides insufficient Pauli coherence to support a magic phase.
However, our framework also inform us on avenues to overcome such an obstacle. To achieve a stable magic phase transition in zero-rate codes, one has to consider coherent errors that provide larger Pauli coherence. This can simply be achieved for example by our choice of unitaries $u(\theta,\mathbf{n})$, Eq.~\eqref{eq:initstate_magic}. 
Alternatively, if one keeps to coherent $Z$ errors, then one can stabilize a magic phase by having a finite code rate $r$ (which was reported by Refs.~\cite{Niroula_PT_2024,sierant2026theorymagicphasetransitions}). 
We can understand why this works within our framework:
by reducing the number of measurements, the predicted threshold condition shifts to $R_{PC}[u(\theta)] = 1-r < 1$. This threshold {\it can} now be crossed with sufficiently large $Z$ rotations~\cite{Niroula_PT_2024,sierant2026theorymagicphasetransitions}. 
Specifically, for a code rate $r=1/2$, Ref.~\cite{sierant2026theorymagicphasetransitions} numerically estimated the critical angle to be $\theta_c\simeq 0.63(2)$ (using different observables each of which yields slightly different estimates). This is consistent with the theoretical prediction of $\theta_c = 0.676$ arising from our framework, obtained by solving  $R_{PC}[u(\theta_c)] = 1/2$. 

These results showcase the power of our resource localization theory in providing insights beyond its original scope of deep thermalization, namely in the field of quantum error correction. 
While our theoretical framework applies naturally to random codes, it will be interesting to apply the lens of Pauli coherence localizability to understand magic phases in more practically relevant, non-random codes~\cite{Cheng_Emergent_2025}.

\subsection{Non-Gaussianity} \label{sec:nongaussianity}

As a second example of a QRT whose block structure arises in the Heisenberg picture, we consider fermionic non-Gaussianity, which as we will show below exhibits SL behavior. 

Fermionic non-Gaussianity quantifies the deviation of a quantum state from the class of fermionic Gaussian states~\cite{Hebenstreit_AllPure_2019,3yx4-1j27,PhysRevA.97.062337,PhysRevA.98.052350,PhysRevA.97.062337,PhysRevA.98.052350}. While the theory can be expressed in terms of qubits (via e.g., the Jordan-Wigner mapping), we will adopt the fermionic language in this section. 
We consider a system of $2N$ Majorana modes $\{\gamma_i\}$---Hermitian operators obeying the canonical anticommutation relation $\{ \gamma_i, \gamma_j\} = 2\delta_{ij}$. A fermionic Gaussian state is one where the covariance matrix 
\begin{equation}
    M_{ab} = -\frac{i}{2} \bra{\psi} [\gamma_a,\gamma_b] \ket{\psi}
\end{equation}
completely fixes all higher-point correlation functions via Wick's theorem: 
\begin{equation}
    \bra{\psi} \gamma_I \ket{\psi} = i^{|I|/2} \mathrm{Pf}(M|_I),
\end{equation}
where $I$ is any subset of indices containing an even number of indices (the expectation vanishes for an odd number), $M|_I$ is the restriction of $M$ to the rows/columns in the multi-index $I$, and $\mathrm{Pf}$ is the Pfaffian. 

In this QRT, the resource-free operations are fermionic Gaussian unitaries, also known as matchgates in the qubit language~\cite{valiant2002quantum,10.1098/rspa.2008.0189,PhysRevA.65.032325}. A fermionic Gaussian unitary $U$ acts like an orthogonal rotation $\mathcal{O}\in SO(2N)$ on the Majorana operators, i.e.,  $U\gamma_i U^{\dagger}=\sum_j \mathcal{O}_{ij} \gamma_j$. A key feature of such orthogonal rotations is that they preserve the \emph{degree} of Majorana monomials, defined as the cardinality of the multi-index $I$ in a Majorana monomial $\gamma_I$. 
To see this fact, consider a Majorana monomial $\gamma_I$, with $I = \{i_1,\dots i_r\}$ (indices are all distinct and taken in increasing order). Since the $4^N$ Majorana monomials form a basis of operator space, we may expand $U\gamma_I U^{\dagger} = \sum_J c_J \gamma_J$, where the sum in principle runs over all $4^N$ multi-indices $J$. However, since 
\begin{equation} 
U\gamma_I U^\dagger = \prod_{i\in I} U\gamma_i U^\dagger = \prod_{i\in I} \sum_j \mathcal{O}_{ij} \gamma_j,
\end{equation}
it is clear that no monomials of degree $|J|>|I|$ are generated. 
Conversely, monomials of strictly lower degree ($|J|<|I|$) also do not appear. 
If they did, then by using 
\begin{equation}
    c_J = \Tr(\gamma_J U\gamma_IU^\dagger) \neq 0
\end{equation}
and the cyclicity of the trace, it would follow that $U^\dagger \gamma_J U$ contains a monomial of degree $|I|>|J|$, which as we showed is impossible. Therefore, the Majorana degree is an invariant of fermionic Gaussian dynamics. 
Specifically, a monomial $\gamma_I = \gamma_{i_1}\cdots\gamma_{i_r}$ evolves as
\begin{equation}
U\,\gamma_{i_1}\cdots\gamma_{i_r}\,U^\dagger
= \sum_{j_1<\cdots<j_r} \det(\mathcal{O}_{I,J})\,
\gamma_{j_1}\cdots\gamma_{j_r},
\end{equation}
with $J = (j_1,\dots j_r)$ and $\mathcal{O}_{I,J}$ the relevant submatrix of $\mathcal{O}$. 
In other words, the operator space splits into a direct sum of Majorana monomials of fixed degree:
\begin{equation} 
\mathcal{B}_r=\mathrm{Span}(\{\gamma_I:\ |I|=r\}).
\end{equation} 
Superposition between these blocks is necessary but insufficient for an operator to be non-Gaussian. 

Since the natural block decomposition arises in operator space, this QRT fits within our operator-evolution setup. Similar to the case of non-stabilizerness, we introduce a local product basis of operator space with tunable resource. Concretely, we choose the input operator on 4 Majorana modes
\begin{equation} 
e^{-i(\theta/2)\gamma_0 \gamma_1\gamma_2\gamma_3} = \cos \frac{\theta}{2} \mathbb{I} - i\sin \frac{\theta}{2} \gamma_0\gamma_1\gamma_2\gamma_3, \label{eq:fermionic_ng_input}
\end{equation}
corresponding to a unitary of tunable strength $\theta$ that can be used to control the injected resource (the unitary is non-Gaussian if $\theta/\pi \notin \mathbb{Z}$). 
We then evolve the input operator in the Heisenberg picture under a random fermionic Gaussian unitary $U$, followed by measuring it with a free measurement on subsystem $B$---a projection onto a fixed Majorana monomial on $B$. 

As with non-stabilizerness, assigning a concrete physical meaning to this mathematical prescription requires a mapping of the operator space to the state space of a doubled system. 
We vectorize fermionic operators by using the Majorana Bell state $\sket{\Phi} \sbra{\Phi} = \prod_j \frac12 (\mathbb{I} + i\gamma_{j,L} \gamma_{j,R})$, with $L$ and $R$ representing the ``left'' and ``right'' legs of a fermionic ladder. This is a pure Gaussian state on two copies of the system with the property that\footnote{
This is to compared with the qubit Bell state which is invariant under $U\otimes U^\ast$.
} $U_L U_R \sket{\Phi} = \sket{\Phi}$ for any fermionic Gaussian unitary $U$ on a single copy. 
We apply the input unitary Eq.~\eqref{eq:fermionic_ng_input} to the left system, then evolve both sides under the same random fermionic Gaussian unitary and measure the Majorana bilinears $i\gamma_{j,L} \gamma_{j,R}$ for all $j\in B$ (which is a Gaussian operation). 
Finally, we quantify non-Gaussianity of the post-measurement state on $A$ using fermionic antiflatness (FAF)~\cite{3yx4-1j27},
\begin{equation}
R_{NG}(\ket{\Psi}) = N - \tfrac{1}{2}\Tr(M^T M),
\end{equation}
where $M$ is the covariance matrix.

\begin{figure}
    \centering
    \includegraphics[width=1.0\linewidth]{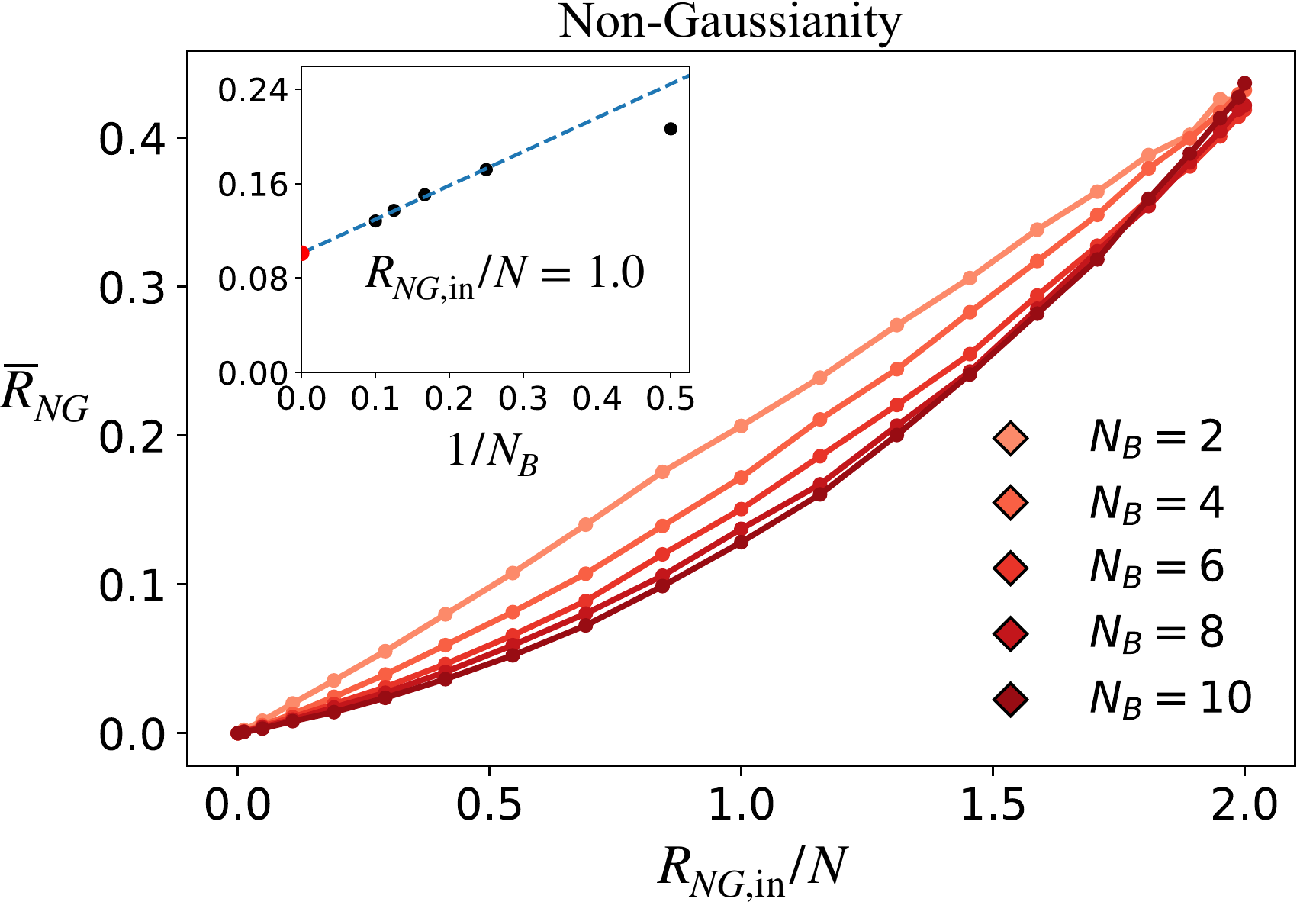}
    \caption{\justifying Absence of localizability threshold in non-Gaussianity. $N_A$ is fixed to be $2$ qubits, i.e., 4 Majorana modes. The curves show the ensemble-average Fermionic antiflatness (FAF) $\overline{R}_{NG}$ of the PE as a function of the input FAF density $R_{NG,\text{in}}/N$. The free evolution $U$ is implemented as a circuit of local Gaussian unitaries of depth $12N$, which gives converged results. Each point is averaged over $10240$ realizations of the dynamics. Inset: finite-size extrapolation at $R_{NG,\mathrm{in}}/N=1.0$ indicates that non-Gaussianity persists in the thermodynamic limit $N_B\to\infty$.
    }
    \label{fig:non_gau}
\end{figure}

Numerical results are shown in Fig.~\ref{fig:non_gau}.
The non-Gaussianity $\overline{R}_{NG}$ of the PE is found to evolve smoothly as a function of the initial resource density $R_{NG,\text{in}}/N$, with no indication of a robust Gaussian phase. We note there is a finite-size downward drift of the PE resource $\overline{R}_{NG}$, but careful analysis (Fig.~\ref{fig:non_gau} inset) indicates saturation to a positive value whenever the input resource is non-zero.

The absence of a Gaussian phase is consistent with our general theory, Sec.~\ref{sec:generaltheory}, and can be explained as follows: since the number of blocks is only polynomial in $N$ (precisely $2N+1$ blocks, of which $O(N^{1/2})$ make up almost all the operator space), the resource is predicted to fall in the SL category. Furthermore, the Majorana degree blocks are formally analogous to the charge blocks encountered in the case of $U(1)$-asymmetry, Sec.~\ref{sec:asymmetry}, which also showed SL behavior. The block dimensions are exactly the same upon substituting $N\mapsto 2N$, and the relationship of block labels (bitstring Hamming weight and Majorana degree, respectively) to the local measurements is also identical. It follows that measurements are generally unable to sharpen an initial Majorana degree superposition into a single Majorana degree block, and so ``Majorana degree coherence'' is generically present in the PE. While this is not sufficient to establish non-Gaussianity, we expect that upon lifting the block-coherence obstruction, the PE becomes resourceful, much like the case of magic. 
The observation of SL behavior in non-Gaussianity, Fig.~\ref{fig:non_gau}, confirms this expectation.

\section{Conclusion and Discussion}
\label{sec:discuss}

\subsection{Summary}
 
We have developed a general theory for deep thermalization in the presence of diverse physical constraints, viewed with the unifying lens of quantum resource theories---i.e., as limitations on the quantum information processing power of the underlying dynamics. 
We showed that the form of the projected ensemble in constrained dynamics depends crucially on the ability of measurements to concentrate the quantum resource content of a global many-body state into a small local subsystem, a property known as resource localizability, which in turn depends sensitively on the nature of the quantum resource theory (QRT) in question.

Our theory yields that many QRTs fall into two distinct classes: smoothly localizable (SL) and threshold localizable (TL) QRTs. 
For SL resources, the resource content of the PE scales smoothly with the overall resource content of the parent many-body state; specifically, the system achieves deep thermalization to wavefunction distributions which smoothly interpolate  between a resourceless ensemble (at zero injected global resource) to the maximally-entropic, resourceful Haar ensemble (with maximal injected global resource). This is the setting in which the maximum entropy principles of deep thermalization hold. For TL resources, in sharp contrast, the resource content of the PE jumps abruptly from zero to near-maximal at a {\it finite} critical density of the input resource, signaling a phase transition between a minimally entropic, resourceless wavefunction distribution and a maximally-entropic, resourceful wavefunction distribution. 
This discontinuity in the behavior of the PE constitutes a breakdown of the maximum entropy principles of deep thermalization, and can be understood as a ``deep ergodicity breaking'' transition. 

We identified the mechanism  behind resource localizability as an information-theoretic phenomenon that we termed ``block sharpening'': viewing each resource as coherence between suitable blocks in (state or operator) Hilbert space, the resource content of the PE is determined by the ability of measurements to collapse, or ``sharpen'', an initial superposition among distinct blocks.
Assuming sufficient scrambling by the random free dynamics, block sharpening can be mapped to the mathematical problem of random subset intersection  and thus solved analytically, underpinning the range of our predictions for different QRTs.
Our analysis predicts that QRTs with extensive ``block probability entropies'' [Eq.~\eqref{eqn:Renyi}], which capture the spread of distributions of block sizes in the QRT,  belong to the TL class; while QRTs with subextensive and intensive block probability entropies  belong to the SL class. 
These predictions are borne out numerically in all cases investigated, with notable results including a novel magic phase transition in zero-rate quantum error correcting codes and the absence of a transition in non-Gaussianity.  

\subsection{Implications}

Beyond their fundamental interest, our results are immediately relevant to protocols in quantum information science. 
In particular, recent works~\cite{Huang_2025_certifying,du2025certifying,varikuti2025resources} have proposed using projected ensembles as tools for the certification of quantum states and quantum resources. 
The idea is simple: rather than addressing a large many-body state directly, one first forms the PE on a local subsystem via partial measurements, and then aims to certify the desired property in the smaller projected states.
This is nontrivial due to the inherent randomness in the PE, which typically requires auxiliary computational resources.
The reason why this route can nonetheless be advantageous is because quantum resources are often diagnosed by global nonlinear observables whose estimation may be very costly, scaling unfavorably with system size; thus reducing the relevant system from a global, extensive many-body system to a local, intensive one presents clear potential benefits.

However, the success of such a certification protocol is predicated {entirely} upon the desired property of the global state being {\it localizable} on the local subsystem.
Our results on resource localizability in deep thermalization hence carry direct implications for this family of protocols and possible extensions thereof. Indeed, as we have seen, the PE does not always accurately reflect the resource content of the parent many-body state. 
In particular, for TL resources, we have identified a robust {\it resource-free} phase of the PE which persists up to a finite threshold density of resource in the parent state. In such a phase, certification of the resource from the PE is impossible\footnote{More carefully, our theory applies to states that are globally scrambled by a free unitary, i.e., typical states with the given resource content. Exceptions to this result may be possible in atypical states.}. 
Above threshold, certification of the resource becomes possible, but its quantitative estimation remains impossible, since all values above threshold yield the same PE (Haar-random). 

For SL resources though, certification is possible. Moreover, the smooth relationship between the resource content of parent state and PE states suggests an interesting direction: to develop protocols that not only certify presence of the resource, but also provide {\it quantitative} bounds on its density. Protocols of this kind are nontrivial due to the intrinsic randomness of measurement outcomes which restricts the observer's ability to query a given state multiple times. 
However, prior work~\cite{PRXQuantum.5.020347,PRXQuantum.5.030311} has shown that general bounds on nonlinear properties of post-measurement states, such as entanglement and negativity, can be systematically improved with better classical models of the experiment. 
Developing analogous bounds for other SL resources would provide a new way to efficiently quantify resources in many-body states, whose investigation we leave to future work. 

\subsection{Outlook}

While our theoretical framework is very broad, encompassing disparate QRTs from coherence to magic and imaginarity, it is still not completely general. 
Specifically, it applies to QRTs that can be mapped to subspace coherence~\cite{aberg2006quantifyingsuperposition,mani_2024_subspacecoherence} for a certain subspace decomposition of the Hilbert space (or more generally, resources where a type of subspace coherence is necessary, e.g., Pauli coherence for magic).
As we have shown, this category includes all resources of current interest in many-body physics, with the notable exception of entanglement. 
Entanglement does not reduce to any obvious subspace decomposition of either state or operator space, thus evading the ``block sharpening'' picture that underlies our theory. 
Furthermore, the set of free operations for the QRT of entanglement is given by local operations and classical communication (LOCC)~\cite{chitambar2019quantum}; its restriction to unitary operations gives the subgroup $\mathcal{G}_R = U(2)^{\otimes N}$ of tensor-product unitaries. This group does not ``scramble'' information in the way that other resource-free subgroups $\mathcal{G}_R$ encountered in this work do, and does not represent the late-time limit of any nontrivial dynamics. As a consequence, the QRT of entanglement does not fit naturally in the theoretical setting of this work.

Nevertheless, it is interesting to note the existence of similar phenomenology. A natural knob to tune the entanglement of an input state is the depth $t$ of a quantum circuit applied to some product state $\ket{0}^{\otimes N}$. (Note that in this case the circuit is considered a part of the {\it input state}, not of the free ``scrambling'' dynamics.) 
It is well known~\cite{Napp_2022_efficient,Bao_2024_finite,McGinley_2025_Measurement} that the PE undergoes a sharp transition at a constant depth $t$ between a phase with short-range measurement-induced entanglement (MIE) and a phase with long-range MIE, sometimes referred to as a teleportation transition.
The transition is most clearly seen by considering two local, far-apart subsystems $A_1$ and $A_2$, measuring their complement $B = \overline{A_1 \cup A_2}$, and asking about entanglement between $A_1$ and $A_2$ (which are separable before the measurements due to light cone arguments): at depth $t < t_c$ $A_1$ and $A_2$ remain separable, while above threshold they become highly entangled. 
This transition can be seen as an entanglement localizability transition, indicating TL behavior. 
A more thorough understanding of entanglement localizability in relation to our present theory and to measurement-induced entanglement transitions more broadly is left to future work. 

The above discussion on entanglement raises another exciting direction for generalizations of our results. Namely, rather than focusing on the PE for a single local subregion $A$, it would be interesting to analyze PEs of spatially separated subregions $\{ A_i \}$. This physical setting presents interesting questions not only about the overall amount of resource in the PE, but also about its multipartite structure~\cite{korbany_2025_longrangemagic,PhysRevA.111.052443,zhang_extensive_2026}. For example, are there distinct phases where the PE states possess a resource, but only share it among $k$ of the parties? This question is sensitive to locality, and thus requires going beyond global free operations (approximating a late-time limit) and instead considering finite-time dynamics. 
This further motivates other interesting questions relating to the time scales for equilibration to the various limiting forms of the PE uncovered in this work. While we exclusively focused on infinite-time behavior, the issue of dynamics is a very natural and interesting direction for future work. 

Another direction for future generalizations of our theory is in connection with the phenomenon of Hilbert space fragmentation---the emergence of a block decomposition of Hilbert space from imposing local constraints such as the conservation of higher multipole moments of a charge, or other kinetic constraints, which heavily constrain dynamics~\cite{sala2020fragmentation,Moudgalya2022fragmentation,iadecola2025symmetryfragmentation}. It would be interesting to study the resource theory of ``fragment coherence'' and explore the possibility of deep thermalization transitions under local Hamiltonian or circuit dynamics in models with this property.

Lastly, it remains an outstanding problem to better understand the phenomenon of ``deep ergodicity breaking'' in quantum many-body dynamics. In this work we identified a number of exceptions to the maximum-entropy paradigm of deep thermalization, all of which are understood in terms of the inability of quantum resource to be localized, and explained by an underlying common ``block sharpening'' mechanism. It would be interesting to identify other qualitatively distinct physical mechanisms  evading deep thermalization, which would amount to different universality classes of deep ergodicity breaking.

\acknowledgments{
We thank Yimu Bao, Sam Garratt, Michael Gullans, Andreas Ludwig, Romain Vasseur, Tianci Zhou, and Yuzhen Zhang for useful discussions.
X.~F. was supported by a TQI Postdoctoral Fellowship from the Texas Quantum Institute at UT Austin. 
W.~W.~H.~is supported by the Singapore National Research Foundation (NRF) Fellowship NRF-NRFF15-2023-0008 and through the National Quantum Office, hosted in A*STAR, under its Centre for Quantum Technologies Funding Initiative (S24Q2d0009). 
M.~I.~is supported by the U.S. Department of Energy, Office of Science, Office of Advanced Scientific Computing Research under Award Number DE-SC0025615. 
This research was supported in part by grant NSF PHY-2309135 to the Kavli Institute for Theoretical Physics (KITP). 
}

\bibliography{bibliography}

\appendix 

\section{Compatibility of local and global subspace coherence QRTs \label{app:blocks}}

In Sec.~\ref{sec:generaltheory}, we consider subspace coherence QRTs defined by a block decomposition of the $N$-qubit Hilbert space 
\begin{equation}
    (\mathbb{C}^2)^{\otimes N} = \bigoplus_i \mathcal{B}_i^{(N)}
\end{equation}
and assume that such a resource can be defined consistently for any system size $N$, so that the question of resource localization in subsystems is meaningful. 
This requirement imposes some constraints on the structure of the QRT, specifically, on the block decompositions $\{\mathcal{B}_i^{(N)}\}$ across distinct system sizes $N$. 

To address this question systematically, let us define the equivalence relation among $N$-bit strings of ``belonging to the same block'': $\ket{\mathbf z}\sim \ket{\mathbf z'}$ if and only if $\ket{\mathbf{z}}, \ket{\mathbf{z'}} \in  \mathcal{B}_i^{(N)}$ for some block $i$ of the appropriate $N$-qubit Hilbert space; 
then, we demand that 
\begin{equation}
\ket{\mathbf z}\sim \ket{\mathbf z'} \iff \ket{\mathbf z}\ket{0} \sim \ket{\mathbf z'}\ket{0} \iff \ket{\mathbf z}\ket{1} \sim \ket{\mathbf z'}\ket{1}.
\label{eq:selfconsistency_blocks}
\end{equation}
This ensures that no resource is gained or lost by adding or removing incoherent qubit states. 

It is easy to verify that Eq.~\eqref{eq:selfconsistency_blocks} holds for all examples of subsystem coherence QRTs used in this work:
\begin{itemize} 
\item {\bf Coherence.} Each $\mathbf{z}$ forms a block unto itself, so $\ket{\mathbf z} \sim \ket{\mathbf z'} \iff \mathbf z = \mathbf z'$. Eq.~\eqref{eq:selfconsistency_blocks} follows immediately. 
In this case the local QRT is simply coherence for the local subsystem. 

\item {\bf $U(1)$ asymmetry.} Blocks $\mathcal{B}_Q^{(N)}$ are defined by the Hamming weight $Q = \sum_i z_i$. Clearly this satisfies Eq.~\eqref{eq:selfconsistency_blocks}. The local QRT is again $U(1)$-asymmetry for the local $U(1)$ symmetry action, with blocks labeled by the local Hamming weight $Q_A$, satisfying $Q = Q_A + Q_B$. 

\item {\bf $\mathbb{Z}_2$ asymmetry.} Analogous to the $U(1)$ case, upon taking the Hamming weight parity: $\pi = (-1)^{\sum_i z_i}$. The local QRT is $\mathbb{Z}_2$ asymmetry for the local $\mathbb{Z}_2$ symmetry action, with blocks labeled by the local parity $\pi_A$, satisfying $\pi = \pi_A \pi_B$. 

\item {\bf Syndrome coherence.} Each parity check defines an independent $\mathbb{Z}_2$ symmetry; this is effectively the QRT of $\mathbb{Z}_2^k$ asymmetry, with $k = \sigma N$. Blocks are labeled by the syndrome bits $\boldsymbol{\pi} \in \{0,1\}^k$, given by the action of the parity check matrix on each bitstring: $\boldsymbol{\pi} = H \mathbf z$. The local QRT is obtained by keeping only the relevant bits of $\boldsymbol{\pi}$, i.e., the relevant columns of $H$. The local syndromes obey $\boldsymbol{\pi} = \boldsymbol{\pi}_A  + \boldsymbol{\pi}_B$, with the addition taken modulo 2. For a random choice of parity check matrix $H$, the restriction $H|_A$ has rank $\min(N_A,\sigma N)$ with high probability; thus if $N_A < \sigma N$, $H|_A$ can be brought (by row reduction) into diagonal form, $(H|_A)_{ij} =\delta_{ij}$ for $0\leq i<\sigma N$, $0\leq j < N_A$. Then $\boldsymbol{\pi}_A = \mathbf z_A$, and the local QRT reduces to coherence. 
\end{itemize}

Another mathematical aspect of the block decomposition with important physical consequences is the ``fusion'' of blocks.
If we take a composite bitstring $\ket{\mathbf z} = \ket{\mathbf z_A} \ket{\mathbf z_B} \in \mathcal{B}_i^{(N)}$, we may ask which blocks $\ket{\mathbf z_A}$ and $\ket{\mathbf z_B}$ came from in their respective subsystem Hilbert spaces. If the answer is always unique given the block $\mathcal{B}_i^{(N)}$, we say the QRT has {\it injective block fusion}. An example of this class is coherence, where each $\mathbf z$ labels its own block, and so $\mathcal{B}_{\mathbf z}^{(N)} = \mathcal{B}_{\mathbf z_A}^{(N_A)} \otimes \mathcal{B}_{\mathbf z_B}^{(N_B)}$: the local blocks can be uniquely inferred from the global block. 
Conversely, a resource has {\it non-injective block fusion} if there exist states such that $\ket{\mathbf z_A}\ket{\mathbf z_B} \sim \ket{\mathbf z'_A} \ket{\mathbf z'_B}$ with $\ket{\mathbf z_A} \not\sim \ket{\mathbf z'_A}$ and $\ket{\mathbf z_B} \not\sim \ket{\mathbf z'_B}$. This means, informally, that the block label admits ``local quantum fluctuations''. 
The prototypical example is $U(1)$ charge, where even within a global charge sector $Q$, local subsystems may host different charges $Q_A$, $Q_B$ subject to the constraint that $Q_A + Q_B = Q$. 

Non-injective block fusion is necessary for the emergence of so-called {\it entanglement asymmetry}~\cite{ares2023entanglement,Liu_2024_symmetryrestoration,Ares_Entanglement_2025}: the fact that a local reduced density matrix of a global symmetric pure state may be asymmetric (or more specifically, {\it weakly symmetric}~\cite{Buca_2012,Lee_Quantum_2023,Lessa_Strong_2025,Lee_symmetryprotected_2025}, i.e., a classical mixture of states in different charge sectors).
Another facet of the same phenomenon is that non-injective block fusion allows measurements to inject resource in the PE of an otherwise free state. 
Indeed, in a QRT with injective block fusion (like coherence), any free pure state on $AB$ lives in a tensor product of local blocks, $\mathcal{B}_i^{(N_A)} \otimes \mathcal{B}_j^{(N_B)}$; any projective measurement on $B$, whether free or resourceful, leaves behind a projected state in block $\mathcal{B}^{(N_A)}_i$, which is free. 
In contrast, in a QRT with non-injective block fusion, a resourceful measurement on $B$ can produce a superposition of blocks on $A$---the resource in the measurement basis on $B$ can ``teleport'' to $A$. 
As a prototypical example, the two-qubit Bell state $(\ket{00} + \ket{11})/\sqrt 2$ is free with respect to the QRT of $\mathbb{Z}_2$ asymmetry (i.e., it lives in the even-parity sector); yet measuring one qubit in the $X$ basis yields a resourceful $\ket{\pm}$ state on the other qubit. 

In this work we frequently refer to the resource injected by the measurements---see e.g., Fig.~\ref{Fig:1}. Strictly speaking, this language is appropriate only for QRTs with non-injective block fusion. However, we adopt it more generally also for QRTs with injective block fusion, like coherence or Pauli coherence.
This choice is justified for two reasons:
(i) measurements in a free basis are more effective at destroying the resource, so one may say that resourceful measurements ``inject resource'' relative to the baseline of free measurements;
(ii) the resource thresholds we derive are symmetric under the exchange of input state and measurement basis, so that it is natural to describe both with the same language. Furthermore, in the replica theory developed in Ref.~\cite{us_2026_replica} for coherence and magic, the ``arrow of time'' is irrelevant, so that the initial state and measurement basis enter the theory exactly on the same footing.

\section{Conversion from mixed basis to tilted basis \label{app:basis}}

In Sec.~\ref{sec:generaltheory}, we formulated a general theory of resource localizability in deep thermalization for subspace-coherence QRTs focusing on initial states and measurement bases within the mixed-basis model. We expect this theory to apply to more general states. In this section, we explain how to convert the quantitative  predictions of Sec.~\ref{sec:generaltheory} 
to states belonging to the tilted-basis model, where the initial state and measurement basis is composed of a uniform product state wherein each qubit is rotated from a free local basis by angles $\theta_0,\theta_m$ respectively. Concretely, we explain here how to achieve a quantitative correspondence between the parameters of the mixed-basis model $\alpha_0,\alpha_m$ and the parameters of the tilted-basis models $\theta_0,\theta_m$, allowing us to directly utilize the threshold conditions derived in the former for the latter. 
This section largely follows the End Matter discussion of Ref.~\cite{liu2025coherence}.

We take the initial state to be of the form
\begin{equation}
    |\Psi_0\rangle\,=\,(\cos (\theta_0/2)|0\rangle + e^{i\phi_0}\sin(\theta_0/2)|1\rangle)^{\otimes N},\label{tilted_basis_psi0}
\end{equation}
Similarly, measurements on $B$ are performed in a uniform local basis specified by the Bloch-sphere direction $\hat{n} = (\sin\theta_m\cos\phi_m, \sin\theta_m\sin\phi_m, \cos\theta_m)$.
Recall  how our general theory of resource localizability in Sec.~\ref{sec:generaltheory} proceeds: we mapped the structure of wave-function amplitudes in the PE to a random subset-intersection problem involving the initial state,  measurement state, and scrambling unitary. The key quantity is the probability $\mathsf{Prob}(\mathcal{I}\cap\mathcal{F}=\emptyset)$, where $\mathcal{I}$ and $\mathcal{F}$ denote the sets of blocks populated by the initial state and by the measurement basis, respectively. For the mixed-basis model, the analysis is simple: since any  active bitstring present in either state appears with uniform weight, the probability is controlled by the count of the number of distinct global bitstrings appearing in $\ket{\Psi_0}$ and $\bra{\Psi_m}$, and hence by the parameters $\alpha_0$ and $\alpha_m$. By contrast, in the tilted-basis model,  the initial state  Eq.~\eqref{tilted_basis_psi0} and the measurement basis have in general nontrivial nonuniform
amplitudes over computational-basis strings, so a na\"ive counting argument breaks down. 

Nevertheless, the precise analysis can still apply, by realizing that it is the  {\it effective} number of global bitstrings in each state that matters (and not the cardinality of non-zero bit-strings---this is typically maximal). To see this, notice that the initial state of the tilted-basis model admits a decomposition into states with different Hamming-weights 
\begin{equation}
    \ket{\Psi}\propto\sum_{h=0}^{N}
    \binom{N}{h}^{1/2}
    \left(\tan\frac{\theta_0}{2}\right)^h
    \ket{S_h},
\end{equation}
where $\ket{S_h} = \binom{N}{h}^{-1/2} \sum_{\mathbf z: |\mathbf z|=h} \ket{\mathbf z}$ denotes a normalized subset state of Hamming weight $h$. Then the wave-function amplitude $\propto \binom{N}{h}^{1/2} \tan^h(\theta_0/2)$  is sharply peaked near $h = N\sin^2(\theta_0/2)$ with fluctuations of order $O(\sqrt{N})$. Therefore, in the thermodynamic limit, the tilted-basis state is effectively supported on a typical Hamming-weight shell containing $\binom{N}{N\sin^2(\theta_0/2)}$ global bit strings. 
Stirling's approximation gives
\begin{align}
\binom{N}{N\sin^2(\frac{\theta_0}{2}) }\sim \frac{1}{\sqrt{2\pi \sin^2(\frac{\theta_0}{2})\cos^2(\frac{\theta_0}{2})N}} 2^{H(\sin^2(\frac{\theta_0}{2})) N}
\end{align}
where
\begin{align}
H(p) = -p \log_2 p -(1-p) \log_2 (1-p)
\end{align}
is the binary entropy function written with the logarithm in base 2.  
Comparing to the active number of bit-strings of the initial state in the mixed-basis model, $2^{\alpha_0 N}$, we see that an equivalence of active bit-strings between the two models is achieved asymptotically if we equate
\begin{align}
H(\sin^2( \theta_0/2)) = \alpha_0.
\end{align}
By similar reasoning, we equate for the measurement basis,
\begin{align}
H(\sin^2( \theta_m/2)) = \alpha_m.
\end{align}

Consequently, the threshold criterion of our theory $\alpha_0+\alpha_m = \alpha_{\mathrm{crit}}$, derived in Sec.~\ref{sec:generaltheory}, becomes 
\begin{equation}
H\left(\sin^2(\theta_0/2)\right)+H\left(\sin^2(\theta_m/2)\right) = \alpha_\text{crit},
\end{equation}
for the tilted-basis model.
Above, $\alpha_{\mathrm{crit}}$ is still  defined in Eq.~\eqref{eqn:alpha_crit} in terms of the second R\'enyi entropy $H_2(\{ p_i\})$ of the block probabilities (also called the collision entropy), set by the QRT in question.

The above analysis shows that the {\it location} of the critical point associated with resource localizability does not depend on the precise  parametrization or microscopic details of the injected resource across different models of initial states and measurement bases. 
Nevertheless, such details could change the {\it nature} of the critical point in terms of its scaling behavior and critical exponents, etc.
An upcoming work Ref.~\cite{us_2026_replica} will explore this in more detail.

\section{Obstruction to deep thermalization purification transition under generic dynamics\label{app:purity}}

Here we present an obstruction to the formation of a purifying phase (zero mixedness) in deep thermalization under generic dynamics that generate a unitary $2$-design, see Sec.~\ref{sec:mixedness}. 

We consider an initial state $\rho_0$ with purity $P_0$. After evolution by a unitary $U$ drawn from a unitary $2$-design, we measure the subsystem $B$ using single-qubit POVM elements $M_a\equiv\frac{\mathbb{I}+(-1)^a\beta O}{2}$, where $O$ is some Pauli operator and $0\leq\beta\leq1$ (this weak measurement model can be viewed as a ``tilted basis'' model with continuous parameter $\beta$). For a measurement outcome string $m = (a_1,\dots,a_{N_B})$, we denote $M(m)=\bigotimes_{i=1}^{N_B}M_{a_i}$. The (unnormalized) post-measurement state on subsystem $A$ is $\rho_{m, A}=\Tr_B\left[U\rho_0 U^\dagger M(m)\right]$, with outcome probability $p_m = \Tr(\rho_{m,A})$. 

We define the averaged post-measurement purity as
\begin{align}\label{eq:purity_def}
\overline{P}
= \mathbb{E}_{U\sim\mathrm{2-design}}
\sum_m \frac{\mathrm{Tr}\left(\rho_{m,A}^2\right)}{p_m}.
\end{align}

Using Weingarten calculus for a unitary $2$-design, one can easily obtain
\begin{align}
    &\mathbb{E}_Up_m = \frac{1}{d_B},\\
\mathbb{E}_{U}\Tr(\rho_{m,A}^2)=\frac{1}{d_B^2}&\left[\frac{1}{d_A}+P_0\left(\frac{1+\beta^2}{2}\right)^{N_B}\right]+O\left(\frac{1}{d_B^2d}\right).
\end{align}

To bound $\overline P$, we use concentration of the outcome probabilities $p_m$ around $1/d_B$.
We introduce a control parameter $0<\varepsilon<1$ and define the typical set
\begin{align}
    \mathcal T(U)=\left\{m:\;p_m(U)\ge \frac{1-\varepsilon}{d_B}\right\}.
\end{align}
We decompose
\begin{align}
    \sum_m\frac{\Tr(\rho_{m,A}^2)}{p_m}=\sum_{m\in\mathcal T(U)}\frac{\Tr(\rho_{m,A}^2)}{p_m}+\sum_{m\notin\mathcal T(U)}\frac{\Tr(\rho_{m,A}^2)}{p_m}.
\end{align}

For $m\in\mathcal T(U)$, we have
\begin{align}
    \frac{1}{p_m}\le\frac{d_B}{1-\varepsilon},
\end{align}
and hence
\begin{align}
    \sum_{m\in\mathcal T(U)}\frac{\Tr(\rho_{m,A}^2)}{p_m}\le\frac{d_B}{1-\varepsilon}\sum_m\Tr(\rho_{m,A}^2).
\end{align}

For $m\notin\mathcal T(U)$, we use the trivial bound
\begin{align}
    \frac{\Tr(\rho_{m,A}^2)}{p_m}\le p_m,
\end{align}
and hence
\begin{align}
    \sum_{m\notin\mathcal T(U)}\frac{\Tr(\rho_{m,A}^2)}{p_m}\le\sum_{m\notin\mathcal T(U)} p_m\leq\frac{1-\varepsilon}{d_B}\left(d_B-|\mathcal{T}(U)|\right).
\end{align}

Using the second moment of a $2$-design, one can obtain
\begin{align}
    \mathrm{Var}_U(p_m)=\mathbb{E}_U[p_m^2]-\mathbb{E}_U[p_m]^2 =O\left(\frac{(1+\beta^2)^{N_B}P_0}{d_B^2 d}\right).
\end{align}
Chebyshev's inequality then gives
\begin{align}
    \mathrm{Pr}_{U}\left(\left|p_m-\frac{1}{d_B}\right|\geq\frac{\varepsilon}{d_B}\right)\leq O\left(\frac{(1+\beta^2)^{N_B}P_0}{\varepsilon^2 d}\right),
\end{align}
and hence
\begin{align}
    \mathbb E_U\left(d_B-|\mathcal{T}(U)|\right)\leq d_BO\left(\frac{(1+\beta^2)^{N_B}P_0}{\varepsilon^2 d}\right).
\end{align}

Combining both contributions, we have
\begin{align}\label{eq:final_bound_concentration}
\overline P&\le\frac{1}{1-\varepsilon}\left[\frac{1}{d_A}+P_0\left(\frac{1+\beta^2}{2}\right)^{N_B}\right]\nonumber\\
&+O\left(\frac{1}{d}\right)+O\left(\frac{(1+\beta^2)^{N_B}P_0}{\varepsilon^2d}\right).
\end{align}
Choosing $\varepsilon=1/3$ and taking $N_B\to\infty$ with $N_A$ fixed, for any weak measurement $\beta<1$, or for asymptotically small initial purity as $N$ increases, we have
\begin{align}
    \overline P\leq \frac{3}{2d_A}.
\end{align}

Therefore, for any fixed subsystem size $N_A$, the averaged post-measurement purity remains strictly smaller than $1$ in the thermodynamic limit, ruling out the formation of a purifying phase in deep thermalization under generic $2$-design dynamics, except at the singular projective measurement point $\beta=1$ with constant initial purity.

\section{Transition in $U(1)$-asymmetry from nonlocal input states \label{app:u1}}

As discussed in Sec.~\ref{sec:generaltheory}, TL behavior can still arise in the $U(1)$-asymmetry resource theory if the initial state is not a product state, but instead is allowed to have long-range entanglement in the form of a ``cat-like'' superposition of $N^{\beta_0}$ random bitstrings. 
Taking the measurements on $B$ to be in the $Z$ basis except for $\beta_m \log_2(N)$ leaves a number of compatible bitstrings scaling as $N^{\beta_m}$. Our theory predicts a resourceless (charge-sharp) phase when
\begin{equation}
    \beta_0+\beta_m < \frac{1}{2}.
\end{equation}
(The right hand side comes from the scaling $\sim N^{-1/2}$ of the second R\'enyi block entropy.)

To test this prediction, we consider the following setup. The initial state is chosen to be an equal-weight superposition of $N^{\beta_0}$ random bitstrings. The dynamics is given by
$U=U_\pi \bigoplus_Q V_Q$,
as in Sec.~\ref{sec:asymmetry}, with $U_\pi$ a random transposition of each pair of blocks $Q\leftrightarrow N-Q$. 
We then measure the entire subsystem $B$ in the $z$ basis, corresponding to $\beta_m=0$. In this case, our theory predicts a threshold at $\beta_0=1/2$.

\begin{figure}
    \centering
    \includegraphics[width=1.0\linewidth]{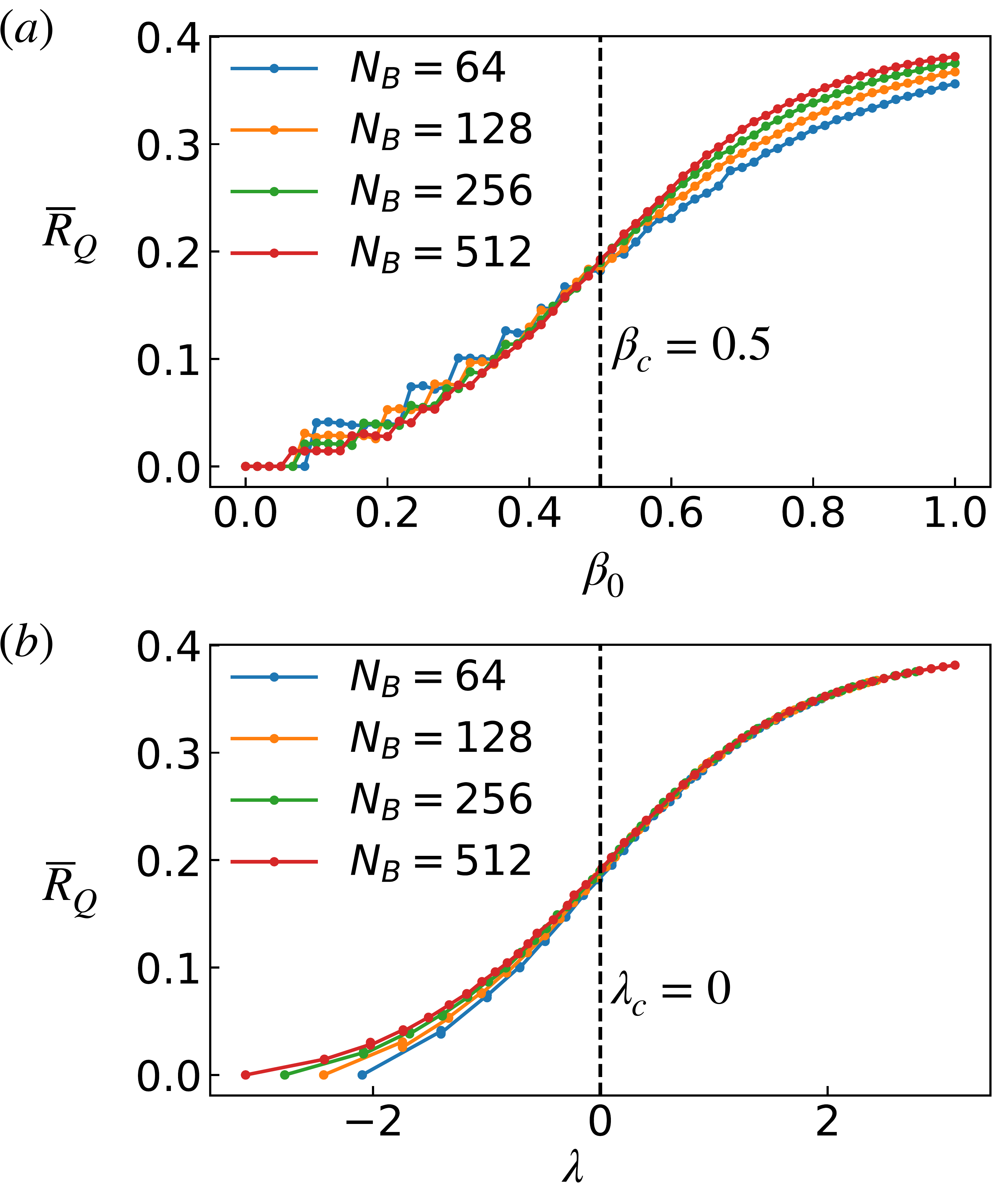}
    \caption{\justifying  
    Hidden threshold in $U(1)$-asymmetry. We choose $N_A=2$ and the initial state as equal weight superposition of $N^{\beta_0}$ random bitstrings. Each data point is obtained by averaging over $2000$ initial states and $60000$ realizations of the dynamics for each initial state.
    (a) Average charge variance of the PE states, $\overline{R}_Q$, versus $\beta_0$. The predicted threshold value $\beta_c = 0.5$ is indicated by the vertical dashed line. The plateaus observed at small $\beta_0$ are due to the rounding of $N^{\beta_0}$ to an integer. 
    (b) Collapse of $\overline{R}_{Q}$ as a function of the scaling variable $\lambda := (\beta_0-1/2)\ln(N)$. The critical point $\lambda_c = 0 $ is shown by the black dashed line.}
    \label{fig:U1_transition}
\end{figure}

Because of the relatively slow threshold behavior (the power-law scaling $\sim N^{\beta_0 + \beta_m - \beta_{\rm crit}}$ instead of the exponential scaling $\sim e^{(\alpha_0 + \alpha_m - \alpha_{\rm crit})N}$ encountered in extensive resources), revealing this hidden threshold requires simulating much larger system sizes. To this end, instead of an exact wavefunction simulation, we adopt the ``generalized Scrooge ensemble'' (GSE) ansatz for the PE from Ref.~\cite{chang2025deep} to compute the charge variance of PE states. 
Developed for the case of $U(1)$-symmetric dynamics, the ansatz depends only on the charge distribution $p(Q)$ of the input state and the charge distributions $p(Q_B|\nu)$ of all measurement basis states $\{\ket{\Phi_\nu}_B\}$ on $B$; here however, because the dynamics includes also the charge-non-conserving transpositions $Q\leftrightarrow N-Q$, the input distribution $p(Q)$ is transformed differently in each realization of the dynamics. We randomly sample many such realizations and generate the GSE prediction for each one. 

In Fig.~\ref{fig:U1_transition}(a), we show the numerical results for the charge variance averaged over different choices of random initial states and random realizations of the dynamics as a function of the parameter $\beta_0\in[0,1]$. Different system sizes show a crossing at the predicted value $\beta_0 = 1/2$. 
Expressed as a function of the scaling variable $\lambda := (\beta_0 - 1/2) \ln(N)$, the curves show a good scaling collapse [Fig.~\ref{fig:U1_transition}(b)], supporting the prediction of our theory (Sec.~\ref{sec:generaltheory}).

\section{Finite-size analysis of the magic phase transition}
\label{appendix:data-collapse}

\begin{figure}
    \centering
    \includegraphics[width=1.0\linewidth]{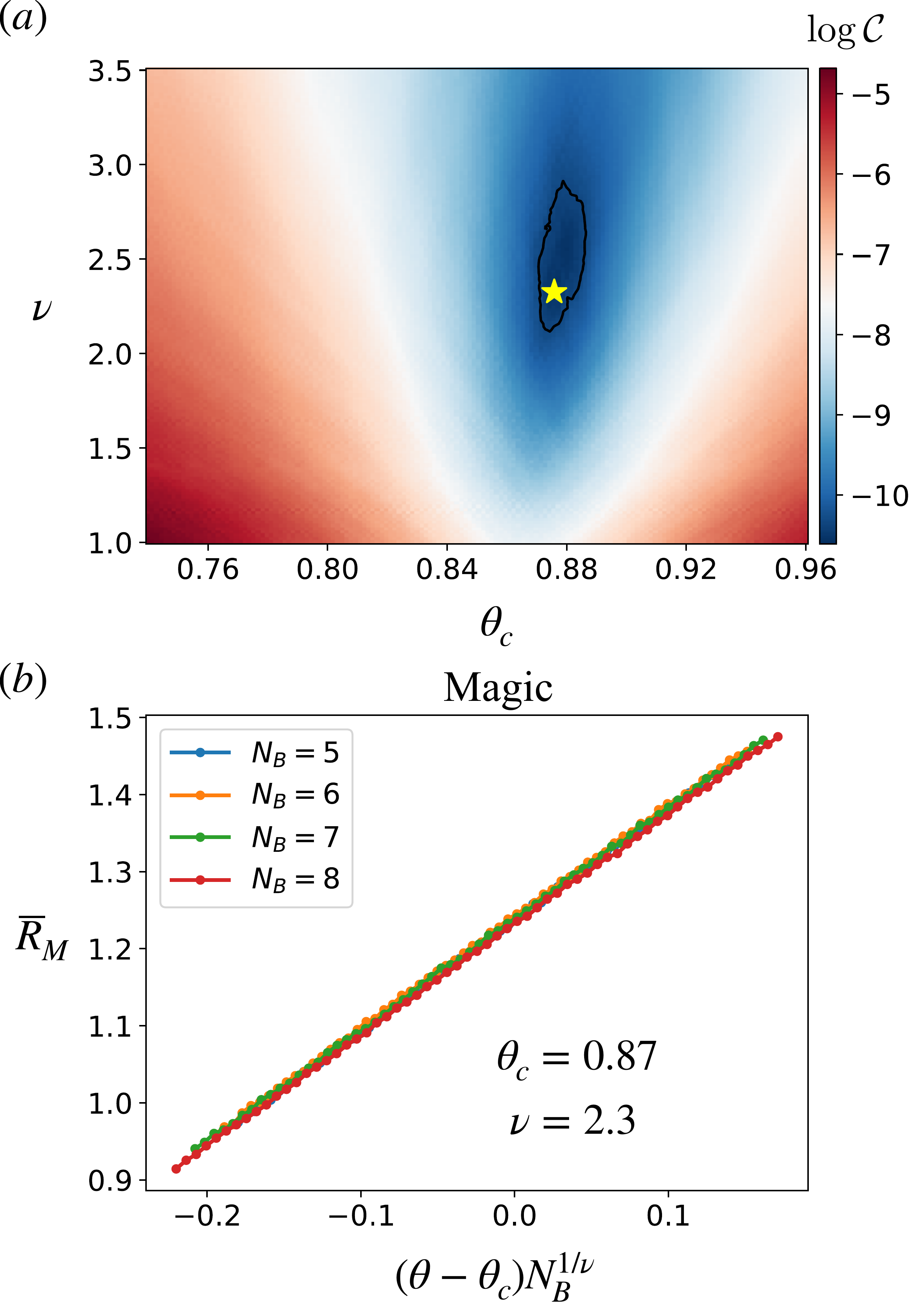}
    \caption{\justifying 
 (a) Color map of the cost function $\mathcal{C}(\theta,\nu)$. The dashed contour marks the region satisfying $\mathcal{C} \le 1.3\, \mathcal{C}_{\mathrm{min}}$, which provides an estimate of the parameters $\theta_c=0.87^{+0.1}_{-0.1}$ and $\nu=2.3^{+0.6}_{-0.2}$. (b) Data collapse of $\overline{R}_{M}$ as a function of $(\theta-\theta_c) N_B^{1/\nu}$, using the fitted parameters indicated in the plot.}
    \label{fig:magic_collapse}
\end{figure}

Here we present details on the calculation of the critical point $\theta_c$ and the critical exponent $\nu$ for the magic phase transition, Sec.~\ref{sec:magic}, via finite-size data collapse. 
As illustrated in Sec.~\ref{sec:magic}, we fix the unit vector $\mathbf{n}$ along the $[1,1,1]$ direction ($\omega = \arccos(\sqrt{3}/3)$ in the parameterization of Fig.~\ref{fig:bell_phase_diagram}). We assume the scaling ansatz
\begin{equation}
R_M(\theta, N_B) = F\!\left[(\theta - \theta_c)\, N_B^{1/\nu}\right],
\label{eq:scaling_ansatz}
\end{equation}
where $F(\cdot)$ is an unknown scaling function. We first construct the rescaled variables
\begin{equation}
x_i = (\theta_i - \theta_c)\, N_{B,i}^{1/\nu}, 
\qquad 
y_i = R_M(\theta_i, N_{B,i}),
\end{equation}
where $i = 1,\dots,n$ labels all data points across different system sizes. The data are then sorted according to $x_i$. We define the cost function
\begin{equation}
\mathcal{C}(\theta_c, \nu) = \frac{1}{n-2} \sum_{i=2}^{n-1} \left( y_i - \bar{y}_i \right)^2,
\label{eq:cost_function}
\end{equation}
where $\bar{y}_i$ is a local linear interpolation of neighboring points,
\begin{equation}
\bar{y}_i = 
\frac{(x_{i+1} - x_i)\, y_{i-1} - (x_{i-1} - x_i)\, y_{i+1}}
{x_{i+1} - x_{i-1}}.
\label{eq:interpolation}
\end{equation}
This procedure measures how well the data collapse onto a single smooth curve without assuming an explicit functional form of $F$. The optimal parameters $(\theta_c, \nu)$ are obtained by minimizing $\mathcal{C}(\theta_c, \nu)$ over a suitable parameter range. To estimate uncertainties, we identify the region in parameter space satisfying
\begin{equation}
\mathcal{C} \leq \kappa\, \mathcal{C}_{\min},
\end{equation}
with $\kappa=1.3$. The spread of $(\theta_c, \nu)$ within this region provides an estimate of the fitting error. In Fig.~\ref{fig:magic_collapse}(a), we show the color plot of $\mathcal{C}$, which shows the range of values of $\theta_c = 0.87(1)$ and $\nu = 2.3_{-0.2}^{+0.6}$ (the latter has strongly asymmetric error bars). 
In Fig.~\ref{fig:magic_collapse}(b), we show the data collapse using the best values $\theta_c = 0.87$ and $\nu = 2.3$.

These results compare well with theoretical predictions. 
Eq.~\eqref{eq:magic_crit} predicts the critical point $\theta_c \simeq 0.900$, while in Ref.~\cite{us_2026_replica}, a statistical mechanics analysis predicts the critical exponent $\nu=2$. 
The mismatch between theoretical values of $(\theta_c,\nu)$ and numerical results indicates a slow convergence of the estimates with system size, consistent with the findings of Ref.~\cite{us_2026_replica} for coherence.

\section{Magic transition in zero-rate QEC codes}
\label{appendix:noisy_QEC}

\begin{figure}
    \centering
    \includegraphics[width=1.0\linewidth]{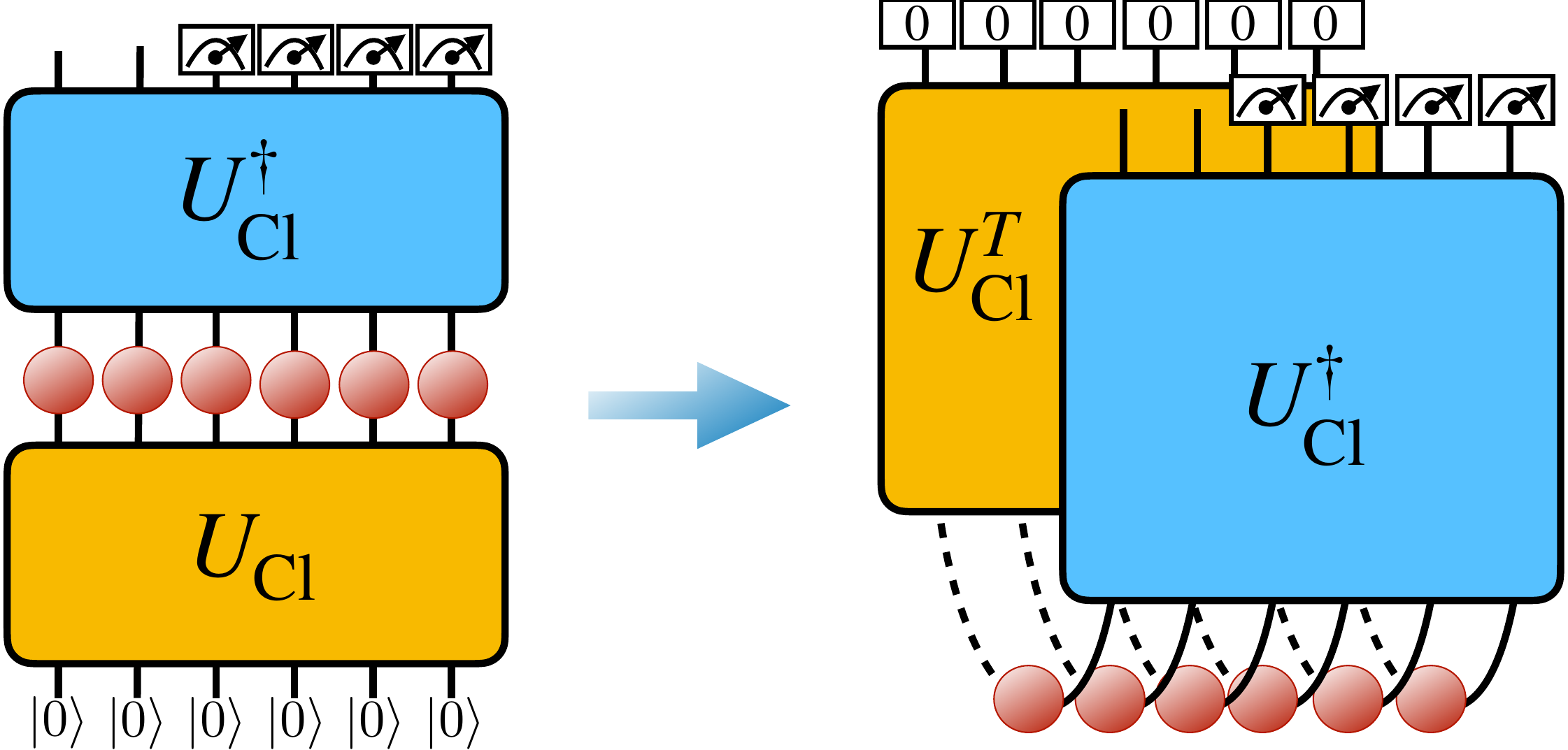}
    \caption{\justifying Mapping a noisy quantum error-correction (QEC) protocol to the operator-evolution picture. The QEC input state $\ket{0}^{\otimes N}$ is interpreted as a forced measurement on the right side. The initial state $\sket{\Psi_0}=\sket{u(\theta)}^{\otimes N}$ in our framework corresponds to the Choi state of the noisy gates $u(\theta)^{\otimes N}$.}
    \label{fig:normal_to_fold}
\end{figure}

As discussed in Sec.~\ref{sec:magic}, our framework for quantum resource-induced deep thermalization transitions naturally applies to magic transitions in encoding-decoding protocols~\cite{Niroula_PT_2024,sierant2026theorymagicphasetransitions}. In this appendix, we make this connection explicit and show how a magic transition emerges in this setting.

We consider an $N$-qubit system initialized in $\ket{0}^{\otimes N}$. A random global Clifford unitary $U$ is first applied, which plays the role of an encoding circuit. The system is then subjected to a layer of coherent noise described by\footnote{The reason for choosing this gate instead of a $Z$-axis rotation is that, as shown in the main text, the latter cannot provide enough Pauli coherence except at the Clifford point $\theta=\pi/2$.} $u(\theta)^{\otimes N}$, where $u(\theta)=\exp\left(-i\frac{\theta}{2}\frac{X+Y+Z}{\sqrt{3}}\right)$. Finally, we apply the decoding unitary $U^\dagger$ and measure the bath subsystem $B$ (i.e., the syndrome) in the computational basis. 
Our object of interest is the ensemble of post-measurement states on subsystem $A$ in the limit $N_B\to\infty$, in particular the average magic of the resulting pure states $\ket{\psi_A}$. Note that we keep $N_A$ finite, representing a zero encoding rate: $r:= N_A/N \to 0$. 

The above setup can be easily mapped to a setup of deep thermalization of Sec.~\ref{sec:magic}, formulated in the Heisenberg picture.
This is illustrated in Fig.~\ref{fig:normal_to_fold}, where the circuit described above is mapped to the model studied in Sec.~\ref{sec:magic} by appropriately ``folding'' the tensor network. The key modification is that the entire right side of the system undergoes forced measurement (i.e., projection) onto the $\ket{0}$ state, reflecting the fixed $\ket{0}^{\otimes N}$ input state of the QEC circuit.
In this representation, the initial state $\sket{\Psi_0}=\sket{u(\theta)}^{\otimes N}$ is the Choi state of the coherent error channel $u(\theta)^{\otimes N}$. 

As we explained in the main text, it is helpful to understand the magic transition through the lens of Pauli coherence. 
Since $B$ is measured in the computational basis, with possible outcome states $\ket{01} \propto\sket{\mathbb{I}} + \sket{Z} $ and $\ket{10} \propto \sket{X}-i\sket{Y}$ (note the right bit is forced to be zero), the Pauli coherence density of the measurement basis is $\alpha_m=1$. 
Thus our general theory predicts a threshold at Pauli coherence density
\begin{equation}
    R_{PC}[u(\theta)] = 1\label{eq:sand_critical},
\end{equation}
with $R_{PC}$ defined in Eq.~\eqref{eqn:PC_op}.

\begin{figure}
    \centering\includegraphics[width=1\linewidth]{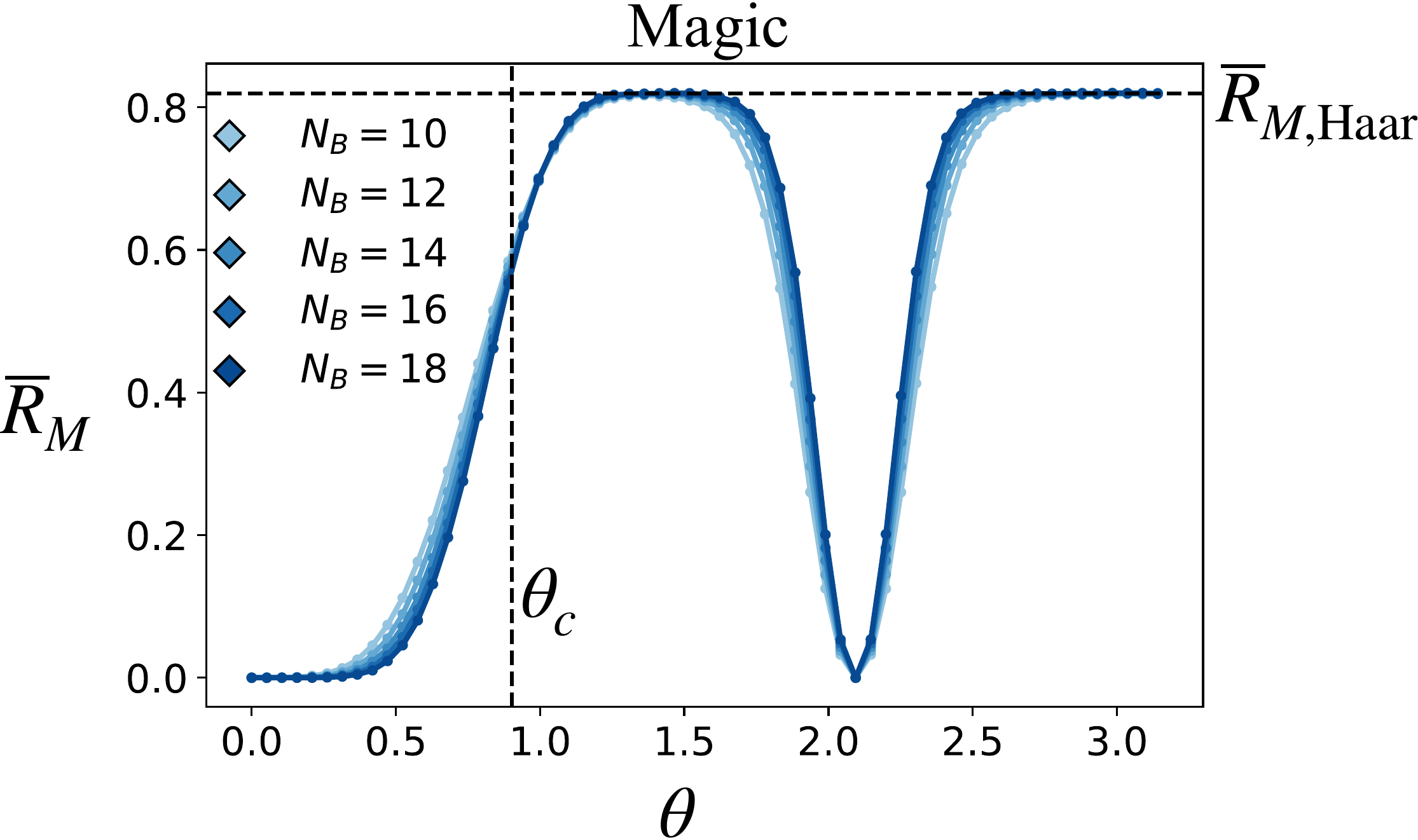}
    \caption{\justifying Average magic of the post-measurement ensemble (PE) for a random Clifford QEC subject to coherent noise. We take the coherent noise $u(\theta,\mathbf n) = e^{-i\frac{\theta}{2} \mathbf{n} \cdot \boldsymbol{\sigma} }$ with the unit vector $\mathbf{n} \propto [1,1,1]/\sqrt{3}$. We fix $N_A=2$. Each data point is averaged over $10240$ realizations. The vertical dashed line indicates the critical point $\theta_c$ predicted by Eq.~\eqref{eq:sand_critical}. The horizontal dashed line marks the average magic of a Haar-random ensemble on subsystem $A$. The approach to this value in the Pauli coherent phase,  except at the isolated Clifford point $\theta = 2\pi/3$, supports the emergence of Haar randomness.}
    \label{fig:sand_finite}
\end{figure}

Below the threshold (when $R_{PC}[u(\theta)] < 1$), our theory predicts a PE made of Pauli operators on $A$, $\sket{P}$. As a result, since the right copy of $A$ is projected onto $\bra{0}^{\otimes N_A}$, the post-measurement state $\ket{\psi}_A$ is a product state in the computational basis, and therefore carries zero magic. This is again the classical bitstring ensemble. 
Above the threshold (when $R_{PC}[u(\theta)] > 1$), by contrast, the resulting PE approaches a Haar-random ensemble on subsystem $A$. This picture is confirmed numerically in Fig.~\ref{fig:sand_finite}, where the average magic increases sharply across the predicted threshold and saturates near the Haar value in the coherent phase. We also observe a slow drift of the apparent critical point, which we attribute to finite-size effects.
From the QEC perspective, this transition can be interpreted as a sharp transition from a \emph{correctable phase}, in which errors are effectively Pauli-like, to a \emph{design phase}, in which errors become Haar-random and hence uncorrectable, in line with recent studies of noisy QEC~\cite{Cheng_Emergent_2025,yan_nonlinear_2026}.

\end{document}